\documentclass{iopart}
\usepackage[left=2cm,top=2cm,right=2cm]{geometry} 

\usepackage{xcolor}
\usepackage{cancel}
\usepackage{dsfont}

\usepackage{iopams}

\usepackage{tikz}
\usetikzlibrary{snakes}
\usetikzlibrary{shapes,arrows}

\xdefinecolor{mylinkcolor}{rgb}{0,0,0.5}
\newcommand{\text}[1]{{\rm #1}}

\newcommand{\LO}  { LO   } %
\newcommand{\nlo} { NLO  } %next-to-leading order~
\newcommand{\nnlo}{ NNLO } %next-to-next-to-leading order~

\newcommand{\eqref}[1]{{(\ref{#1})}}

\newcommand{\smi}{s_{-}}
\newcommand{\spl}{s_{+}}

\newcommand{\pb}[2]{\left\{#1,#2\right\}}

\def\vct#1{{\mathbf{#1}}}
\def\defdby{:=} % LHS is defined by RHS
\def\bydefd{=:} % RHS is defined by LHS

\def\eanl{\\}
\def\neanl{\nonumber\\}

\newcommand{\HAM}[2]{	{\rm H}_{\rm {#1}}^{\rm {#2}}}

\newcommand{\cInv}[1]{ c^{-#1}}

\newcommand{\omfeta}{\sqrt{1-4\eta}}

\newcommand{\diffq}[2]{\frac{{\rm d} #1}{{\rm d}#2}}
\newcommand{\diffql}[2]{{\rm d}#1/{\rm d}#2}
\newcommand{\pdiffq}[2]{\frac{\partial #1}{\partial #2}}

\def\canmomp{p}
\def\canmom{\hat{\canmomp}}
\def\vcanmom{\hat{\vct{\canmomp}}}

\newcommand{\spin}[3]{\hat{S}_{#1\, (#2)(#3)}}
\newcommand{\spinHC}[3]{\tilde{S}_{#1\, (#2)(#3)}}

\newcommand{\SpinUD}[5]{\hat{S}^{~\, #2 \, #3}_{#1\, #4 \, #5}}

\newcommand{\mom}[2]{{\canmom}_{#1\,#2}}

\newcommand{\nxa}[2]{{n}_{#1}^{#2}}
\newcommand{\nunit}[1]{\nxa{12}{#1}}

\newcommand{\nun}[1]{\nunit{#1}} % carries only vector index
\newcommand{\vnunit}{{\vnxa{12}}}
\newcommand{\vmom}[1]{{\vcanmom}_{#1}}
\newcommand{\vnxa}[1]{{\vct{n}}_{#1}}

\newcommand{\scpm}[2]{(#1 \cdot #2)}

\newcommand{\vspin}[1]{\hat{\vct{S}}_{#1}}

\newcommand{\rel}{r}
\newcommand{\dotrel}{\dot{r}}

\def\cInvMT{\cInv{}{}}

\newcommand{\vel}[1]{v^{#1}} % velocity (unexpanded)
\newcommand{\order}[2]{{\cal O}({#1}^{#2})}
\newcommand{\MeAno}{{\cal M}}
\newcommand{\MeMo}{{\cal N}}

\newcommand{\EccAno}{{\cal E}}
\newcommand{\NRG}{|E|}

\newcommand{\ang}{L}
\newcommand{\vAng}{\vct{L}}
\newcommand{\Ang}{L}
\newcommand{\EccNewton}{e_{\rm N}}%{\sqrt{1-2\NRG \ang^2}}
\newcommand{\EccNewtonSq}{e_{\rm N}^2}%{(1-2\NRG \ang^2)}

\newcommand{\ORS}{\sqrt{1-e_r^2}}
\newcommand{\SumChi}	{ \left( \chi_1 + \chi_2 \right) }
\newcommand{\DiffChi}	{ \left( \chi_1 - \chi_2 \right) }

\newcommand{\SPAT}[3]{\left[#1 \cdot \left(#2 \times #3\right) \right]}
\newcommand{\OutXPnn}{	{\cal P}^{(\times,+)}_{nn}}
\newcommand{\OutXPvv}{	{\cal P}^{(\times,+)}_{vv}}
\newcommand{\OutXPnv}{	{\cal P}^{(\times,+)}_{nv}}

\newcommand{\SoneNunit}{ \left( {\hat S}_{1m} \nunit{m} \right) }
\newcommand{\SoneVel}{ \left( {\hat S}_{1m} v^m \right) }
\newcommand{\SoneSq}{ \left( {\hat S}_{1m} {\hat S}_1^m \right) }

\newcommand{\StwoNunit}{ \left( {\hat S}_{2m} \nunit{m} \right) }
\newcommand{\StwoVel}{ \left( {\hat S}_{2m} v^m \right) }
\newcommand{\StwoSq}{ \left( {\hat S}_{2m} {\hat S}_2^m \right) }

\newcommand{\SoneStwo}{ \left[ {\hat S}_{1m} {\hat S}_2^m \right] }
\newcommand{\SoneStwoTensor}{ \left( \spin{1}{j}{k} \spin{2}{j}{k} \right) }  %S_{1jk} {S_2}^{jk} \right) }

\newcommand{\SoneTensorSq}{  \left( \spin{1}{j}{k} \spin{1}{j}{k}  \right)   }

\newcommand{\NunitSoneQNunitStwoQ}{ \left( \nunit{j}\nunit{k} \spin{1}{j}{q} \spin{2}{k}{q} \right) }
\newcommand{\NunitSoneQNunitSoneQ}{ \left( \nunit{j}\nunit{k} \spin{1}{j}{q} \spin{1}{k}{q} \right) }

\newcommand{\PvecSoneQPvecStwoQ}{ \left( p^{j} p^{k} \spin{1}{j}{q} \spin{2}{k}{q} \right) } 
\newcommand{\PvecSoneQPvecSoneQ}{ \left( p^{j} p^{k} \spin{1}{j}{q} \spin{1}{k}{q} \right) }

\newcommand{\NvecSoneQPvecSoneQ}{ \left( \nunit{j} \spin{1}{j}{q} p^{k} \spin{1}{k}{q} \right) }
\newcommand{\NvecSoneQPvecStwoQ}{ \left( \nunit{i} \spin{1}{i}{q} p^{j} \spin{2}{j}{q} \right) }
\newcommand{\NvecStwoQPvecSoneQ}{ \left( \nunit{i} \spin{2}{i}{k} p^{j} \spin{2}{j}{k} \right) }

\newcommand{\NunitPvec}{\scpm{\vct{p}}{\vnunit} }
\newcommand{\PvecSq}{ {(\vct{p}^2)} }

\newcommand{\NunitPvecSumSpinTensor}{ \left( \nunit{i} p^{j} \hat{\Sigma}_{ij} \right) }
\newcommand{\NunitPvecDiffSpinTensor}{ \left( \nunit{i} p^{j} \hat{\Delta}_{ij} \right) }
\newcommand{\NunitPvecSoneTensor}{ \left( \nunit{i} p^{j} \spin{1}{i}{j} \right) }
\newcommand{\NunitPvecStwoTensor}{ \left( \nunit{i} p^{j} \spin{2}{i}{j} \right) }

\newcommand{\default}[1]{#1} % Aus default koennen andere Kommandos gemacht werden - die passenden Stellen sind bekannt.

\begin{document}

\hypersetup{
	pdftitle={Post-Newtonian Approximate Aligned Spins and Dissipative Dynamics},
	pdfauthor={Manuel Tessmer, Johannes Hartung and Gerhard Sch\"afer}
}

\title[Post-Newtonian Approximate Aligned Spins and Dissipative Dynamics]{Aligned Spins: Orbital Elements, Decaying Orbits, and Last Stable Circular Orbit to high post-Newtonian Orders
% Generalisation and Extension of ``dynamics and gravitational waves from compact binaries with
% aligned spins''. Higher order in spin dynamics and decaying orbits
}
\author{M Tessmer, J Hartung, and G Sch\"afer}
%\affiliation{Theoretisch-Physikalisches Institut, Friedrich-Schiller-Universit\"at, Max-Wien-Platz\ 1, 07743 Jena, Germany}
\address{Theoretisch-Physikalisches Institut, Friedrich-Schiller-Universit\"
		at Jena, Max-Wien-Platz 1, 07743 Jena, Germany}
\ead{\mailto{m.tessmer@uni-jena.de}}
\date{\today}

\begin{abstract}
In this article the quasi-Keplerian parameterisation for the case that spins and orbital angular momentum
in a compact binary system are aligned or anti-aligned with the orbital angular momentum vector
is extended to
3PN point-mass, next-to-next-to-leading order spin-orbit, next-to-next-to-leading order spin(1)-spin(2), and next-to-leading order spin-squared
dynamics in the conservative regime. In a further step, we use the expressions for the 
radiative multipole moments with spin to leading order
linear and quadratic in both spins to
compute radiation losses of the orbital binding energy
and angular momentum.
Orbital averaged expressions for the decay of energy and eccentricity are provided.
An expression for the last stable circular orbit is given in terms of the angular velocity type variable $x$.
\end{abstract}

\pacs{04.25.-g, 04.25.Nx, 97.80.-d}

\newcommand{\vSone}{\vct{S}_{1}}
\newcommand{\vStwo}{\vct{S}_{2}}

\newcommand{\CqOne}{{C}_{\rm Q1}}
\newcommand{\CqTwo}{{C}_{\rm Q2}}
\newcommand{\ScSOneNunit}{ \scpm{   \vSone }{  \vnunit   }    }
\newcommand{\ScSTwoNunit}{ \scpm{\vStwo}{\vnunit} }

\newcommand{\ScSOneVel}{ \scpm{\vSone}{\vct{v}}}
\newcommand{\ScSTwoVel}{ \scpm{\vStwo}{\vct{v}}}

\newcommand{\ScSOneSq}{ \scpm{\vSone}{ \vSone } }
\newcommand{\ScSTwoSq}{ \scpm{\vStwo}{ \vStwo } }

\newcommand{\ScSOneSTwo}{ \scpm{\vSone}{\vStwo} }

\newcommand{\ScNunitSOneCrossVel}{  \SPAT{\vnunit}{\vSone}{\vct{v}} }
\newcommand{\ScNunitSTwoCrossVel}{  \SPAT{\vnunit}{\vStwo}{\vct{v}} }

\newcommand{\VelSq}{ {v^2} }
\newcommand{\RadVel}{  {\dot{r}}  }

\maketitle
% \newcommand{\rmd}{{\rm d}}

% \tableofcontents

\section{Introduction}
% \remark{ich wuerde die abkuerzungen auf ein minimum reduzieren, ADM, PN (bei 1PN, usw.) und das wars auch schon}
Recent progresses in the post-Newtonian treatment of compact binary systems  with spinning
components call for an extension of the known parametric solutions to the binary dynamics
to include latest spin interaction terms as well as radiative dynamics.
As computer resources are recently unable to generate thousands of orbits,
which are interesting to ELISA (which may be, optimistically, able in a few decades
to see gravitational waves of orbital periods of a few hours), it is desirable
to have accurate and efficient gravitational wave templates for an analysis of the
detector data.
In a prequel paper \cite{Tessmer:Hartung:Schafer:2010} reasons
can be found why to regard compact binary systems with aligned
spins with regard to their implications to gravitational wave
data analysis.
This article will aim to incorporate higher-order terms to the
in the orbital and spin dynamics to the quasi-Keplerian
parameterisation for compact binaries with aligned spins. These
may become interesting when the binaries come to the final stage
of their life before merger.

Let us state why we concentrate on the case of ``up-up'', ``down-down''
or any mixed alignments of the spins $\vct{S}_1$ and $\vct{S}_2$ with respect to the orbital angular
momentum $\vAng$.
In case that the spins are not aligned, we have to deal with
spin precession equations, whose analytic solutions are not known in general.
A special treatment of, for example, canonical transformations with the help
of Lie series is required to shift oscillatory parts of the precession 
equations of motion to a sufficiently high order in (a special choice of) the perturbation smallness parameter.
That would exceed the aim of this article and will be treated 
in a forthcoming publication.

We confirm our goal to deal with aligned spins by stating that
those sources are the ``loudest'' sources of gravitational waves in the sense of 
Ref. \cite{Reisswig:Husa:Rezzolla:Dorband:Pollney:Seiler:2009}
and are of the high physical importance, because a number of effects
as already listed in Ref. \cite{Tessmer:Hartung:Schafer:2010}
arrange it so that the final configuration of the spins is that of alignment.
At last, aligned spins are, despite the complicated expressions, 
still treatable with the
help of the quasi-Keplerian parameterisation in an analytical manner.

The present publication will provide analytical expressions
for the elements of the quasi-Keplerian parameterisation (QKP)
to all the conservative orders we listed in the abstract, and 
it will also provide first time derivatives of selected orbital
elements due to gravitational wave emission
incorporating leading-order spin-orbit,
spin(1)-spin(2) and spin-squared dynamics.

Let us give a reason why we included S$^2$-effects, although
they may be negligible for neutron stars having spin parameters%
\footnote{$a \defdby S/m$.}
$a \lesssim 0.1 $.
In contrast, the spin parameters of black holes are allowed to be in the
region $a \lesssim 1$, and the spin of non-compact objects 
-- like the sun -- are even larger, say $a \sim 4$ and {\em do}
affect the binary motion.
As we regard rotational deformation, we {\em have to}
include those kinds of effects. Furthermore, although
these interactions are weak compared to point-particle
contributions, they are even stronger than spin(1)-spin(2)
interactions and will lead to modifications in the
long-term
evolution of gravitational-wave signals.

The mathematical context of the orbital elements describing the
motion in the orbital plane will be given 
in Eqs. (\ref{Eq::QKP_radial}) -- (\ref{Eq::vDef}).
For better readability of the article, we list the most important
terms in a small table below.

\begin{table}[hc!]
\label{Tab::Quantities}
\hspace{2.0cm}
{\scriptsize
\begin{tabular}[c]{ l | c | r | r  }
% \hline
Quantity  & Description & Defined in & Result  \\
\hline \hline
	$c^{-1}$		& \dotfill Power counting for post-Newtonian orders, mostly set to 1 & & \\
	$n$PN			& \dotfill $n^{\rm th}$ post-Newtonian order,  $\order{c}{-2n}$ & & \\
	$\delta_{\rm{S}}$	& \dotfill Spin power counting, mostly set to 1 & & \\
% 	-				&	\dotfill set to 1 enables leading order tail contributions\\
	$\chi_a$		& \dotfill Projection of object $a$'s spin onto $\vct{e}_z$: $\chi_a:=(\vct{S}_a\cdot\vct{e}_z )$ & & \\
	$\eta$			& \dotfill Symmetric mass ratio: $\eta:= m_1 m_2/(m_1+m_2)^2$ & & \\
	$x$			& \dotfill Quantity related to orbital angular velocity & & \\
	$|E|$ 			& \dotfill Absolute value of binding energy 				& & \\
	$\ang$ 			& \dotfill Angular momentum of orbit, $\ang:=|\vct{\ang}|$		& & \\
	$\MeAno$		& \dotfill Mean anomaly  						& Eq. (\ref{Eq::QKP_radial}) & \\
	$\MeMo$			& \dotfill Mean motion or radial angular velocity, respectively		& Eq. (\ref{Eq::QKP_Kepler}) &  Eq. (\ref{Eq::Period}) \\
	$\EccAno$		& \dotfill Eccentric anomaly 						& Eq. (\ref{Eq::QKP_Kepler}) & \\
	$v$			& \dotfill True anomaly 						& Eq. (\ref{Eq::vDef}) 	     & \\
	$\phi$			& \dotfill Elapsed phase as function of $\EccAno$			& Eq. (\ref{Eq::QKP_phase})  & \\
	$\vct{S}_a$		& \dotfill Spin vector of object $a$ 					& & \\
	$S_{a (i) (j)}$		& \dotfill Spin tensor of object $a$ , $S_{a (i) (j)} = \frac{1}{2}\epsilon_{ijk}S^k_a$	(harmonic {\em and} canonical) & Eq. (\ref{Eq::SpinTensorToSpinVector}) & \\
	$C_{Q a}$		& \dotfill Quadrupole constant of object $a$ 				& & \\
	${\cal I}_{i_N}$	& \dotfill Mass-type multipole moments					& Eqs. (\ref{Eq::Mass2})-(\ref{Eq::Mass6}) & \\
	${\cal J}_{i_N}$	& \dotfill Current-type multipole moments				& Eqs. (\ref{Eq::Current2})-(\ref{Eq::Current5})& \\
\hline
% & {\bf Results} & given in \\
% \hline
	$a_r$			& \dotfill Semimajor axis						& Eq. (\ref{Eq::QKP_radial}) & Eq. (\ref{Eq::ar})	\\
	$e_{r}$			& \dotfill Radial eccentricity						& Eq. (\ref{Eq::QKP_radial}) & Eq. (\ref{Eq::erSq}) \\
	$e_t$			& \dotfill Time eccentricity 						& Eq. (\ref{Eq::QKP_Kepler}) & Eq. (\ref{Eq::etSq})	\\
	$P \bydefd \frac{2\pi}{\MeMo}$ &  \dotfill Radial period					& Eq. (\ref{Eq::QKP_radial}) & Eq. (\ref{Eq::Period}) \\
	$\Phi$			& \dotfill Total Phase elapsed between 2 successive periastron passages	& Eq. (\ref{Eq::QKP_phase})  & Eq. (\ref{Eq::TotalPhase}) \\
	$e_{\phi}$		& \dotfill Phase eccentricity						& Eq. (\ref{Eq::vDef}) 	     & Eq. (\ref{Eq::ephiSq}) \\
\hline
	$\langle \frac{\rmd \NRG}{\rmd t} \rangle$ 
				& \dotfill Orbital-averaged decay of energy 		& Eq. (\ref{Eq::dNRGdt})		& Eq. (\ref{Eq::dEdt_avraged})  \\
	$\langle \frac{\rmd e_r}{\rmd t} \rangle$ 
				& \dotfill Orbital-averaged decay of eccentricity	& Eq. (\ref{Eq::dNRGdt})	& Eq. (\ref{Eq::derdt_avraged}) \\
% \hline
\end{tabular}
}
\caption{List of used quantities in this article: definitions and where to find results.}
\end{table}

\noindent
The other orbital elements of the Kepler equation (\ref{Eq::QKP_Kepler}), namely
$F_v, F_{2v}, F_{3v}$, and $F_{v-u}$ can be found
in Eqs. 
(\ref{Eq::Fv}),
(\ref{Eq::F2v}),
(\ref{Eq::F3v}),
and (\ref{Eq::Fvmu}),
while
the further elements of the orbital phase, (\ref{Eq::QKP_phase}),
$G_{2v}$, $G_{3v}$, $G_{4v}$, and $G_{5v}$
can be found in  Eqs.
(\ref{Eq::G2v}),
(\ref{Eq::G3v}),
(\ref{Eq::G4v}), and
(\ref{Eq::G5v})
in order of appearance.

Let us, for convenience, state some milestones in the recent
literature.
% \remark{Alte Hamiltonians auf vorhergehendes Paper verweisen; NNLO-Sachen hier halt noch zitieren sind ja nur drei oder so}
For literature on point-mass Hamiltonians through 3PN and leading-order
spin-squared and spin-orbit Hamiltonians, we refer the reader to
the introduction of our previous paper \cite{Tessmer:Hartung:Schafer:2010}
and start from there.
Note that there were two recent publications concerning the 4PN conservative point-mass dynamics
(see \cite{Foffa:Sturani:2012} for the 4PN Lagrangean and
\cite{Jaranowski:Schafer:2012} for the Hamiltonian in the center-of-mass frame, both as preliminary 
results up to order $G^2$).

% The point-mass (PM) Hamiltonian through 2PN has first been derived in
% \cite{Schafer:1985}
% and can be found to 3PN order in
% \cite{Jaranowski:Schafer:1998, Damour:Jaranowski:Schafer:2001}.
% The leading-order spin-orbit Hamiltonian (labeled LO-SO) has been first derived in 
% \cite{Barker:OConnell:1970},
% together with the leading-order spin(1)-spin(2) (labeled LO-S1S2) Hamiltonian, 
% deduced from quantum theory of gravitation.
The next-to-leading order (NLO) spin-orbit (SO) contributions have been derived in 
\cite{Damour:Jaranowski:Schafer:2008:1} and later in \cite{Hartung:Steinhoff:2010}.
Next-to-next-to-leading order (NNLO) spin-orbit and spin(1)-spin(2) (S(1)S(2)) Hamiltonians are recently derived in 
\cite{Hartung:Steinhoff:2011:1,Hartung:Steinhoff:2011:2} and the corresponding Lagrangian potential 
via the effective field theory formalism in \cite{Levi:2011} and will also be used in our calculation.
The leading-order S(1)S(2) Hamiltonians are available in \cite{Barker:OConnell:1975}
and extended to 
% next-to-leading order
\nlo
in \cite{Steinhoff:Hergt:Schafer:2008:2}.
Spin-squared dynamics (S(1)$^2$, S(2)$^2$) depend on the model -- or more precisely on the equation of state --
of the matter of the constituents of the binary. Depending on its rotational velocity
and its stiffness, the included body will deform and self-induce a spin quadrupole moment
that will start interacting with the binary orbit and re-couple to the gravitational field.
The proportionality factor $C_{{\rm Q}a}$ will represent this issue. It varies from one of black holes
to four for neutron stars (such that it is related to the
constant $\lambda_a$ of object $a$ in Ref.~\cite{Tessmer:Hartung:Schafer:2010}
via $C_{{\rm Q}a}= -2\lambda_a$)
and thus characterises how a body resists rotational deformation.
References \cite{Steinhoff:Hergt:Schafer:2008:1, Hergt:Steinhoff:Schafer:2010:1}
provide the 
% next-to-leading order
\nlo
interactions of this type.

Anyway, as the spins do not precess, the quasi-Keplerian parameterisation 
\cite{Damour:Deruelle:1985, Memmesheimer:Gopakumar:Schafer:2004} 
can be employed to obtain a parametric solution to the dynamics.
When the spins are not aligned, their orientation is not constant. 
There exist several publications about precessing spins, see e.g.
\cite{Apostolatos:1995} for the case of ``simple precession''
(which means circular orbits together with the fact that the angle between
the total angular momentum $\vct{J}=\vAng+\vSone+\vStwo$ and $\vAng$ is conserved),
\cite{Schafer:Wex:1993,Schafer:Wex:1993:err, Konigsdorffer:Gopakumar:2005}
for the case that only one body is spinning or the masses are equal,
and \cite{Tessmer:2009} for circular binaries with unequal masses.

As gravitational waves carry away energy and angular momentum, the
semimajor axis and the orbital eccentricity will suffer a slow decay
in the validity regime of the post-Newtonian approximation.
The radiative losses of compact binaries have been extensively discussed
in the literature. Reference \cite{Thorne:1980} gives a general expression
for the losses due to gravitational waves in terms of the mass and
current-type multipole moments of the binary.
This has been elaborated in general in \cite{Blanchet:1995}
and applied to non-spinning compact binaries in \cite{Gopakumar:Iyer:1997}
through 2PN point particle dynamics, 
and, recently, in \cite{Arun:Blanchet:Iyer:Sinha:2009} to 3PN point particle dynamics.
In Section \ref{Sec::RadiationReaction} we give expressions
for the decay of the orbital energy and the radial eccentricity,
deduced from the spin dependent multipole moments given in \cite{Thorne:1980}
and \cite{Porto:Ross:Rothstein:2010}.

% \newpage
Since we used different approximation levels for the last stable circular orbit calculations,
the quasi-Keplerian parameterisation, and the radiative dynamics 
they will be applied exclusively in the corresponding sections.
This is due to the fact that the conservative and dissipative effects are
known for the spin up to different approximation levels.
% Employing the shorthand notation of the introduction,
% the set of relevant Hamiltonians is given by,
% \begin{eqnarray}
%  \HAM{total}{} =&&
% 	\HAM{N}{PM}
% \nonumber \\ &+&
% 	\HAM{1PN}{PM}
% +	\HAM{2PN}{PM}
% +	\HAM{3PN}{PM}
% \nonumber \\ &+&
% 	\HAM{LO}{SO}
% +	\HAM{NLO}{SO}
% +	\HAM{NNLO}{SO}
% \nonumber \\ &+&
% 	\HAM{LO}{S1S2}
% +	\HAM{NLO}{S1S2}
% +	\HAM{NNLO}{S1S2}
% \nonumber \\ &+&
% 	\HAM{LO}{S1sq}
% +	\HAM{NLO}{S1sq}
% %+	{\HAM{NNLO}{S1sq}}
% \,.
% \end{eqnarray}
The 
% next-to-next-to-leading order
\nnlo
spin-orbit Hamiltonian completes the knowledge
of the dynamics of binary black holes
up to and including 3.5PN for maximally rotating objects. 
For general compact objects like neutron stars the leading
order spin(1)$^3$ Hamiltonians are still missing.
Since the NNLO S(1)S(2) Hamiltonian is at 4PN if both
objects are rapidly rotating, the full post-Newtonian approximate dynamics up to and
including 4PN requires further efforts. 
% We state that for general
% compact objects the leading order spin(1)$^4$ Hamiltonian is only known for black holes
% and has to be generalised to include neutron stars as well.
% Furthermore, the NNLO spin(1)$^2$ Hamiltonian 
% -- which is also at 4PN if the object is maximally rotating 
% and maybe even stronger than NNLO S(1)S(2) -- 
% is completely unknown.
% \remark{aus technischem paper entnommen, umformulieren!}

The structure of the paper is as follows. First of all the determination of the last stable circular orbit and a resummed 
binding energy will be discussed in Section \ref{sec:LSO}. Afterwards in Section \ref{sec:QKP} a quick summary of
the quasi-Keplerian parameterisation for eccentric orbits and the appropriate orbital elements will be given. In Section \ref{sec:dissipative}
the dissipative dynamics and energy and angular momentum loss will be discussed. Finally, in Section \ref{sec:conclusions}
the conclusions and future applications will be provided.
The reader will find short
subsection of the rescaling of several quantities to simplify equations and discussions in the appendix.
As well, we provided a general discussion about the stability of the chosen configuration.

\section{Last Stable Circular Orbit}\label{sec:LSO}

To determine the last stable circular orbit our starting point is the binding energy for circular orbits.
Either one can extract it from a Lagrangian potential constructed from a given metric or one can get the binding
energy from a Hamiltonian in the center-of-mass frame for circular motion. This means in the 
center-of-mass frame Hamiltonian the $p_r = \scpm{\vnunit}{\vmom{}}$ component of the linear momentum must be set to zero
and $\vmom{}^2 = \ang^2/r^2$ holds. For a system with spins we further set the spins to
a configuration in which they are aligned with the orbital angular momentum. For an energy $E(r, \ang)$ the circular orbit is given by
the perturbative solution of
\begin{eqnarray}
 \pdiffq{E}{r}(r,\ang) &=& 0\,,
\end{eqnarray}
for $\ang$. Then $E(r,\ang(r))$ is the binding energy at the circular orbit $r$. Since the radial coordinate
depends on a certain gauge we have to transform it into a 
gauge invariant quantity
(invariant under a large class of gauge transformations, see \cite{Blanchet:Buonanno:Faye:2006, Blanchet:Buonanno:Faye:2006:err, Blanchet:Buonanno:Faye:2006:err:2})
% (invariant up to  a change in the spin supplementary condition) 
given by
\begin{eqnarray}
 x &=& \left(\pdiffq{E}{\ang}(r,\ang(r))\right)^{2/3}\,,
\end{eqnarray}
which is related to the orbital angular velocity, see Table~\ref{Tab::Quantities} in the
introduction and \ref{Sec::DimlessQuantities}. For the Schwarzschild spacetime 
the binding energy for a circular orbit is given by
\begin{eqnarray}
\label{Eq::ESchw}
 E_{\text{Schw}}(x) &=& \frac{1-2x}{\sqrt{1-3x}} - 1\,.
\end{eqnarray}
For a test-spin in a stationary Kerr spacetime
we can also write down this expression for orbital momentum aligned Kerr spin and test-spin to linear order in test-spin
(see \cite{Rasband:1973,Hojman:Hojman:1977,Suzuki:Maeda:1998} for the appropriate potentials\footnote{In the given literature there
are a few typos which were corrected in the appendix of \cite{Steinhoff:Puetzfeld:2012}.}), namely
\begin{eqnarray}
 E_{\text{Kerr,TS}}(x) &=& -1
 - \frac{2}{\sqrt[6]{-a + x^{-3/2}}\sqrt{a + x^{-3/2} - 3 \sqrt[3]{-a + x^{-3/2}}}} \nonumber\\
&& + \frac{x^{-3/2}}{\sqrt{-a + x^{-3/2}}\sqrt{a + x^{-3/2} - 3 \sqrt[3]{-a + x^{-3/2}}}}\nonumber\\
&& + S \frac{\left(-a + \sqrt[3]{-a + x^{-3/2}}\right)x^{3/2}}{\sqrt{-a + x^{-3/2}}\sqrt{a + x^{-3/2} - 3 \sqrt[3]{-a + x^{-3/2}}}\left(a x^{3/2}-1\right)}\,.
\end{eqnarray}
In the post-Newtonian case one can also write down this binding energy,
but we will not provide it here since the expression
is very lengthy and gives no further deep insights into the calculation (see e.g. \cite{Steinhoff:Puetzfeld:2012} for very
recent results). We briefly note that it is a polynomial in $\sqrt{x}$ which corresponds to a post-Newtonian expansion.
From the binding energy for circular orbits $E(x)$ we constructed a quantity $e(x)$ via 
\begin{eqnarray}
  e(x) & = & \left[\frac{(\eta E(x) + 1)^2 -1}{2\eta}+1\right]^2 - 1\,,\label{eq:Etoe} 
\end{eqnarray}
(see e.g. \cite{Damour:Jaranowski:Schafer:2000:3}).
We refer to $e(x)$ as ``modified binding energy'' in contrast to ``binding energy'' in case of $E(x)$.
The modified binding energy $e(x)$ has for a test-mass ($\eta = 0$) moving in a Schwarzschild spacetime 
a polynomial structure in $x$ in the numerator and denominator.
In contrast, this is not true for the binding energy $E(x)$, see Eq.~(\ref{Eq::ESchw}).
For the modified binding energy of a test-spin moving
in the equatorial plane of a stationary Kerr black hole this is not true either,
see below in Eq.~(\ref{Eq::eKerrTS}).
The relation between the $\chi_1$, $\chi_2$ spin
magnitudes and $a$, $S$ is given by $\chi_1 = a$ and $\chi_2 = S/\eta$. Notice that the $1/\eta$ terms are not singular, because 
$1 - \sqrt{1-4\eta} = 4\eta/(1 + \sqrt{1-4\eta})$, which renders these terms well-defined at $\eta=0$. This issue appears due to the fact that $S$ is 
a test-spin and so may not vanish in the limit $\eta\to0$.
Here one has to approximate 
in Kerr spin and test-spin to get a rational structure. The mentioned modified binding energies are given by
\begin{eqnarray}
\label{Eq::eSchwX}
 e_{\text{Schw}}(x) &=& -x \frac{1-4x}{1-3x}\,,\\
\label{Eq::eKerrTS}
 e_{\text{Kerr,TS}}(x) & =& \frac{- \sqrt[3]{- a + x^{-{3}/{2}}} x^{{3}/{2}} 
	    + x^{3} \bigl(a^{2} - 3 a \sqrt[3]{- a + x^{-{3}/{2}}} + 4 \sqrt[3]{- a + x^{-{3}/{2}}}^{2}\bigr)}
	{\left(-1 + a x^{{3}/{2}}\right)\biggl(-1 + x^{{3}/{2}} \Bigl(- a + 3 \sqrt[3]{- a + x^{-{3}/{2}}}\Bigr)\biggl)} \neanl &&
	- 2 S \frac{ x^{3} \left(-a + \sqrt[3]{-a + x^{-3/2}}\right)\left(-1 + 2 \sqrt[3]{-a + x^{-3/2}}x^{3/2}\right) }
	  {\left(-1 + a x^{3/2}\right)^{2}\left(-1 + \left[-a + 3 \sqrt[3]{-a + x^{-3/2}}\right]x^{3/2}\right)}\,.
\end{eqnarray}
In the approximation in Kerr spin and test-spin the modified binding energy reads
\begin{eqnarray}
\label{Eq::eApproxKerrTS}
     e_{\text{approx. Kerr,TS}}(x) &=& -x\frac{
      {1 - 4x} 
      + \frac{8}{3}a x^{3/2} - a^2 x^2 + \frac{8}{3}a x^{5/2} - \frac{10}{9}a^2 x^3 + \frac{4}{9} a^2 x^4}
      {{1-3x} + 4ax^{5/2} - a^2 x^3 - \frac{2}{3}a^2 x^4 {+2 S x^{5/2} - 2 a S x^3 + 2 S x^{7/2} - \frac{2}{3} a S x^4}}\,,
\end{eqnarray}
where one can still identify the terms coming from the test-mass motion in a Schwarzschild spacetime. (Notice
that the test-spin parts in the denominator were implemented by using the geometric series at first order, i.e. $1/(1-x) \approx 1+x$,
to implement the correct pole structure coming from the test-spin.) 
In summary, by using an approximation in Kerr spin and test-spin,
we were able to construct a rational function of $\sqrt{x}$ appearing in Eq.~(\ref{Eq::eApproxKerrTS})
similar to the Schwarzschild case Eq.~(\ref{Eq::eSchwX}).
This rational function in $\sqrt{x}$ can be taken as a starting point
to interpolate between modified binding energy for a test-spin in Kerr spacetime
and
post-Newtonian approximate expression for a gravitating mass orbiting another gravitating mass.

\subsection{Construction of Binding Energy}
% \remark{ordentliche struktur; wie kommt man dahin? graphen! vergleichbar mit \cite{Steinhoff:Puetzfeld:2012}}
% \remark{Hamiltonians fuer circulare bahnen und aligned spins auch in \cite{Steinhoff:Puetzfeld:2012} angegeben, zitieren!}
%\remark{auch graph aus dem oppurg vortrag fuer $x_{\rm LSO}$ verwursten!}
After having an initial guess for the rational modified binding energy we tuned all parts 
of the numerator by $\eta$-dependent coefficients and matched the Taylor expansion
in $\sqrt{x}$ with the post-Newtonian approximated $e(x)$ obtained from the ADM-Hamiltonians 
(3PN point-mass, 
% next-to-next-to-leading order 
\nnlo
%  (NNLO)
spin-orbit, 
% next-to-next-to-leading order
\nnlo
S(1)S(2), 
% next-to-leading order
\nlo
spin-squared).
% Since we only had linear corrections in the test-spin to start with, we tuned the denominator by adding linear-in-test-spin 
% contributions with the opposite sign to get the correct singularity structure of the modified binding energy
% $e(x)$. 
These considerations lead to the expression
\begin{eqnarray}
\fl
 e^{}_{\text{tuned Kerr,TS}}(x) &= -x \biggl[
   1
   -\left(4+\frac{\eta}{3}\right) x
   + a \left(-\frac{2 \eta }{3}+\frac{4}{3}+\frac{4}{3} \sqrt{1-4 \eta}\right) x^{3/2}
   + S\left(-\frac{2}{3}+\frac{4}{3\eta }(1 - \sqrt{1-4 \eta })\right) x^{3/2} \neanl \fl & \quad
   + \left(
	\frac{47 \eta}{12}
	-2 a S
	+a^2 C_{Q1} \left( \eta-\frac{1}{2}-\frac{1}{2}  \sqrt{1-4 \eta }\right)
   \right) x^2 \neanl \fl & \quad
   + a \left(
      \frac{4 \eta^2}{9} - \frac{91 \eta}{18} + \frac{10}{3}
     -\sqrt{1-4 \eta } \left(
       \frac{43\eta}{18} + \frac{2}{3}
     \right)
    \right)x^{5/2} \neanl \fl & \quad
   +S \left(
    \frac{4 \eta}{9}-\frac{55}{18} + \frac{43}{18} \sqrt{1 - 4\eta} + \frac{2}{3\eta} (\sqrt{1-4\eta} - 1)
   \right) x^{5/2} \neanl \fl & \quad
   +\left(
      \frac{\eta ^3}{81}
      -\frac{103 \eta^2}{36}
      -\frac{205 \pi ^2 \eta }{96}
      +\frac{3679 \eta}{72}
    \right) x^3
   +a S \left(\frac{4 \eta}{9}+\frac{10}{3}\right) x^3 \neanl \fl & \quad
   + a^2 \left(
	C_{Q1}\left(-\frac{4 \eta ^2}{3}+\frac{14 \eta}{3}-\frac{7}{6}\right)
	-\frac{10 \eta^2}{9}-\frac{5 \eta }{18}+\frac{1}{9}	
      +\sqrt{1-4 \eta }\left(
	C_{Q1} \left(\frac{7 \eta}{3}-\frac{7}{6}\right)
	+\frac{35 \eta}{18}+\frac{10}{9}
      \right)
    \right) x^3 \neanl\fl  & \quad
   +a \left(
      \frac{95 \eta^2}{6} - \frac{197 \eta}{8} - 2
      +\sqrt{1-4\eta}\left(
	\frac{3\eta^2}{2} - \frac{463 \eta}{24} + 2
      \right)
    \right) x^{7/2} \neanl \fl & \quad
   +S \left(
      \frac{91 \eta}{6}-\frac{559}{24}
      +\sqrt{1-4 \eta } \left(-\frac{3 \eta}{2}+\frac{463}{24}\right)
      +\frac{2}{\eta} (1 - \sqrt{1-4\eta})
    \right) x^{7/2} \neanl \fl & \quad
   +\biggl(
      F(\eta)
      +a S \left(
	  \frac{92 \eta ^2}{27} + \frac{202 \eta}{9} - 8
	  +\frac{8}{3} \sqrt{1-4 \eta }
	  +\frac{16}{3\eta} (1 - \sqrt{1 - 4\eta})
	\right) \neanl \fl & \quad\quad
      +\frac{4}{9} a^2 f_7(\eta)
   \biggr) x^4
 \left.\biggr]\right/
 \biggl[
    1
    -3 x
    +(4 a 
    +2 S) x^{5/2} 
   -(a^2 
   +2 a S) x^3
   +2 S x^{7/2}   
   -\frac{2}{3}( a^2
   + a S )x^4
 \biggr]\,.
\end{eqnarray}
There are no quadratic terms in the test-spin, so there will be no $C_{Q2}$ in the expression. 
However, $S$ is a test-spin and the denominator of $e(x)$ is only valid
in the test-spin limit, hence the results given here are only valid around $\eta\approx0$.
The unknown function $F(\eta)$ has to be fixed by the 
4PN point-mass Hamiltonian later (see \cite{Foffa:Sturani:2012} and
\cite{Jaranowski:Schafer:2012}) and the unknown function $f_7(\eta)$ by the 
\nnlo
S(1)$^2$ Hamiltonian (
$f_7(\eta) \stackrel{ \eta \to 0 }{ \longrightarrow } 1$
is required
to be consistent with the Kerr-limit).

We wish to mention Ref. \cite{LeTiec:Barausse:Buonanno:2012} where self-force corrections to the
binding energy for circular orbits have been computed and also Ref. \cite{Barausse:Buonanno:LeTiec:2012}
where the resulting energy has been compared to post-Newtonian theory and numerical relativity.

\begin{figure}
\begin{center}
\includegraphics{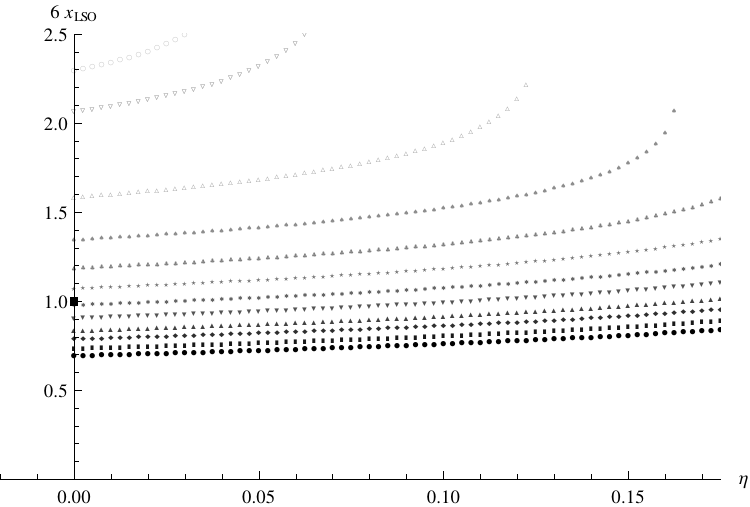}
 \caption{Last stable circular orbit for $S=0.1$ plotted for different symmetric mass ratios $\eta$ and Kerr spins $a$. 
 (From black to light grey $a=-1.0$ to $a=0.84$. The two uppermost plots contain the cases $a=0.8$ and $a=0.84$ respectively. The difference in Kerr spin
 between all other plots is $\Delta a = 0.2$.) The black square ($\blacksquare$) denotes the last stable circular orbit of a testmass orbiting a Schwarzschild black hole.
 Also notice that continuation to large Kerr spins $a$ is
 invalid because the last stable circular orbit will be of the order of magnitude of the Schwarzschild radius which violates
 the post-Newtonian approximation (wide separation). The reader should be reminded that the
 frequency-type quantity $x$
 increases as the radius of a circular orbit decreases.
 }
\label{fig:lso}
\end{center}
\end{figure}

The equation $\diffql{e_{\text{tuned Kerr,TS}}(x)}{x} = 0$ is numerically solved and one can obtain the solution
$x_{{\rm LSO}}$ for certain mass ratios $\eta$, spins and quadrupole constants, see Figure~\ref{fig:lso}.

To link to recent literature, we wish to mention reference
\cite{Damour:Jaranowski:Schafer:2000:3} where the last stable circular orbit 
through third post-Newtonian order for point masses has been computed.
To generalise to certain configurations including spin,
the binding energy of ``last stable spherical orbits'' 
has been derived, for example, in \cite{Damour:2001} in the effective one-body approach
for non-aligned spins.
Compact binaries with spin under 
\nlo
spin-orbit coupling evolving in
circular orbits were studied in \cite{Damour:Jaranowski:Schafer:2008:3}.
We also wish to reference \cite{Barack:Sago:2009} where corrections to
the last stable circular orbit of a Schwarzschild black hole
due to the gravitational self-force have been derived, 
and also \cite{Barack:Sago:2011} where the authors calculated gravitational self-force
corrections to strongly bound eccentric orbits in a Schwarzschild spacetime.

\section{Eccentric Orbits: Calculation Of 
       The Quasi-Keplerian Parameterisation}
  \label{sec:QKP}
\subsection{Included Hamiltonians}
The 
\nnlo
spin-orbit \cite{Hartung:Steinhoff:2011:1},
the
\nlo S(1)S(2) \cite{Steinhoff:Schafer:Hergt:2008}
and 
\nnlo S(1)S(2) \cite{Hartung:Steinhoff:2011:2},
and finally the 
\nlo S(1)$^2$ Hamiltonian \cite{Hergt:Steinhoff:Schafer:2010:1}
in reduced form (in the center of mass)  are listed 
below.
The point-mass and 
\LO spin-orbit, S(1)S(2), and spin-squared Hamiltonians
can be found in \cite{ Tessmer:Hartung:Schafer:2010}.
We define the sums and differences of the two {\em canonical} spin tensors
as
\begin{eqnarray}
 \hat{\Sigma}_{ij} &:=& \spin{1}{i}{j} + \spin{2}{i}{j} \,, \\
 \hat{\Delta}_{ij} &:=& \spin{1}{i}{j} - \spin{2}{i}{j} \,,
\end{eqnarray}
labeled with a ``hat''.
Those satisfying the covariant spin supplementary condition will be
labeled with a ``tilde'', namely
\begin{eqnarray}
 \tilde{\Sigma}_{ij} &:=& \spinHC{1}{i}{j} + \spinHC{2}{i}{j} \,, \\
 \tilde{\Delta}_{ij} &:=& \spinHC{1}{i}{j} - \spinHC{2}{i}{j} \,,
\end{eqnarray}
and the latter will become important for the multipole moment expressions
taken from the literature.
The reduced Hamiltonians read
\begin{eqnarray}
\fl
\HAM{SO}{NNLO} &=&
 \frac{\delta_S}{c^6} \Biggl\{
\frac{1}{\rel^3}
\Biggl[
\left(\frac{275 \eta ^2}{32}-8
   \eta \right) \NunitPvec^2
   \NunitPvecSumSpinTensor+\sqrt{1-4 \eta }
   \NunitPvecDiffSpinTensor \biggl(\frac{-\eta}{32} 
   (45 \eta + 256) \NunitPvec^2
\neanl \fl && %===================================
-\frac{\eta}{32} (39 \eta +314)
   \PvecSq \biggr)+\frac{\eta}{32} (73 \eta -206)
   \NunitPvecSumSpinTensor
   \PvecSq
\Biggr]
% \neanl \fl && %===================================
+\frac{1}{\rel^2}
\Biggl[
\frac{-15}{32} \eta ^2 (2 \eta
   -1) \NunitPvec^4 \NunitPvecSumSpinTensor
\neanl \fl && %===================================
-\frac{3}{32}
   \eta  \left(6 \eta ^2-11 \eta +4\right) \NunitPvec^2
   \NunitPvecSumSpinTensor \PvecSq
% \neanl \fl && %===================================
+\sqrt{1-4 \eta }
   \NunitPvecDiffSpinTensor \biggl(\frac{15 \eta ^2
   \NunitPvec^4}{32}
\neanl \fl && %===================================
+\frac{3}{32} \eta  (9 \eta -4)
   \NunitPvec^2 \PvecSq+\frac{1}{16} \eta  (22 \eta -9)
   \PvecSq^2\biggr)
% \neanl \fl && %===================================
-\frac{1}{16} \eta  \left(5 \eta ^2-3 \eta
   +2\right) \NunitPvecSumSpinTensor
   \PvecSq^2
\Biggr]
\neanl \fl && %===================================
+\frac{1}{\rel^4}
\biggl[
\frac{1}{8} \left(-2 \eta ^2+33
   \eta +42\right) \NunitPvecSumSpinTensor+\frac{21}{4} \sqrt{1-4
   \eta } (\eta +1) \NunitPvecDiffSpinTensor
\Biggr]
\Biggr\}\,,
\eanl \fl %======================================
%======================================
%======================================
\HAM{S(1)S(2)}{NLO}
&=&
\frac{\delta_S^2}{c^4}  
\Biggl\{
\frac{1}{{\rel}^3}
\Biggl[_3
-\frac{15}{2} \eta ^2
   {\NunitPvec}^2 {\NunitSoneQNunitStwoQ}
% \neanl \fl && %===================================
   +{\SoneStwoTensor}
   \frac{\eta^2}{2}
   \left(3 {\NunitPvec}^2+ {\PvecSq}\right)
\neanl \fl && %===================================
-\frac{3}{2} \eta ^2
   {\NunitSoneQNunitStwoQ} {\PvecSq}+\frac{\eta^2}{4}
   {\PvecSoneQPvecStwoQ}
\neanl \fl && %===================================
+{\NunitPvec}
   \frac{3}{4}
   \Biggl(_1\left(
     (2 \eta +1) \eta 
   + \sqrt{1-4 \eta
   } \eta \right) {\NvecSoneQPvecStwoQ}
\neanl \fl && %===================================
+\left(
    \eta (2\eta +1)
   - \sqrt{1-4 \eta } \eta \right)
   {\NvecStwoQPvecSoneQ}\Biggr)_1
\neanl \fl && %===================================
   -\frac{3}{4} (\eta +2) \eta 
   {\NunitPvecSoneTensor}
   {\NunitPvecStwoTensor}
\Biggr]_3
\neanl \fl && %===================================
+\frac{12 \eta }{{\rel}^4}
   {\NunitSoneQNunitStwoQ}-3 \eta 
   {\SoneStwoTensor}
\Biggr\}\,,
\eanl \fl %========================================================================================
%========================================================================================
%========================================================================================
\HAM{S(1)S(2)}{NNLO}
&=&
 \frac{\delta_S^2}{c^6}
\Biggl\{
\frac{1}{{\rel}^4}
\Biggl[_4
\frac{1}{4} \eta ^2
   {\NunitSoneQNunitStwoQ} \left(303 {\NunitPvec}^2+77
   {\PvecSq}\right)+
   \frac{15}{4} \eta ^2 {\PvecSoneQPvecStwoQ}
\neanl \fl && %===================================
+\frac{1}{4} \eta  \left(12 (4 \eta +3)
   {\NunitPvecSoneTensor} {\NunitPvecStwoTensor}-\eta 
   {\SoneStwoTensor} \left(52 {\NunitPvec}^2+23
   {\PvecSq}\right)\right)
\neanl \fl && %===================================
+{\NvecSoneQPvecStwoQ}
   \left(-\frac{3}{2} \sqrt{1-4 \eta } \eta  (\eta +3)
   {\NunitPvec}-\frac{1}{2} \eta  (44 \eta +9)
   {\NunitPvec}\right)
\neanl \fl && %===================================
+{\NvecStwoQPvecSoneQ} \left(\frac{3}{2}
   \sqrt{1-4 \eta } \eta  (\eta +3) {\NunitPvec}-\frac{1}{2} \eta 
   (44 \eta +9)
   {\NunitPvec}\right)
\Biggr]_4
\neanl \fl && %===================================
+\frac{1}{{\rel}^3}
\Biggl[_3
{\NvecSoneQPvecStwoQ} \biggl(\frac{3}{16} \eta  {\NunitPvec} \left(20 \eta ^2
   {\NunitPvec}^2+\left(6 \eta ^2+4 \eta -3\right)
   {\PvecSq}\right)
\neanl \fl && %===================================
   +\frac{9}{16} \sqrt{1-4 \eta } \eta  (2 \eta -1)
   {\NunitPvec} {\PvecSq}\biggr)
   +{\NvecStwoQPvecSoneQ}
   \biggl(\frac{3}{16} \eta  {\NunitPvec} \Bigl(20 \eta ^2
   {\NunitPvec}^2
\neanl \fl && %===================================
   +\left(6 \eta ^2+4 \eta -3\right)
   {\PvecSq}\Bigr)+\frac{9}{16} \sqrt{1-4 \eta } (1-2 \eta ) \eta 
   {\NunitPvec} {\PvecSq}\biggr)
\neanl \fl && %===================================
   +\frac{1}{16} \eta ^2
   {\PvecSoneQPvecStwoQ} \left((4 \eta +1) {\PvecSq}-6 \eta 
   {\NunitPvec}^2\right)
\neanl \fl && %===================================
+\frac{1}{16} \eta  
   \biggl(_1 30 \eta ^2
   {\NunitPvec}^4 {\SoneStwoTensor}+{\PvecSq} \biggl(2 \eta 
   (3 \eta -2) {\PvecSq} {\SoneStwoTensor}
\neanl \fl && %===================================
   -3 \left(4 \eta ^2+17
   \eta -6\right) {\NunitPvecSoneTensor}
   {\NunitPvecStwoTensor}\biggr)
\neanl \fl && %===================================
   -6 \eta  {\NunitPvec}^2 (5 \eta
   {\NunitPvecSoneTensor} {\NunitPvecStwoTensor}-2 (\eta -1)
   {\PvecSq} {\SoneStwoTensor})
   \biggr)_1
\neanl \fl && %===================================
   -\frac{3}{8} \eta ^2
   {\NunitSoneQNunitStwoQ} \left(35 \eta  {\NunitPvec}^4+10
   (\eta -1) {\NunitPvec}^2 {\PvecSq}+(3 \eta -2)
   {\PvecSq}^2\right) 
\Biggr]_3
\neanl \fl && %===================================
+\frac{1}{\rel^5}
\biggl[ _5
\frac{3}{4} \eta  (10
   \eta +7) {\SoneStwoTensor}-\frac{1}{4} \eta  (79 \eta +105)
   {\NunitSoneQNunitStwoQ} 
\biggr]_5
\Biggr\}
\,,
\eanl \fl 
%==============================================================
%==============================================================
%==============================================================
\HAM{S(1)^2}{NLO}
&=&
\frac{\delta_S^2}{c^4}
\Biggl\{
 \frac{1}{\rel^3}
\Biggl[_3
{\NunitPvec}^2
   {\NunitSoneQNunitSoneQ} \left(\frac{15}{8} \CqOne \eta  (2
   \eta -1)-\frac{15}{8} \CqOne \sqrt{1-4 \eta } \eta
   \right)
\neanl \fl && %===================================
   +{\SoneTensorSq} \biggl(_1 {\NunitPvec}^2
   \left(-\frac{3}{16} \eta  (\CqOne (4 \eta +2)+\eta
   -4)-\frac{3}{8} (\CqOne-2) \sqrt{1-4 \eta } \eta
   \right)
\neanl \fl && %===================================
   +{\PvecSq} \left(\frac{3}{8} \sqrt{1-4 \eta }
   ((\CqOne-2) \eta +\CqOne)+\frac{1}{16} (\CqOne (6-2 \eta
    (2 \eta +3))+3 (\eta -4) \eta )\right)\biggr)_1
\neanl \fl && %===================================
   +{\NunitPvec}
   {\NvecSoneQPvecSoneQ} \left(\frac{3}{4} \eta  (-2 \CqOne
   (\eta -2)+\eta -3)+\frac{3}{4} (4 \CqOne-3) \sqrt{1-4 \eta }
   \eta \right)
\neanl \fl && %===================================
   +{\NunitSoneQNunitSoneQ} {\PvecSq}
   \left(\frac{1}{8} (3 \CqOne (\eta +3) (2 \eta -1)-3 (\eta -4)
   \eta )-\frac{3}{8} \sqrt{1-4 \eta } (\CqOne (\eta +3)-4 \eta
   )\right)
\neanl \fl && %===================================
   +{\PvecSoneQPvecSoneQ} \left(-\frac{1}{4} \eta  (3
   \CqOne+\eta -3)-\frac{3}{4} (\CqOne-1) \sqrt{1-4 \eta } \eta
   \right)
\Biggr]_3
\neanl \fl && %===================================
+\frac{1}{\rel^4}
\Biggl[_4
{\NunitSoneQNunitSoneQ}
   \left(\frac{1}{4} (3 \CqOne (\eta +1)-16 \eta +6)+\frac{1}{4}
   \sqrt{1-4 \eta } (\CqOne (9 \eta +3)-4 \eta
   +6)\right)
\neanl \fl && %===================================
   +{\SoneTensorSq} \left(\frac{1}{4} (-\CqOne (\eta
   +1)+3 \eta -1)+\frac{1}{4} \sqrt{1-4 \eta } (-\CqOne (3 \eta
   +1)+\eta -1)\right)
\Biggr]_4
\Biggr\}\,,
\eanl \fl %===================================
\label{Eq::HAMS2SqNLO_switch_particle}
\HAM{S(2)^2}{NLO}
&=& \HAM{S(1)^2}{NLO} \quad (1 \leftrightarrow 2) \,,
\end{eqnarray}
where Equation (\ref{Eq::HAMS2SqNLO_switch_particle}) follows from the fact that $\vnunit$
always appears in a quadratic form and the sign has no influence.
Note that the 
NLO S(1)$^2$ potentials have also been computed in
\cite{Porto:Rothstein:2008:2, Porto:Rothstein:2008:2:err} and the
NLO S(1)S(2) potentials in
\cite{Porto:Rothstein:2008:1, Porto:Rothstein:2008:1:err}
with the help of the effective field theory.
We will incorporate these interactions into the
quasi-Keplerian parameterisation in the subsequent subsection.

\subsection{Geometrical meaning of the elements of the quasi-Keplerian parameterisation}
The quasi-Keplerian parameterisation is the basis of the calculation of the radiation losses. 
Having polar coordinates ($r,\phi$) for the plane
of motion characterised by $\vAng$ and the spin as $\vct{S}_a = \chi_a \vAng/\ang$ for the $a$th object at initial
instant of time, we list the elements of the parameterisation
schematically without stating technical details of the computation. These can be found
in, e.g., \cite{Damour:Deruelle:1985,Memmesheimer:Gopakumar:Schafer:2004}, \cite{ Tessmer:Hartung:Schafer:2010}
and references therein.
This parameterisation describes 
the radial distance $\rel$ 
and the elapsed orbital phase $\phi$
as a function of the eccentric anomaly $\EccAno$
and {\em implicitly} provides $\EccAno$ as a function of time
via the {\em Kepler equation}, see Eq. (\ref{Eq::QKP_Kepler}).
A pictorial description of this parameterisation can be found in Figure~\ref{Fig::Kepler}
below.%
\footnote{
It is obvious that the integral for the {\em radial period} contains
a fifth-order polynomial ${\cal P}_5$ rather than a third-order one, as it has
been stated in Eq. (4.38) of \cite{ Tessmer:Hartung:Schafer:2010}. This is only
part of the description and will
not affect the correctness of the result.
}
Symbolically, the QKP looks as follows:
\begin{eqnarray}
\label{Eq::QKP_radial}
 r		&=& a_r \left( 1- e_r \cos \EccAno \right)\,, \\
\label{Eq::QKP_Kepler}
\MeMo (t-t_0)	&=& \EccAno - {e_t}	
	\sin \EccAno	+ F_{v-\EccAno}(v-\EccAno)	+{F_v} \sin v 
			+ F_{2v}\sin(2v) + F_{3v} \sin(3v)	\label{eq:KE}\,,\\
\label{Eq::QKP_phase}
 \frac{2\pi}{\Phi}(\phi-\phi_0)	&=& v + G_{2v}\sin (2v) 
+ G_{3v}\sin (3v)+ G_{4v} \sin (4v) + G_{5v} \sin (5v)\,, \\
\label{Eq::vDef}
v						&:=& 2 \arctan 
\left[ \sqrt{\frac{1+e_\phi}{1 - e_\phi }} \tan \frac{\EccAno}{2} \right]\,.
\end{eqnarray}
Let us state the importance of these expressions.
To express the gravitational waves emitted by the binary as
expressions of the elapsed time one also requires to implement the orbital
positions and velocities as functions of time, which one might do
with the help of th QKP.
Further, to get a more or less explicit time dependency of the radiation reaction
equations of energy and angular momentum,
it is needed to re-express the luminosity and angular momentum loss
-- which are given in general terms of $\vct{v}^2$, $\dotrel$, and $r$ --
with the help of Equations (\ref{Eq::QKP_radial}) -- (\ref{Eq::vDef}).
Note that the Kepler equation \eref{eq:KE} cannot be inverted by $t-t_0$
without use of infinite series.

\begin{figure}[!hc]
\begin{center}
\includegraphics[scale=0.5]{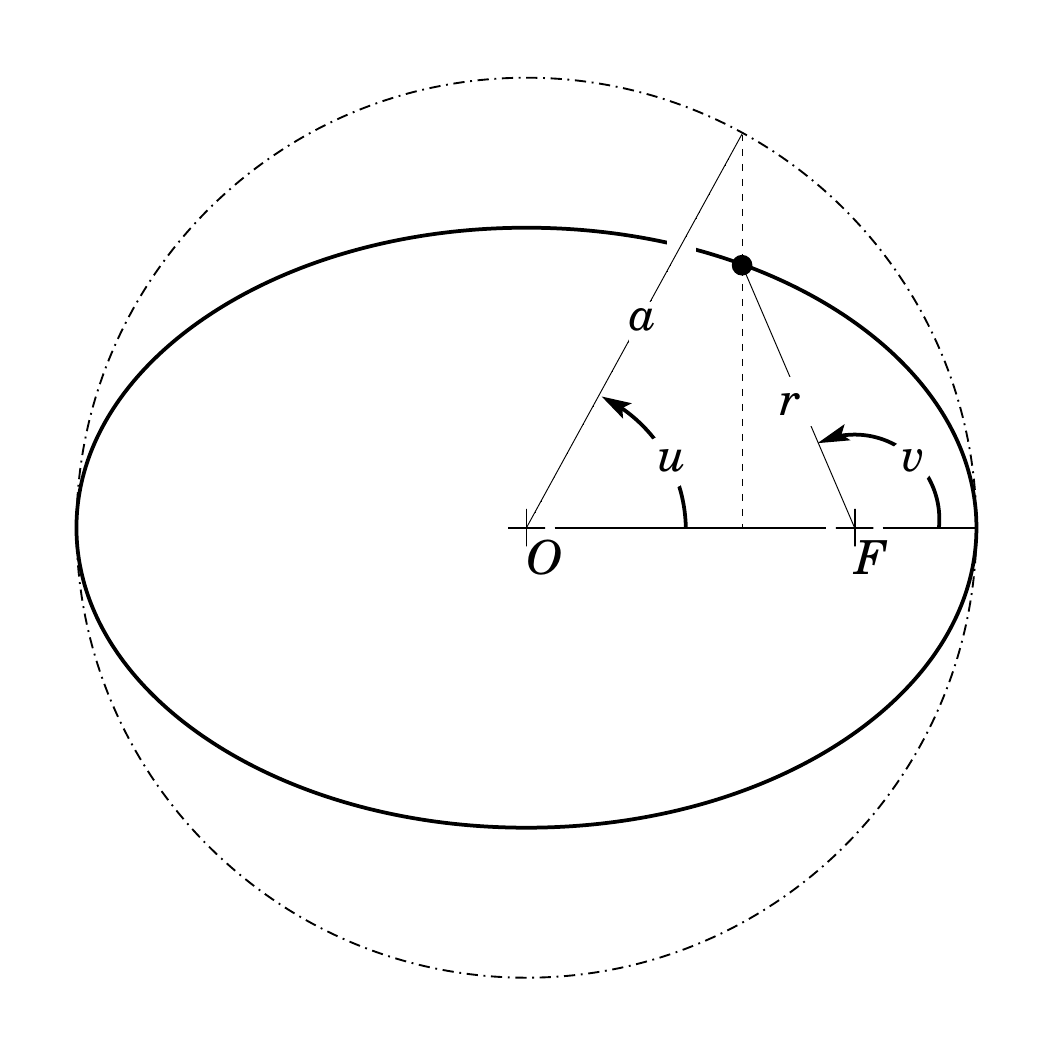}
 \caption{	The motion of the reduced mass (black dot) on an ellipse in the Newtonian case.
		$O$ denotes the origin and $F$ is one focus of the ellipse.
		Note that $v$ is {\em not} identical to the phase $\phi$ in the post-Newtonian case
		and loses its meaning as the angle between $\vnunit{}$ and $\vct{e}_x$.
		The area enclosed by the ellipse, the $x$-axis and the vector $\vct{r}$ in the first
		quadrant equals the quantity $\MeMo(t-t_0)$. This figure is taken from 
		\cite{Memmesheimer:Gopakumar:Schafer:2004} and modified appropriately.
 }
\label{Fig::Kepler}
\end{center}
\end{figure}

\section{Results for the orbital elements}
\subsection{Semimajor axis and eccentricities}
Let us define $\EccNewton$ as the ``Newtonian'' value
of the orbital eccentricity,
\begin{equation}
\fl \EccNewton := \sqrt{1-2\NRG \ang^2} \,.
\end{equation}
Then, the orbital elements read as follows.

\begin{eqnarray}
\label{Eq::ar}
\fl
a_r &=&
\frac{1}{2 \NRG}
%  \nonumber \\ \fl && %========================================
+\frac{1}{c^2}
   \Biggl\{
   \frac{\eta -7}{4}
   \Biggr\}
+\frac{1}{c^4}
   \Biggl\{
    \frac{11 \eta -17}{4\Ang^2}
   +\frac{1}{8} \left(\eta ^2+10 \eta +1\right)
   \NRG
   \Biggr\}
\nonumber \\ \fl && %========================================
+\frac{\delta_S}{c^2 \Ang }
   \Biggl\{
   \sqrt{1-4 \eta } \left(\chi _1-\chi _2\right)+\left(1-\frac{\eta }{2}\right)
   \left(\chi _1+\chi _2\right)
   \Biggr\}
\nonumber \\ \fl && %========================================
+  \frac{\delta_S^2}{c^2 \Ang^2}
   \Biggl\{
    \frac{1}{4} \CqOne (2 \eta -\omfeta-1)\chi_1^2
   +\frac{1}{4} \CqTwo (2 \eta +\omfeta-1) \chi_2^2-\eta 
   \chi_2 \chi_1
   \Biggr\}
\nonumber \\ \fl && %========================================
+
\frac{\delta_S}{c^4}
   \Biggl\{
   \frac{1}{\Ang^3}
   \Biggl[_3
   {\DiffChi} \left(8-\frac{9 \eta }{4}\right)
   {\omfeta}+\left(\eta ^2-\frac{39 \eta }{4}+8\right)
   {\SumChi}
   \Biggr]_3
\nonumber \\ \fl && %========================================
   +\frac{\NRG}{8 \Ang}
   \left[
   {\DiffChi}
   (5 \eta -8) {\omfeta}+\left(-6 \eta ^2+19 \eta -8\right)
   {\SumChi}
   \right]
\Biggr\}
\nonumber \\ \fl && %========================================
+
\frac{\delta_S^2}{c^4}
\Biggl\{
\frac{\chi_1^2}{2} 
   \Biggl(
   \frac{2   }{{\Ang}^4}
   \biggl[_4
   {\CqOne} \left(\frac{1}{4} \left(-6
   \eta ^2+41 \eta -18\right)+\left(\frac{5 \eta
   }{4}-\frac{9}{2}\right) {\omfeta}\right)
   -2 \left(\eta ^2-9 \eta
   +3\right)+(6-6 \eta ) {\omfeta}
%    -2 \left(\eta ^2-\eta
%    +1\right)+6 (\eta -1) {\omfeta}-4 (1-4 \eta
   )
   \biggr]_4
\nonumber \\ \fl && %========================================
   +\frac{2 \NRG}{{\Ang}^2} \left({\CqOne}
   \left((\eta -1)^2+{\omfeta}\right)-\frac{3 \eta 
   {\omfeta}}{2}+\frac{1}{2} (\eta -3) \eta
   \right)
   \Biggr)
\nonumber \\ \fl && %========================================
   +\frac{\chi_2^2}{2} 
   \Biggl(
   \frac{2   }{{\Ang}^4}
   \biggl[_4
   {\CqTwo} \left(\frac{1}{4} \left(-6
   \eta ^2+41 \eta -18\right)+\left(\frac{9}{2}-\frac{5 \eta
   }{4}\right) {\omfeta}\right)-2 \left(\eta ^2-9 \eta
   +3\right)+(6-6 \eta ) {\omfeta}
   \biggr]_4
\nonumber \\ \fl && %========================================
   +\frac{2\NRG }{{\Ang}^2}
   \left({\CqTwo} \left((\eta
   -1)^2-{\omfeta}\right)+\frac{3 \eta 
   {\omfeta}}{2}+\frac{1}{2} (\eta -3) \eta
   \right)
   \biggr)
\nonumber \\ \fl && %========================================
-\left[
\frac{\eta  (\eta +32)}{{\Ang}^4}-\frac{(\eta -4) \eta 
   \NRG}{{\Ang}^2} \right] \chi_1 \chi_2
\Biggr\}
\nonumber \\ \fl && %========================================
+
\frac{1}{c^6}\Biggl\{
   \frac{1   }{4{\Ang}^4}
   \biggl(
   -16 \eta ^2+\left(281-5 \pi ^2\right) \eta -134
   \biggr)
   +\frac{1{\NRG}}{{\Ang}^2}
   \left(\frac{9 \eta ^2}{2}-\frac{1}{96} \left(2212+3 \pi
   ^2\right) \eta +9\right)
\nonumber \\ \fl && %========================================
+\frac{1}{16} \biggl(
   \eta^3-2 \eta ^2-3 \eta +1 \biggr) {\NRG}^2
\Biggr\}
\nonumber \\ \fl && %========================================
+
\frac{\delta_S}{c^6}
\Biggl\{
   \frac{\NRG^2}{\Ang} 
   \left(\frac{1}{16} \sqrt{1-4 \eta } \eta
    (13 \eta -3) \left(\chi _1-\chi _2\right)-\frac{1}{16} \eta 
   \left(14 \eta ^2-21 \eta +5\right) \left(\chi _1+\chi
   _2\right)\right)
\nonumber \\ \fl && %==================================
   +\frac{\NRG}{\Ang^3}
   \left(\frac{1}{4}
   \left(12 \eta ^3-129 \eta ^2+318 \eta -130\right) \left(\chi _1+\chi
   _2\right)-\frac{1}{4} \sqrt{1-4 \eta } \left(29 \eta ^2-149 \eta
   +130\right) \left(\chi _1-\chi _2\right)\right)
\nonumber \\ \fl && %===========================
+\frac{1}{\Ang^5}
\left(
\sqrt{1-4 \eta } \left(4 \eta ^2-69
   \eta +97\right) \left(\chi _1-\chi _2\right)+\left(-\eta ^3+\frac{79
   \eta ^2}{2}-\frac{325 \eta }{2}+97\right) \left(\chi _1+\chi
   _2\right)
\right)
   \Biggr\}
\nonumber \\ \fl && %========================================
+\frac{\delta_S^2}{c^6}
\chi_1 \chi_2
\Biggl\{-\frac{\eta ^2 \NRG^2}{2\Ang^2} (\eta -1) 
   -\frac{\eta \NRG}{2 \Ang^4}  \left(24 \eta ^2+191 \eta -373\right)
   +\frac{\eta }{\Ang^6} \left(13 \eta^2+125 \eta -581\right)
   \Biggr\}
\,,
\end{eqnarray}

\begin{eqnarray}
\label{Eq::etSq}
\fl
e_t^2
&=&
\EccNewton^2
\neanl  \fl && %========================================
+
\frac{1}{c^2}
\Biggl\{
\frac{{\delta_S}^2}{\Ang^2} \NRG \left(\chi_1^2 \left(\CqOne
   (1-2 \eta )+\CqOne \sqrt{1-4 \eta }\right)+4 \chi_1
   \chi_2 \eta +\chi_2^2 \left(\CqTwo (1-2 \eta
   )-\CqTwo \sqrt{1-4 \eta
   }\right)\right)
\neanl  \fl && %========================================
+\Ang^2 (17-7 \eta )
   \NRG^2
   +\frac{{\delta_S} \NRG}{\Ang}
   \left(2 (\eta -2)
   (\chi_1+\chi_2)-4 \sqrt{1-4 \eta }
   (\chi_1-\chi_2)\right)
   +4 (\eta -1) \NRG
\Biggr\}
\neanl \fl && %========================================
+
\frac{1}{c^4}
\Biggl\{
\Ang^2 \left(-16 \eta ^2+47 \eta -112\right)
   \NRG^3
   +\frac{\NRG}{\Ang^2} (17-11 \eta )
% \neanl \fl &&
+\frac{6\NRG^{3/2}}{\Ang} \sqrt{2}
   \EccNewtonSq (2 \eta -5) 
   +2
   \left(5 \eta ^2+\eta +2\right) \NRG^2
\neanl \fl && %==================================
+\delta_S \,  \Biggl[
   \frac{4 \sqrt{2}}{\Ang^2}
   \EccNewtonSq \left(\eta ^2-8 \eta +6\right) \NRG^{3/2}
   (\chi_1+\chi_2)
\neanl \fl && %========================================
+16 \sqrt{2-8 \eta }
   (\eta -3) \NRG^{5/2}
   (\chi_1-\chi_2)
\neanl \fl && %========================================
+\NRG
   \biggl(
   \frac{1}{\Ang^3}
   \left(
   \left(-4 \eta ^2+39 \eta -32\right)
   (\chi_1+\chi_2)+\sqrt{1-4 \eta } (9 \eta -32)
   (\chi_1-\chi_2)
   \right)
\neanl  \fl && %========================================
-\frac{8}{\Ang^2} (\eta -3)
   (\chi_1-\chi_2) 
   \sqrt{(2-8 \eta ) \NRG}
\biggr)
\neanl \fl && %==================================
+\frac{\NRG^2}{\Ang} \left(\left(16
   \eta ^2-\frac{159 \eta }{2}+62\right)
   (\chi_1+\chi_2)+\frac{1}{2} \sqrt{1-4 \eta } (124-59 \eta
   ) (\chi_1-\chi_2)\right)
\Biggr]
\Biggr\}
\neanl \fl && %========================================
+\frac{\delta_S^2}{c^4}
\Biggl\{
   \frac{\NRG^{3/2}}{\Ang^3} \Bigl(\sqrt{2} \chi_1^2
   \left(\CqOne \EccNewtonSq \left(-6 \eta ^2+33 \eta
   -14\right)+\EccNewtonSq \left(-4 \eta ^2+29 \eta
   -8\right)\right)
\neanl \fl && %==================================
+4 \sqrt{2} \chi_1 \chi_2
   \EccNewtonSq (\eta -15) \eta +\sqrt{2} \chi_2^2
   \left(\CqTwo \EccNewtonSq \left(-6 \eta ^2+33 \eta
   -14\right)+\EccNewtonSq \left(-4 \eta ^2+29 \eta
   -8\right)\right)\Bigr)
\neanl \fl && %========================================
+\frac{\NRG^2}{\Ang^2}
   \Biggl(\chi_1^2 \left(-\frac{19}{2} \CqOne \left(2 \eta ^2-7
   \eta +3\right)+\sqrt{1-4 \eta } \left(\frac{19}{2} \CqOne (\eta
   -3)+26 \eta -16\right)-9 \eta ^2+58 \eta -16\right)
\neanl \fl && %========================================
+\chi_2^2 \left(-\frac{19}{2}
   \CqTwo \left(2 \eta ^2-7 \eta +3\right)+\sqrt{1-4 \eta }
   \left(-\frac{19}{2} \CqTwo (\eta -3)-26 \eta +16\right)-9 \eta
   ^2+58 \eta -16\right)
\neanl \fl && %========================================
+2 \chi_1 \chi_2 \eta  (10 \eta -73)
\Biggr)
\neanl \fl && %========================================
+\NRG
   \Biggl(
\frac{1}{\Ang^4}
\Biggl[\chi_1^2 \left(\CqOne \left(6 \eta ^2-41 \eta
   +18\right)+\sqrt{1-4 \eta } (\CqOne (18-5 \eta )-24 (\eta -1))+8
   \left(\eta ^2-9 \eta +3\right)\right)
\neanl \fl && %========================================
+\chi_2^2 \left(\CqTwo \left(6 \eta ^2-41 \eta
   +18\right)+\sqrt{1-4 \eta } (\CqTwo (5 \eta -18)+24 (\eta -1))+8
   \left(\eta ^2-9 \eta +3\right)\right)
\neanl \fl && %========================================
+4 \chi_1 \chi_2 \eta (\eta +32)
\Biggr]
\neanl \fl && %========================================
+\frac{\sqrt{2}}{\Ang^3}
\biggl[_3
   \chi_1^2 (\CqOne (5 \eta -14)+13 \eta -8) \sqrt{\NRG-4 \eta  \NRG}
  +\chi_2^2 (\CqTwo (14-5 \eta )-13 \eta +8) \sqrt{\NRG-4 \eta  \NRG}
\biggr]_3
\Biggr)
\neanl \fl && %========================================
+\frac{\NRG^{5/2}}{\Ang}
   \left(\chi_1^2 \sqrt{2-8 \eta } (\CqOne (28-10 \eta )-26
   \eta +16)+\chi_2^2 \sqrt{2-8 \eta } (\CqTwo (10 \eta
   -28)+26 \eta -16)\right)
\Biggr\}
\neanl \fl && %=============================
+\frac{1}{c^6}
\Biggl\{
\frac{\NRG}{\Ang^4}
   \left(16 \eta ^2+\left(5 \pi ^2-281\right) \eta +134\right)
   -\frac{\NRG^{3/2}}{24 \sqrt{2}\Ang^3}
   \left(
   1440 \eta ^2+\left(123 \pi^2-13952\right) \eta +10080
   \right)
\neanl \fl && %=============================
+\frac{\NRG^2}{\Ang^2}
   \left(
   -\frac{167 \eta^2}{2}+\left(\frac{3127}{6}-\frac{39 \pi ^2}{8}\right) \eta
   -\frac{561}{2}
   \right)
   +\Ang^2
   \left(-30 \eta ^3+\frac{385 \eta ^2}{4}-250 \eta +660\right)
   \NRG^4
\neanl \fl && %=============================
-3 \sqrt{2} \Ang \left(46 \eta ^2-193 \eta
   +265\right) \NRG^{7/2}
   +\frac{\NRG^{5/2}}
         {12 \sqrt{2}\Ang}
   \left(3240 \eta ^2+\left(123 \pi^2-19964\right) \eta +16380\right)
\neanl \fl && %=============================
+\left(20 \eta ^3+42 \eta ^2-135 \eta +46\right)
   \NRG^3
\Biggr\}
\neanl \fl && %=============================
+{\frac{1}{c^6}\delta_S}
\Biggl\{
\frac{\NRG^2}{\Ang^3}
   \biggl(\frac{1}{2} \sqrt{1-4 \eta } \left(171 \eta ^2-1211 \eta
   +1384\right) (\chi_1-\chi_2)
\neanl \fl && %=============================
+\left(-34 \eta ^3+\frac{955
   \eta ^2}{2}-\frac{2653 \eta }{2}+692\right)
   (\chi_1+\chi_2)\biggr)
\neanl \fl && %=============================
+\frac{\NRG^{5/2}}{\Ang^2}
   \left(\sqrt{2} \left(41 \eta ^3-665 \eta ^2+1844 \eta -966\right)
   (\chi_1+\chi_2)-2 \sqrt{2-8 \eta } \left(50 \eta ^2-394
   \eta +483\right)
   (\chi_1-\chi_2)\right)
\neanl \fl && %=============================
+\NRG
   \biggl(
   \frac{1}{\Ang^5}
   \left[
   \left(
   4 \eta ^3-158 \eta ^2+650 \eta -388\right)
   (\chi_1+\chi_2)
   -4 \sqrt{1-4 \eta } \left(4 \eta ^2-69 \eta
   +97\right) (\chi_1-\chi_2)
   \right]
\neanl \fl && %=============================
+\frac{3}{\Ang^4} \left(7
   \eta ^2-99 \eta +140\right) (\chi_1-\chi_2) \sqrt{(2-8
   \eta ) \NRG}
\biggr)
\neanl \fl && %=============================
+ \frac{3 \sqrt{2}}{\Ang^4}
 \left(-2 \eta ^3+69 \eta ^2-239 \eta +140\right)
   \NRG^{3/2}
   (\chi_1+\chi_2)
%==
\neanl \fl && %=============================
+\frac{\NRG^3}{\Ang}
   \left(\left(68 \eta ^3-\frac{847 \eta ^2}{2}+835 \eta -498\right)
   (\chi_1+\chi_2)-\frac{1}{2} \sqrt{1-4 \eta } \left(243
   \eta ^2-770 \eta +996\right)
   (\chi_1-\chi_2)\right)
\neanl \fl && %=============================
+\NRG^{7/2}
   \biggl(
	4 \sqrt{2-8 \eta } \left(27 \eta ^2-133 \eta +177\right)
   (\chi_1-\chi_2)
\neanl \fl && %=============================
+2 \sqrt{2} \left(-27 \eta ^3+277 \eta
   ^2-650 \eta +354\right) (\chi_1+\chi_2)
	\biggr)
\Biggr\}
\neanl \fl 
&& %============================= 
%=============================
+{\frac{1}{c^6}\delta_S^2}
\Biggl\{
-\frac{2 \sqrt{2} \chi_1 \chi_2
   \eta  \left(15 \eta ^2-442 \eta +939\right)
   \NRG^{7/2}}{\Ang}
+\frac{\chi_1 \chi_2 \eta 
   \left(40 \eta ^2-751 \eta +1364\right)
   \NRG^3}{\Ang^2}
\neanl  \fl && %=============================
+\frac{\sqrt{2} \chi_1 \chi_2
   \eta  \left(-51 \eta ^2-1306 \eta +3849\right)
   \NRG^{5/2}}{\Ang^3}
+\frac{10 \chi_1 \chi_2
   \eta  \left(11 \eta ^2+136 \eta -354\right)
   \NRG^2}{\Ang^4}
\neanl  \fl && %=============================
+\frac{6 \sqrt{2} \chi_1
   \chi_2 \eta  \left(7 \eta ^2+83 \eta -315\right)
   \NRG^{3/2}}{\Ang^5}
-\frac{4 \chi_1 \chi_2 \eta  \left(13 \eta ^2+125 \eta
   -581\right) \NRG}{\Ang^6}
\Biggr\}
\end{eqnarray}

\begin{eqnarray}
\fl
 e_{\phi}^2
&=&
\EccNewton^2
% \neanl \fl && %=================================
+
\frac{1}{c^2}
\Biggl\{
12 \NRG
+
\Ang^2 (\eta -15)
   \NRG^2
\Biggr\}
% \neanl \fl && %=================================
%=================================
%=================================
%=================================
+\frac{\delta_S}{c^2} 
\Biggl\{
   \Ang \NRG^2
   \left(8 \sqrt{1-4 \eta } (\chi_1-\chi_2)+(8-4 \eta
   ) (\chi_1+\chi_2)\right)
\neanl \fl && %=================================
+\frac{\NRG }{\Ang} \left(4
   (\eta -2) (\chi_1+\chi_2)-8 \sqrt{1-4 \eta }
   (\chi_1-\chi_2)\right)
\Biggr\}
\neanl \fl && %=================================
%=================================
%=================================
%=================================
+\frac{\delta_S^2}{c^2}
\Biggl\{
\frac{\NRG}{\Ang^2} \left(\chi_1^2
   \left(\CqOne (3-6 \eta )+3 \CqOne \sqrt{1-4 \eta
   }\right)+12 \chi_1 \chi_2 \eta +\chi_2^2
   \left(\CqTwo (3-6 \eta )-3 \CqTwo \sqrt{1-4 \eta
   }\right)\right)
\neanl \fl && %=================================
+\NRG^2
   \left(\chi_1^2 \left(\CqOne (8 \eta -4)-4
   \CqOne \sqrt{1-4 \eta }\right)-16 \chi_1
   \chi_2 \eta +\chi_2^2 \left(\CqTwo (8 \eta
   -4)+4 \CqTwo \sqrt{1-4 \eta
   }\right)\right)
\Biggr\}
\neanl \fl && %=================================
+\frac{1}{c^4}
\Biggl\{
\Ang^2 \left(-\frac{3 \eta ^2}{2}+15 \eta -80\right)
   \NRG^3
   +\frac{\NRG}{\Ang^2}
   \left(
   -\frac{15 \eta ^2}{8}-29 \eta
   +51\right)
   +\left(\frac{9 \eta^2}{2}+44 \eta -8\right) \NRG^2
\Biggr\}
\neanl \fl && %=================================
%=================================
%=================================
%=================================
+{\frac{\delta_S}{c^4}}
\Biggl\{
(\chi_1+\chi_2) \left(
   -\frac{3\NRG}{2 \Ang^3} \left(\eta^2-71 \eta
   +64\right)
   +\frac{4\NRG^2}{\Ang} \left(\eta^2-36 \eta +17\right)
   +\Ang(80-31 \eta ) \NRG^3
   \right)
\neanl \fl && %=================================
\quad +(\chi_1-\chi_2)
   \left(
   \frac{3\NRG}{2\Ang^3} \sqrt{1-4 \eta } (11 \eta -64) 
   -\Ang \sqrt{1-4 \eta } (\eta -80)
   \NRG^3
   +\frac{2\NRG^2}{\Ang} (34-15 \eta ) \sqrt{1-4 \eta }
   \right)
\Biggr\}
\neanl \fl && %=================================
%=================================
%=================================
%=================================
+{\frac{\delta_S^2}{c^4} }
\Biggl\{
% \neanl \fl && %=================================
\chi_1^2 
\biggl(\frac{\NRG}{\Ang^4}
   \Biggl(\CqOne
   \left(\frac{51 \eta ^2}{4}-\frac{1261 \eta
   }{8}+66\right)+\sqrt{1-4 \eta } \left(\CqOne
   \left(66-\frac{205 \eta }{8}\right)-\frac{291 \eta
   }{4}+72\right)
\neanl \fl && %=================================
   +\frac{187 \eta ^2}{8}-\frac{867 \eta
   }{4}+72\Biggr)
% \neanl \fl && %=================================
+\frac{\NRG^2}{\Ang^2}
   \Biggl(\CqOne \left(-30 \eta ^2+221 \eta
   -\frac{171}{2}\right)
\neanl \fl && %=================================
   +\sqrt{1-4 \eta } \left(\CqOne
   \left(50 \eta -\frac{171}{2}\right)+103 \eta
   -56\right)-\frac{69 \eta ^2}{2}+215 \eta
   -56\Biggr)
\neanl \fl && %=================================
+\NRG^3 \left(\CqOne
   \left(7 \eta ^2+\frac{103 \eta }{2}-28\right)+\sqrt{1-4 \eta
   } \left(\CqOne \left(-\frac{9 \eta }{2}-28\right)+9 \eta
   -48\right)-\frac{5 \eta ^2}{2}+105 \eta
   -48\right)\biggr)
\neanl \fl && %=================================
+\chi_2^2 \Biggl(\frac{\NRG}{\Ang^4}
   \Biggl(\CqTwo \left(\frac{51 \eta ^2}{4}-\frac{1261 \eta
   }{8}+66\right)+\sqrt{1-4 \eta } \left(\CqTwo
   \left(\frac{205 \eta }{8}-66\right)+\frac{291 \eta
   }{4}-72\right)
\neanl \fl && %=================================
+\frac{187 \eta ^2}{8}-\frac{867 \eta}{4}+72
   \Biggr)
\neanl \fl && %=================================
+\frac{\NRG^2 }{\Ang^2}
   \Biggl(\CqTwo \left(-30 \eta ^2+221 \eta
   -\frac{171}{2}\right)+\sqrt{1-4 \eta } \left(\CqTwo
   \left(\frac{171}{2}-50 \eta \right)-103 \eta
   +56\right)
\neanl \fl && %=================================
-\frac{69 \eta ^2}{2}+215 \eta
   -56\Biggr)
% \neanl \fl && %=================================
+\NRG^3 \biggl(\CqTwo
   \left(7 \eta ^2+\frac{103 \eta }{2}-28\right)+\sqrt{1-4 \eta
   } \left(\CqTwo \left(\frac{9 \eta }{2}+28\right)-9 \eta
   +48\right)
\neanl \fl && %=================================
-\frac{5 \eta ^2}{2}+105 \eta -48\biggr)\Biggr)
% \neanl \fl && %=================================
+\chi_1 \chi_2 \left(
   \frac{\eta \NRG}{4 \Ang^4} (85 \eta +1484)
   -\frac{\eta \NRG^2}{\Ang^2} (9\eta +290)
   -\eta(19 \eta +292) \NRG^3
   \right)
\Biggr\}
\neanl \fl && %=================================
%=================================
%=================================
%=================================
+{\frac{1}{c^6}}
\Biggl\{
   \frac{\NRG}{64\Ang^4}
   \left(-70 \eta ^3+3440 \eta ^2+\left(1325 \pi^2-65436\right) \eta +27776\right)
   -\frac{3}{2} \Ang^2 \left(\eta ^3+13 \eta
   ^2-80 \eta +248\right) \NRG^4
\neanl \fl && %=================================
+\frac{\NRG^2}{16 \Ang^2} \left(-10 \eta
   ^3-2689 \eta ^2+\left(26860-581 \pi ^2\right) \eta
   -2456\right) 
\neanl \fl && %=================================
+\frac{1}{48}
   \left(234 \eta ^3+4536 \eta ^2+\left(2764+3 \pi ^2\right)
   \eta -16032\right) \NRG^3
\Biggr\}
\neanl \fl && %=================================
%=================================
%=================================
%=================================
+{\frac{\delta_S}{c^6}  } 
\Biggl\{
(\chi_1-\chi_2) \biggl(-\frac{\NRG}{8
   \Ang^5}
\sqrt{1-4 \eta }
   \left(161 \eta ^2-7968 \eta +10336\right)
\neanl \fl && %=================================
+\frac{\NRG^2 }{16 \Ang^3} 
\sqrt{1-4 \eta } \left(1297 \eta
   ^2-27988 \eta +13568\right) 
\neanl \fl && %=================================
   +\frac{1}{4} \Ang \sqrt{1-4 \eta }
   \left(31 \eta ^2-214 \eta +2152\right)
   \NRG^4+\frac{\sqrt{1-4 \eta }\NRG^3}{4 \Ang} 
   \left(-237 \eta^2 + 404 \eta +3664\right)
   \biggr)
\neanl \fl && %=================================
+(\chi_1+\chi_2)
   \Biggl(\frac{\NRG}{\Ang^5} \left(-\frac{27 \eta ^3}{8}-\frac{4581 \eta
   ^2}{8}+2278 \eta -1292\right)
   +\frac{ \NRG^2}{\Ang^3} \left(\frac{7 \eta
   ^3}{2}+\frac{20425 \eta ^2}{16}-\frac{12041 \eta
   }{4}+848\right)
\neanl \fl && %=================================
+\Ang
   \left(-2 \eta ^3+\frac{111 \eta ^2}{4}-\frac{427 \eta
   }{2}+538\right) \NRG^4
   +\frac{\NRG^3}{\Ang}\left(\frac{\eta^3}{2}-\frac{1577 \eta ^2}{4}-737 \eta +916\right)
   \Biggr)
\Biggr\}
\neanl \fl && %=================================
%=================================
%=================================
%=================================
+{ \frac{\delta_S^2}{c^6}}
\Biggl\{
\frac{\chi_1 \chi_2}{8 \Ang^6} \eta  \NRG 
\bigl(
-4	\Ang^6 (\eta  (63 \eta +220)+5728) \NRG^3
-8	\Ang^4 (5 \eta  (5 \eta +107)+5247) \NRG^2
\neanl \fl && %=================================
+5	\Ang^2 (\eta  (335 \eta +5306)-8552) \NRG
-756	\eta ^2-11919 \eta +61766
\bigr)
\Biggr\}
\,,
\end{eqnarray}
\begin{eqnarray}
\fl
 e_r^2 &=&
\EccNewton^2
+
\frac{1}{c^2} \, 5 \Ang^2 \left\{ (\eta -3) \NRG^2-2 (\eta -6) \NRG \right\}
\neanl \fl && %=================================
+{\frac{\delta_S}{c^2}}
\Biggl\{
\left((\chi_1-\chi_2) \left(8
   \Ang \sqrt{1-4 \eta } \NRG^2-\frac{8 \sqrt{1-4
   \eta }
   \NRG}{\Ang}\right)+(\chi_1+\chi_2)
   \left(\Ang (8-4 \eta ) \NRG^2+\frac{4 (\eta -2)
   \NRG}{\Ang}\right)\right)
\Biggr\}
\neanl \fl && %=================================
+{\frac{\delta_S^2}{c^2}}
\Biggl\{
   \chi_1^2 \left(\frac{\NRG
   \left(\CqOne (2-4 \eta )+2 \CqOne \sqrt{1-4 \eta
   }\right)}{\Ang^2}+\NRG^2 \left(\CqOne (4
   \eta -2)-2 \CqOne \sqrt{1-4 \eta
   }\right)\right)
\neanl \fl && %=================================
+\chi_2^2 \left(\frac{\NRG
   \left(\CqTwo (2-4 \eta )-2 \CqTwo \sqrt{1-4 \eta
   }\right)}{\Ang^2}+\NRG^2 \left(\CqTwo (4
   \eta -2)+2 \CqTwo \sqrt{1-4 \eta }\right)\right)
\neanl \fl && %=================================
+\chi_1 \chi_2 \left(\frac{8 \eta 
   \NRG}{\Ang^2}-8 \eta 
   \NRG^2\right)
\Biggr\}
\neanl \fl && %=================================
+{\frac{1}{c^4} } \Biggl\{ \Ang^2 \left(-4 \eta ^2+55 \eta -80\right)
   \NRG^3+\frac{(34-22 \eta )
   \NRG}{\Ang^2}+\left(\eta ^2+\eta +26\right)
   \NRG^2
\Biggr\}
\neanl \fl && %=================================
+{\frac{\delta_S}{c^4}}
\Biggl\{
   (\chi_1+\chi_2)
   \left(
   \frac{\NRG}{\Ang^3}   \left(-8 \eta ^2+78 \eta -64\right)
   +\Ang \left(6 \eta ^2-49 \eta
   +80\right) \NRG^3
   +\frac{\NRG^2}{\Ang}   \left(10 \eta ^2-70 \eta+4\right)
   \right)
\neanl \fl && %=================================
   +(\chi_1-\chi_2)
   \left(\frac{2\NRG}{\Ang^3} \sqrt{1-4 \eta } (9 \eta -32)
   +\Ang (80-19 \eta ) \sqrt{1-4
   \eta } \NRG^3
   +\frac{4\NRG^2}{\Ang} (1-4 \eta )^{3/2}
   \right)
\Biggr\}
\neanl \fl && %=================================
+{\frac{\delta_S^2}{c^4}}
\Biggl\{
   \chi_1^2 \Biggl(
   \frac{\NRG}{\Ang^4}
   \left(\CqOne \left(12 \eta ^2-82 \eta
   +36\right)+\sqrt{1-4 \eta } (\CqOne (36-10 \eta )-48
   (\eta -1))+16 \left(\eta ^2-9 \eta
   +3\right)\right)
\neanl \fl && %=================================
   +\frac{\NRG^2}{\Ang^2}
   \left(\CqOne \left(-18 \eta ^2+85 \eta
   -37\right)+\sqrt{1-4 \eta } (\CqOne (11 \eta -37)+52
   \eta -8)-20 \eta ^2+68 \eta
   -8\right)
\neanl \fl && %=================================
   +\NRG^3 \left(\CqOne
   \left(-4 \eta ^2+34 \eta -14\right)+\sqrt{1-4 \eta }
   (\CqOne (6 \eta -14)+12 (\eta -4))-2 \left(\eta ^2-54
   \eta +24\right)\right)\Biggr)
\neanl \fl && %=================================
   +\chi_1 \chi_2
   \left(
   \frac{8 \eta  (\eta +32)
   \NRG}{\Ang^4}-\frac{4 \eta  (\eta +31)
   \NRG^2}{\Ang^2}+4 (\eta -50) \eta 
   \NRG^3\right)
\neanl \fl && %=================================
   +\chi_2^2 \Biggl(
   \frac{\NRG}{\Ang^4}
   \left(\CqTwo \left(12 \eta ^2-82 \eta
   +36\right)+\sqrt{1-4 \eta } (\CqTwo (10 \eta -36)+48
   (\eta -1))+16 \left(\eta ^2-9 \eta
   +3\right)\right)
\neanl \fl && %=================================
   +\frac{\NRG^2}{\Ang^2}
   \left(\CqTwo \left(-18 \eta ^2+85 \eta
   -37\right)+\sqrt{1-4 \eta } (\CqTwo (37-11 \eta )-52
   \eta +8)-20 \eta ^2+68 \eta
   -8\right)
\neanl \fl && %=================================
   +\NRG^3 \left(\CqTwo
   \left(-4 \eta ^2+34 \eta -14\right)+\sqrt{1-4 \eta }
   (\CqTwo (14-6 \eta )-12 (\eta -4))-2 \left(\eta ^2-54
   \eta +24\right)\right)\Biggr)
\Biggr\}
\neanl \fl && %=================================
+{\frac{1}{c^6}} \Biggl\{
   \frac{\NRG}{\Ang^4} \left(32 \eta ^2+2 \left(5 \pi ^2-281\right) \eta
   +268\right)
   +\frac{\NRG^2}{\Ang^2}
   \left(-40 \eta^2+\left(\frac{2053}{3}-\frac{59 \pi ^2}{4}\right) \eta
   -49\right) 
\neanl \fl && %=================================
   +\Ang^2
   \left(\eta ^3-\frac{319 \eta ^2}{4}+389 \eta -372\right)
   \NRG^4+\left(-9 \eta ^2-\frac{1}{12} \left(172+3 \pi
   ^2\right) \eta -32\right) \NRG^3
 \Biggr\}
\neanl \fl && %=================================
+{\frac{\delta_S}{c^6}  }
\Biggl\{
(\chi_1-\chi_2) \Biggl(-\frac{8
   \sqrt{1-4 \eta } \left(4 \eta ^2-69 \eta +97\right)
   \NRG}{\Ang^5}+\frac{\sqrt{1-4 \eta } \left(70
   \eta ^2-597 \eta +352\right)
   \NRG^2}{\Ang^3}
\neanl \fl && %=================================
   +\frac{1}{2} \Ang \sqrt{1-4
   \eta } \left(35 \eta ^2-581 \eta +1076\right)
   \NRG^4-\frac{\sqrt{1-4 \eta } \left(27 \eta ^2+254
   \eta -536\right) \NRG^3}{2
   \Ang}\Biggr)
\neanl \fl && %=================================
+(\chi_1+\chi_2) \Biggl(\frac{4
   \left(2 \eta ^3-79 \eta ^2+325 \eta -194\right)
   \NRG}{\Ang^5}+\frac{\left(-20 \eta ^3+388 \eta
   ^2-1171 \eta +352\right)
   \NRG^2}{\Ang^3}
\neanl \fl && %=================================
   +\Ang \left(-2 \eta
   ^3+\frac{195 \eta ^2}{2}-\frac{901 \eta }{2}+538\right)
   \NRG^4+\frac{\left(2 \eta ^3+\frac{269 \eta ^2}{2}-541
   \eta +268\right) \NRG^3}{\Ang}\Biggr)
\Biggr\}
\neanl \fl && %=================================
+{\frac{\delta_S^2}{c^6} }
\chi_1 \chi_2
\Biggl\{
-\frac{8 \eta 
   \left(13 \eta ^2+125 \eta -581\right)
   \NRG}{\Ang^6}+\frac{4 \eta  \left(43 \eta ^2+319
   \eta -592\right) \NRG^2}{\Ang^4}-\frac{2 \eta 
   \left(10 \eta ^2-131 \eta +1008\right)
   \NRG^3}{\Ang^2}
\neanl \fl && %=================================
   +2 \eta  (232 \eta -1097)
   \NRG^4
\Biggr\}
\end{eqnarray}

\subsection{The Elements of the Kepler equation}

\begin{eqnarray}
\label{Eq::Period}
\fl 
P = 2\pi/\MeMo &=& \frac{1}{2 \sqrt{2} \NRG^{3/2}} 
- \cInvMT^{2}  \frac{\eta -15}{8 \sqrt{2} \NRG^{1/2}} 
+ \cInvMT^{4}  \biggl\{
    \frac{3}{\ang} \left(\frac{5}{2} - \eta\right) 
    - \frac{3 \NRG^{1/2}}{64 \sqrt{2}} \left(35 + 30 \eta + 3 \eta^{2}\right) \neanl \fl && \quad
    + \frac{\delta_S}{\ang^2} \left(
	\sqrt{1 -4 \eta} \left( 2\eta-6 \right) \left(\chi_1 - \chi_2\right) 
	-( \eta^{2} - 8 \eta + 6) \left(\chi _1 + \chi _2\right)
    \right) \neanl \fl && \quad
    + \frac{\delta_S^2}{\ang^3} \biggl(
	\chi _1 \chi _2 \eta \left(15  - \eta\right)
	+ \left(\chi _1^{2} - \chi _2^{2}\right) \biggl[
	  \sqrt{1-4\eta} \left(
	    2 - \frac{13}{4} \eta 
	    + (\CqOne + \CqTwo) \left(\frac{7}{4} - \frac{5}{8} \eta\right) 
	  \right) \neanl \fl &&\quad\quad\quad
	  + (\CqOne-\CqTwo) \left(\frac{7}{4} - \frac{33}{8} \eta + \frac{3}{4} \eta^{2}\right) 
	\biggr]
	+ \left(\chi _1^{2} + \chi _2^{2}\right)\biggl[
	  2 - \frac{29}{4} \eta + \eta^{2} \neanl \fl && \quad\quad\quad
	  + \sqrt{1-4\eta} (\CqOne-\CqTwo) \left(\frac{7}{4} - \frac{5}{8} \eta\right) 
	  + (\CqOne+\CqTwo) \left(\frac{7}{4} - \frac{33}{8} \eta + \frac{3}{4} \eta^{2}\right) 
	\biggr]
    \biggr)
   \biggr\} \neanl \fl &&
%===
+ \cInvMT^{6} \biggl\{
    -\frac{5 \NRG^{3/2}}{256 \sqrt{2}}  \left(21 -105 \eta + 15 \eta^{2} + 5 \eta^{3}\right) 
    -\frac{3\NRG}{2\ang} \left(5  -5 \eta + 4 \eta^{2}\right) \neanl \fl && \quad
    + \frac{1}{\ang^3} \left(\frac{105}{2} + \frac{15}{2} \eta^{2} + \eta \left(-\frac{218}{3} + \frac{41}{64} \pi^{2}\right)\right)    
    + \frac{\delta_S}{\ang^4} \biggl[
      \sqrt{1-4\eta} \left(-105 + \frac{297}{4} \eta - \frac{21}{4} \eta^{2}\right) \left(\chi_1 - \chi_2\right) \neanl \fl && \quad\quad
      +\left( -105 + \frac{717}{4} \eta - \frac{207}{4} \eta^{2} + \frac{3}{2} \eta^{3}\right) \left(\chi_1 + \chi_2\right)
      \biggr] 
    + \frac{\NRG}{\ang^2}
      \biggl(
	  \sqrt{1-4\eta} \left(15 - \frac{39}{2} \eta + 6 \eta^{2}\right) \left(\chi _1 - \chi _2\right) \neanl \fl &&\quad\quad
	  + (15  -42 \eta + 27 \eta^{2}  -3 \eta^{3}) \left(\chi _1 + \chi _2\right)
	\biggr)\delta_S
\neanl \fl && \quad\quad %=======
    + \default{ \delta_S^2 \chi_1 \chi_2
      \left(
      \frac{\NRG }{\Ang^3} 3 \eta  (14 \eta -17)
     -\frac{3 \eta}{2 \Ang^5} \left(7 \eta ^2+83 \eta -315\right) \right)
     }
\biggr\}
\,,
\end{eqnarray}
% \remark{Manuel: Hier muss dringend nochmal mit dem Mathematica-Ergebnis verglichen werden! NNLO S1S2 fehlt noch}

\begin{eqnarray}
\label{Eq::Fv}
 \fl
 F_v &=&
{\frac{1}{c^4}}
\Biggl\{
-\frac{\eta  (\eta +4) \sqrt{\NRG^3-2 \Ang^2
   \NRG^4}}{2 \sqrt{2} \Ang}
\Biggr\}
\neanl \fl && %=================================
+{\frac{ \delta_S}{c^4}}
\Biggl\{
	\frac{\EccNewton \NRG^{3/2} ((13 \eta -8)
   (\chi_1+\chi_2)-(\eta +8) \sqrt{1-4\eta}
   (\chi_1-\chi_2))}{2 \sqrt{2} \Ang^2}
\Biggr\}
\neanl \fl && %=================================
+{\frac{1}{c^4} \delta_S}
\frac{\EccNewton}{\Ang^3} 
\Biggl\{
-\frac{\chi_1
   \chi_2 (\eta -4) \eta 
   \NRG^{3/2}}{\sqrt{2}}
\neanl \fl && %=================================
+  \chi_1^2 \NRG^{3/2} \left(\CqOne
   \left(\frac{(8-3 \eta ) {} \sqrt{1-4\eta}}{2
   \sqrt{2}}+\frac{(\eta  (2 \eta -19)+8) {} }{2
   \sqrt{2}}\right)-\frac{3 \eta  {}
   \sqrt{1-4\eta}}{\sqrt{2}}+\frac{(\eta -6) \eta 
   {} }{2 \sqrt{2}}\right)
\neanl \fl && %=================================
+\chi_2^2 {\NRG^{3/2}} \left(\CqTwo
   \left(\frac{(3 \eta -8) {} \sqrt{1-4\eta}}{2
   \sqrt{2}}+\frac{(\eta  (2 \eta -19)+8) {} }{2
   \sqrt{2}}\right)+\frac{3 \eta  {}
   \sqrt{1-4\eta}}{\sqrt{2}}+\frac{(\eta -6) \eta 
  {} }{2 \sqrt{2}}\right)\Biggr\}
\neanl \fl &&
%=================================
%=================================
%=================================
{\frac{1}{c^6}}
\Biggl\{
\frac{\NRG^{3/2} \left(1728 \EccNewton^2+\eta 
   \left(33 \eta ^2+600 \eta +3 \pi ^2-4148\right)\right)}{48
   \sqrt{2} \Ang^3 \EccNewton}+\frac{\Ang
   \eta  \left(23 \eta ^2-4 \eta -64\right) \NRG^{7/2}}{4
   \sqrt{2} \EccNewton}
\neanl \fl && %=================================
-\frac{\eta  \left(105 \eta ^2+627
   \eta +3 \pi ^2-4232\right) \NRG^{5/2}}{24 \sqrt{2}
   \Ang \EccNewton}
\Biggr\}
\neanl \fl && %=================================
+{\frac{\delta_S}{c^6} }
\Biggl\{
(\chi_1-\chi_2)
   \biggl(\frac{\sqrt{2-8 \eta } \left(-19 \eta ^2+2928 \eta
   -2976\right) \NRG^{3/2}}{64 \Ang^4
   \EccNewton}
\neanl \fl && %=================================
+\frac{\sqrt{2-8 \eta } \left(-12 \eta
   ^2-1453 \eta +1544\right) \NRG^{5/2}}{16 \Ang^2
   \EccNewton}+\frac{\sqrt{2-8 \eta } \left(57 \eta ^2+8
   \eta -256\right) \NRG^{7/2}}{16
   \EccNewton}\biggr)
\neanl \fl && %=================================
+(\chi_1+\chi_2)
   \biggl(-\frac{\left(130 \eta ^3+2359 \eta ^2-7232 \eta
   +2976\right) \NRG^{3/2}}{32 \sqrt{2} \Ang^4
   \EccNewton}+\frac{\left(86 \eta ^3+1202 \eta ^2-3711
   \eta +1544\right) \NRG^{5/2}}{8 \sqrt{2} \Ang^2
   \EccNewton}
\neanl \fl && %=================================
-\frac{\left(50 \eta ^3+35 \eta ^2-472 \eta
   +256\right) \NRG^{7/2}}{8 \sqrt{2}
   \EccNewton}\biggr)
\Biggr\}
% \neanl \fl && %=================================
+{ \frac{\delta_S^2}{c^6} }
\Biggl\{
\chi_1 \chi_2
   \biggl(-\frac{3 \eta  \left(26 \eta ^2+103 \eta -828\right)
   \NRG^{3/2}}{8 \sqrt{2} \Ang^5
   \EccNewton}
\neanl \fl && %=================================
+\frac{\eta  \left(71 \eta ^2+363 \eta
   -2388\right) \NRG^{5/2}}{4 \sqrt{2} \Ang^3
   \EccNewton}+\frac{\eta  \left(2 \eta ^2+9 \eta
   +52\right) \NRG^{7/2}}{2 \sqrt{2} \Ang
   \EccNewton}\biggr)
\Biggr\}
\,,
\end{eqnarray}
% \neanl \fl && \quad \qquad %=================================
% {\cInv{6} \delta_S}
% \Biggl\{
% \Biggr\}

% \flushleft
\begin{eqnarray}
\label{Eq::F2v}
\fl
F_{2v}
&=&
{\frac{1}{c^6}}
\Biggl\{
\frac{ \EccNewton^2 \eta  \left(6 \eta ^2+12
   \eta +23\right) \NRG^{3/2}}{8 \sqrt{2} \Ang^3}
\Biggr\}
% \neanl \fl && %=================================
+{ \frac{\delta_S^2}{c^6} }
\Biggl\{
-\frac{\chi_1 \chi_2
   \EccNewton^2 \eta  \left(2 \eta ^2-35 \eta +4\right)
   \NRG^{3/2}}{4 \sqrt{2} \Ang^5}
\Biggr\}
\neanl \fl && %=================================
+{\frac{\delta_S}{c^6} }
\Biggl\{
\frac{\EccNewton^2
   \sqrt{\frac{1}{2}-2 \eta } \left(13 \eta ^2+32 \eta
   -8\right) \NRG^{3/2} (\chi_1-\chi_2)}{8
   \Ang^4}
\neanl \fl && %=================================
-\frac{\EccNewton^2 \left(21 \eta ^3+23
   \eta ^2-48 \eta +8\right) \NRG^{3/2}
   (\chi_1+\chi_2)}{8 \sqrt{2} \Ang^4}
\Biggr\}\,,
\eanl \fl %=================================
%=================================
\label{Eq::F3v}
F_{3v}
&=&
{\frac{1}{c^6}}
\Biggl\{
 \frac{13 \EccNewton \eta ^3
   \NRG^{3/2}}{48 \sqrt{2} \Ang^3}-\frac{13
   \EccNewton \eta ^3 \NRG^{5/2}}{24 \sqrt{2}
   \Ang}
\Biggr\}
%  \neanl \fl && %=================================
+{\frac{\delta_S^2}{c^6}}
\Biggl\{
\chi_1 \chi_2
   \frac{\EccNewton \eta  \left(2 \eta ^2-11 \eta
   +4\right) \NRG^{5/2}}{4 \sqrt{2}
   \Ang^3}
\neanl \fl && %=================================
-\frac{\EccNewton \eta  \left(2 \eta
   ^2-11 \eta +4\right) \NRG^{3/2}}{8 \sqrt{2}
   \Ang^5}
\Biggr\}
\neanl \fl && %=================================
+{ \frac{\delta_S}{c^6} }
\Biggl\{
(\chi_1-\chi_2)
\sqrt{\frac{1}{2}}
\sqrt{1-4\eta}
 \eta ^2
   \left(\frac{9 \EccNewton 
% \sqrt{\frac{1}{2}}
% \sqrt{1-4\eta}
%    \eta ^2 
\NRG^{3/2}}{32 \Ang^4}-\frac{9
   \EccNewton 
% \sqrt{\frac{1}{2}}
% \sqrt{1-4\eta}
%  \eta ^2
   \NRG^{5/2}}{16
   \Ang^2}\right)
\neanl \fl && \qquad \quad %=================================
+(\chi_1+\chi_2)
   \left(\frac{\EccNewton (5-18 \eta ) \eta ^2
   \NRG^{3/2}}{32 \sqrt{2}
   \Ang^4}+\frac{\EccNewton \eta ^2 (18 \eta -5)
   \NRG^{5/2}}{16 \sqrt{2} \Ang^2}\right)
\Biggr\}\,,
\eanl \fl %=================================
\label{Eq::Fvmu}
F_{v-u}
&=&
{\frac{1}{c^4} }
% \Biggl\{
\frac{3 \sqrt{2} (5-2 \eta )
   \NRG^{3/2}}{\Ang}
% \Biggr\}
\neanl \fl && %=================================
+
\NRG^{3/2}
{\sqrt{2}}
\Biggl\{_1
{\frac{\delta_S}{c^4} }
\Biggl\{
\frac{4 
% \sqrt{2-8 \eta }
{\sqrt{1-4\eta}}
   (\eta -3) 
%    \NRG^{3/2}
   (\chi_1-\chi_2)}{\Ang^2}
-\frac{2 
% \sqrt{2}
   \left(\eta ^2-8 \eta +6\right) 
% \NRG^{3/2}
   (\chi_1+\chi_2) }{\Ang^2}
\Biggr\}
% \neanl \fl && %=================================
+{\frac{\delta_S^2}{c^4} }
\Biggl\{
\frac{
% \sqrt{2} 
% \NRG^{3/2}
2\,
\chi_1 \chi_2 
   (15-\eta) \eta }{\Ang^3}
\neanl \fl && %=================================
+\frac{
% \sqrt{2}
 \chi_2^2
%    \NRG^{3/2}
 \left(\CqTwo \left(3 \eta ^2-\frac{33
   \eta }{2}+7\right)+\sqrt{1-4 \eta } \left(\CqTwo
   \left(\frac{5 \eta }{2}-7\right)+\frac{13 \eta
   }{2}-4\right)+2 \eta ^2-\frac{29 \eta
   }{2}+4\right)}{\Ang^3}
\neanl \fl &&  %=================================
+   \frac{
% \sqrt{2} 
% \NRG^{3/2}
 \,
   \chi_1^2 \left(\CqOne \left(3 \eta
   ^2-\frac{33 \eta }{2}+7\right)+\sqrt{1-4 \eta }
   \left(\CqOne \left(7-\frac{5 \eta }{2}\right)-\frac{13
   \eta }{2}+4\right)+2 \eta ^2-\frac{29 \eta
   }{2}+4\right)}{\Ang^3}
\Biggr\}
\Biggr\}_1
\neanl \fl && %=================================
+{\frac{1}{c^6} }
\Biggl\{
\frac{\left(1440 \eta ^2+\left(123 \pi
   ^2-13952\right) \eta +10080\right) \NRG^{3/2}}{48
   \sqrt{2} \Ang^3}-\frac{3 \left(18 \eta ^2-55 \eta
   +95\right) \NRG^{5/2}}{2 \sqrt{2} \Ang}
\Biggr\}
\neanl \fl &&  %=================================
+{\frac{\delta_S}{c^6} }
\Biggl\{
(\chi_1-\chi_2)
   \left(\frac{\sqrt{2-8 \eta } \left(13 \eta ^2-57 \eta
   +75\right) \NRG^{5/2}}{\Ang^2}-\frac{3
   \sqrt{\frac{1}{2}-2 \eta } \left(7 \eta ^2-99 \eta
   +140\right)
   \NRG^{3/2}}{\Ang^4}\right)
\neanl \fl &&  %=================================
+(\chi_1+\chi_2) \left(\frac{3 \left(2 \eta ^3-69 \eta ^2+239 \eta
   -140\right) \NRG^{3/2}}{\sqrt{2}
   \Ang^4}+\frac{\left(-13 \eta ^3+131 \eta ^2-294 \eta
   +150\right) \NRG^{5/2}}{\sqrt{2}
   \Ang^2}\right)
\Biggr\}
\neanl \fl && %=================================
+{\frac{\delta_S^2}{c^6} }
\Biggl\{
\chi_1 \chi_2
   \left(-\frac{3 \sqrt{2} \eta  \left(7 \eta ^2+83 \eta
   -315\right) \NRG^{3/2}}{\Ang^5}-\frac{\eta 
   \left(\eta ^2-198 \eta +429\right)
   \NRG^{5/2}}{\sqrt{2} \Ang^3}\right)
\Biggr\}
\,,
\end{eqnarray}

\subsection{The Elements of the Orbital phase}
\begin{eqnarray}
\label{Eq::TotalPhase}
 \fl \frac{\Phi}{2\pi} - 1 &=& \cInv{2} \biggl\{
  \frac{3}{\ang^2} 
+ \frac{\default{ \delta_S }}{\ang^3} \left[-2(\chi_1 - \chi_2)\sqrt{1-4\eta} + (\chi_1+\chi_2)(\eta-2)\right] 
\neanl \fl && \quad %==========
  +\frac{3 \default{ \delta_S^2 } }{8\ang^4} \biggl[8\chi_1 \chi_2 \eta
    + (\chi_1^2 + \chi_2^2) \left((\CqOne - \CqTwo)\sqrt{1-4\eta} + (\CqOne+\CqTwo)(1-2\eta)\right) \neanl \fl && \quad\quad
    +(\chi_1^2 - \chi_2^2) \left((\CqOne + \CqTwo)\sqrt{1-4\eta} + (\CqOne-\CqTwo)(1-2\eta)\right)
  \biggr]
 \biggr\} \neanl \fl &&
 + \cInv{4} \biggl\{
  \frac{3 \NRG}{\ang^2} \left(\eta - \frac{5}{2}\right) - \frac{15}{4 \ang^4} \left(2\eta-7\right) 
\neanl \fl && \quad %==========
 +\default{ \delta_S } \frac{2 \NRG}{\ang^3} \biggl[2(\chi_1-\chi_2)(3-\eta)\sqrt{1-4\eta} + (\chi_1+\chi_2)(6-8\eta+\eta^2)\biggr] \neanl \fl && \quad %==========
  + \default{ \delta_S } \frac{3}{\ang^5}\biggl[7(\chi_1-\chi_2)\left(\frac{\eta}{4} - 2\right)\sqrt{1-4\eta} - (\chi_1+\chi_2)\left(14-\frac{49}{4}\eta+\frac{\eta^2}{2}\right)\biggr] \neanl \fl && \quad 
  + \default{ \delta_S^2 } \frac{\NRG}{\ang^4} \biggl[3 \chi_1 \chi_2 \eta (\eta - 15) \neanl \fl && \quad\quad
    - (\chi_1^2 + \chi_2^2) \biggl(
      \frac{3}{4} (\CqOne + \CqTwo) \left(7 - \frac{33}{4}\eta + 3 \eta^2\right)
      +\frac{3}{4} (\CqOne - \CqTwo) \sqrt{1-4\eta} \left(7 - \frac{5}{2}\eta\right) \neanl \fl && \quad\quad\quad
      +6 - \frac{87}{4}\eta + 3 \eta^2 
    \biggr)
    - (\chi_1^2 - \chi_2^2) \biggl(
      \frac{3}{4} (\CqOne - \CqTwo) \left(7 - \frac{33}{4}\eta + 3 \eta^2\right) \neanl \fl && \quad\quad\quad
      + \sqrt{1-4\eta} \left(6 - \frac{39}{4}\eta + \frac{3}{4} (\CqOne + \CqTwo) \left(7 - \frac{5}{2}\eta\right)\right)
    \biggr)
  \biggr] \neanl \fl && \quad
  \default{ \delta_S^2 } \frac{15}{\ang^6} \biggl[
    \frac{1}{2}  \chi_1 \chi_2\, \eta (\eta + 21) \neanl \fl && \quad\quad
    + (\chi_1^2 + \chi_2^2) \biggl(
      (\CqOne + \CqTwo) \left(\frac{3}{4} - \frac{27}{16}\eta + \frac{\eta^2}{8}\right)
      +\frac{3}{4}(\CqOne - \CqTwo) \sqrt{1-4\eta} \left(1 - \frac{\eta}{4}\right) 
      +2 - \frac{45}{8}\eta + \frac{ \eta^2 }{2}
    \biggr) \neanl \fl && \quad\quad
    +(\chi_1^2 - \chi_2^2) \biggl(
      (\CqOne - \CqTwo) \left(\frac{3}{4} - \frac{27}{16}\eta + \frac{\eta^2}{8}\right) 
      + \sqrt{1-4\eta} \left(2 - \frac{13}{8}\eta + \frac{3}{4} (\CqOne + \CqTwo) \left(1 - \frac{\eta}{4}\right)\right)
  \biggr]
 \biggr\} \neanl \fl &&
 \cInv{6} \biggl\{
  \frac{3 \NRG^2}{4 \ang^2} (5 - 5\eta + 4\eta^2)
  - \frac{\NRG}{\ang^4} \left(\frac{315}{2} - \left(218 - \frac{123}{64}\pi^2\right) \eta + \frac{45}{2}\eta^2\right) \neanl \fl && \quad
  + \frac{5}{128 \ang^6} (7392 - (8000 - 123 \pi^2) \eta + 336 \eta^2) \neanl \fl && \quad
  + \default{ \delta_S }
\biggl[ \frac{\NRG^2}{\ang^3} \biggl[
    (\chi_1 - \chi_2)\sqrt{1-4\eta}\left(-15 + \frac{39}{2}\eta - 6 \eta^2\right)
    +(\chi_1 + \chi_2)(-15+42\eta-27\eta^2+3\eta^3)
  \biggr]  \neanl \fl && \quad
  + \frac{\NRG}{\ang^5} \biggl[
    (\chi_1 - \chi_2)\sqrt{1-4\eta}\left(420-297\eta+21\eta^2\right)
    +(\chi_1 + \chi_2)(420-717\eta+207\eta^2-6\eta^3)
  \biggr]  \neanl \fl && \quad
  - \frac{1}{2 \ang^7} \biggl[
    +(\chi_1 - \chi_2)\sqrt{1-4\eta}\left(1485 - \frac{5265}{8}\eta + \frac{75}{4}\eta^2\right)\neanl \fl && \quad\quad
    +(\chi_1 + \chi_2)\left(1485-\frac{15165}{4}\eta + 345\eta^2 - \frac{15}{4}\eta^3\right)
  \biggr]
\biggr]
\neanl \fl && \quad\quad %====================================
+{\chi_1 \chi_2} \default{ \delta_S^2 }
\left(
-\frac{315 \eta }{8{\Ang}^8}
   \left(\eta ^2+11 \eta -110\right)
+\frac{15 \eta}{2{\Ang}^6}  \left(7 \eta ^2+83
   \eta -315\right) {\NRG}
+\frac{9}{2 {\Ang}^4} (17-14 \eta ) \eta {\NRG}^2\right)
 \biggr\}\,,
\end{eqnarray}
% \remark{Manuel: Hier muss dringend nochmal mit dem Mathematica-Ergebnis verglichen werden!}

\begin{eqnarray}
\label{Eq::G2v}
\fl
G_{2v}
&=& 
{\frac{1}{c^4}}
\Biggl\{
\frac{ \EccNewton^2 \left(\eta -3 \eta
   ^2\right)}{8 \ang^4}
\Biggr\}
% \neanl \fl && %=================================
+{\frac{1}{c^4} \delta_S}
\Biggl\{
\frac{3 \delta_S \,  \EccNewton^2 \eta 
   \left(\chi_1 \,   \left(\eta -\sqrt{1-4 \eta
   }-1\right)+\chi_2 \,   \left(\eta +\sqrt{1-4 \eta
   }-1\right)\right)}{4 \ang^5}
\Biggr\}
\neanl \fl && %=================================
+{\frac{\delta_S^2}{c^4} }
\Biggl\{
\frac{\chi_1^2
   \left(\frac{1}{16} \CqOne \EccNewton^2 \left(-6
   \eta ^2-19 \eta +6\right)+\sqrt{1-4 \eta }
   \left(\frac{1}{16} \CqOne \EccNewton^2 (6-7 \eta
   )+\frac{3 \EccNewton^2 \eta }{16}\right)+\frac{1}{16}
   \EccNewton^2 (3-2 \eta ) \eta
   \right)}{\ang^6}
\neanl \fl &&   %=================================
+\frac{\chi_2 \,  ^2 \left(\frac{1}{16}
   \CqTwo \EccNewton^2 \left(-6 \eta ^2-19 \eta
   +6\right)+\sqrt{1-4 \eta } \left(\frac{1}{16} \CqTwo
   \EccNewton^2 (7 \eta -6)-\frac{3 \EccNewton^2
   \eta }{16}\right)+\frac{1}{16} \EccNewton^2 (3-2 \eta
   ) \eta \right)}{\ang^6}
\neanl \fl && \quad %=================================
+\frac{\chi_1 \, \chi_2 \, \EccNewton^2 \eta  (2 \eta -5)}{4\ang^6}
\Biggr\}
\neanl \fl && %=================================
+{\frac{1}{c^6}}
\Biggl\{
\frac{-40 \eta ^3-384 \eta ^2+\left(49 \pi
   ^2-1076\right) \eta +256}{256 \ang^6}
\neanl \fl && %=================================
+\frac{\left(-80
   \eta ^3+336 \eta ^2+\left(1192-49 \pi ^2\right) \eta
   -256\right) \NRG}{128 \ang^4}
% \neanl \fl && %=================================
+\frac{\eta 
   \left(24 \eta ^2-40 \eta -11\right) \NRG^2}{16
   \ang^2}
\Biggr\}
\neanl \fl && %=================================
+{\frac{\delta_S}{c^6} }
\Biggl\{
(\chi_1 \,  +\chi_2 \,  )
   \Biggl(\frac{\frac{1}{64} \eta  \left(-212 \eta ^2+119 \eta
   +492\right)-4 \EccNewton^2}{\ang^7}
\neanl \fl && %=================================
+\frac{\eta 
   \left(152 \eta ^2-47 \eta -296\right) \NRG}{16
   \ang^5}
\neanl \fl && %=================================
-\frac{\eta  \left(104 \eta ^2-127 \eta
   +24\right) \NRG^2}{16
   \ang^3}\Biggr)
% \neanl \fl && \qquad %=================================
+(\chi_1 \,  -\chi_2 \,  )
   \Biggl(\frac{\frac{1}{64} \sqrt{1-4 \eta } \eta  (319 \eta
   +172)-4 \EccNewton^2 \sqrt{1-4 \eta
   }}{\ang^7}
\neanl \fl && %=================================
-\frac{\sqrt{1-4 \eta } \eta  (197 \eta
   +136) \NRG}{16 \ang^5}+\frac{3 \sqrt{1-4 \eta }
   \eta  (37 \eta -8) \NRG^2}{16
   \ang^3}
\Biggr)
\Biggr\}
\neanl \fl && %=================================
+{\frac{\delta_S^2}{c^6}}
\Biggl\{
\chi _1 \chi _2 \biggl(\frac{\eta  \left(186 \eta ^2+895 \eta
   +476\right)}{32 \ang^8}-\frac{\eta  \left(80 \eta ^2+379 \eta
   +260\right) \NRG}{8 \ang^6}
\neanl \fl && %=================================
+\frac{\eta  \left(2 \eta
   ^2+137 \eta -46\right) \NRG^2}{8 \ang^4} \biggr)
\Biggr\}
\,,
\\
%=================================
%=================================
%=================================
\label{Eq::G3v}
\fl
G_{3v}
&=&
% \Biggl\{
-
{ \frac{1}{c^4}}
\frac{3 \EccNewton^3 \eta ^2}{32
   \ang^4}
% \Biggr\}
% \neanl \fl &&  %=================================
+{ \frac{\delta_S}{c^4} }
\Biggl\{
\frac{\EccNewton^3
   (\eta -1) \eta  (\chi_1 \,  +\chi_2 \,  )}{8
   \ang^5}-\frac{\EccNewton^3 \sqrt{1-4 \eta } \eta
    (\chi_1 \,  -\chi_2 \,  )}{8 \ang^5}
\Biggr\}
% \neanl \fl && %=================================
+{ \frac{1}{c^4} \delta_S^2}
\Biggl\{
   \frac{\chi_1 \,   \chi_2 \, \EccNewton^3 (\eta -4) \eta }{16\ang^6}
\neanl \fl && %=================================
  +\frac{\chi_1 \,  ^2
   \left(-\frac{1}{32} \CqOne \EccNewton^3 \eta  (2
   \eta +1)+\sqrt{1-4 \eta } \left(\frac{\EccNewton^3
   \eta }{16}-\frac{1}{32} \CqOne \EccNewton^3 \eta
   \right)-\frac{1}{32} \EccNewton^3 (\eta -2) \eta
   \right)}{\ang^6}
\neanl \fl && %=================================
+\frac{\chi_2 \,  ^2 \left(-\frac{1}{32}
   \CqTwo \EccNewton^3 \eta  (2 \eta +1)+\sqrt{1-4
   \eta } \left(\frac{1}{32} \CqTwo \EccNewton^3 \eta
   -\frac{\EccNewton^3 \eta }{16}\right)-\frac{1}{32}
   \EccNewton^3 (\eta -2) \eta
   \right)}{\ang^6}
\Biggr\}
\neanl \fl && %=================================
+{ \frac{1}{c^6}}
\Biggl\{
\frac{\frac{1}{768} \left(220+3 \pi
   ^2\right) \EccNewton^3 \eta +\frac{1}{256}
   \EccNewton \eta ^2 (15 \eta
   +32)}{\ang^6}-\frac{\EccNewton (25 \eta +52)
   \eta ^2 \NRG}{64 \ang^4}+\frac{\EccNewton
   (26 \eta -9) \eta ^2 \NRG^2}{64 \ang^2}
\Biggr\}
\neanl \fl && %=================================
+{ \frac{\delta_S}{c^6} }
\Biggl\{
(\chi_1+\chi_2 )
   \biggl(\frac{\EccNewton \eta  \left(-170 \eta ^2+7 \eta
   +96\right)}{128 \ang^7}+\frac{\EccNewton \eta 
   \left(188 \eta ^2+43 \eta -144\right) \NRG}{64
   \ang^5}
\neanl \fl && %=================================
-\frac{\EccNewton \eta  \left(15 \eta
   ^2-17 \eta +3\right) \NRG^2}{16
   \ang^3}\biggr)
% \neanl \fl && \qquad %=================================
+(\chi_1-\chi_2) \biggl(\frac{3
   \EccNewton \sqrt{1-4 \eta } \eta  (61 \eta +32)}{128
   \ang^7}
\neanl \fl && %=================================
-\frac{3 \EccNewton \sqrt{1-4 \eta } \eta
    (59 \eta +48) \NRG}{64 \ang^5}-\frac{3
   \EccNewton (1-4 \eta )^{3/2} \eta  \NRG^2}{16
   \ang^3}\biggr)
\Biggr\}
\neanl \fl && %=================================
+{ \frac{\delta_S^2}{c^6} }
\chi_1 \,   \chi_2 \,
\Biggl\{
\frac{\frac{1}{8} \EccNewton^3 \eta  \left(3
   \eta ^2-\eta -6\right)+\frac{5}{32} \EccNewton \eta 
   \left(6 \eta ^2+33 \eta
   -2\right)}{\ang^8}
\neanl \fl && %=================================
-\frac{\EccNewton \eta 
   \left(50 \eta ^2+211 \eta +32\right) \NRG}{32
   \ang^6}
% \neanl \fl && \qquad %=================================
-\frac{\EccNewton \eta  \left(5 \eta
   ^2-17 \eta +4\right) \NRG^2}{32 \ang^4}
\Biggr\}
\,,
\eanl \fl %==============================================
%==============================================
\label{Eq::G4v}
G_{4v}&=&
\frac{1}{c^6} \Biggl\{
\frac{\eta  \left(10 \eta ^2+28 \eta
   +5\right)}{128 \ang^6}-\frac{\eta  \left(10 \eta ^2+28
   \eta +5\right) \NRG}{32 \ang^4}+\frac{\eta 
   \left(10 \eta ^2+28 \eta +5\right) \NRG^2}{32
   \ang^2}
\neanl \fl && %======================================
+\delta_S^2 \chi _1 \chi _2
   \left(\frac{\eta  \left(12 \eta ^2-13 \eta +20\right)}{64
   \ang^8}-\frac{\eta  \left(12 \eta ^2-13 \eta +20\right)
   \NRG}{16 \ang^6}+\frac{\eta  \left(12 \eta ^2-13
   \eta +20\right) \NRG^2}{16
   \ang^4}\right)
\neanl \fl && %======================================
+\delta_S \Biggl\{\left(\chi _1+\chi
   _2\right) \left(-\frac{\eta  \left(40 \eta ^2+31 \eta
   -42\right)}{128 \ang^7}+\frac{\eta  \left(40 \eta ^2+31
   \eta -42\right) \NRG}{32 \ang^5}-\frac{\eta 
   \left(40 \eta ^2+31 \eta -42\right) \NRG^2}{32
   \ang^3}\right)
\neanl \fl && %======================================
+\left(\chi _1-\chi _2\right)
   \left(\frac{21 \sqrt{1-4 \eta } \eta  (\eta +2)}{128
   \ang^7}-\frac{21 \sqrt{1-4 \eta } \eta  (\eta +2)
   \NRG}{32 \ang^5}+\frac{21 \sqrt{1-4 \eta } \eta 
   (\eta +2) \NRG^2}{32 \ang^3}\right)
\Biggr\}
\Biggr\}
\,,
\eanl \fl
%==============================================
%==============================================
\label{Eq::G5v}
G_{5v}&=&
\frac{1}{c^6} 
\Biggl\{
\delta_S \left(\frac{3 \DiffChi
   \eta ^2 \sqrt{1-4\eta} \left(1-2 \ang^2
   \NRG\right)^{5/2}}{128 \ang^7}-\frac{3 \eta ^2 (2
   \eta -1) \SumChi \left(1-2 \ang^2
   \NRG\right)^{5/2}}{128 \ang^7}\right)
\neanl \fl && %======================================
+\frac{5
   \eta ^3 \left(1-2 \ang^2 \NRG\right)^{5/2}}{256
   \ang^6}
\Biggr\}
\,.
\end{eqnarray}

\noindent
As the binary system loses energy and angular momentum
via the emission of gravitational waves, the above orbital elements
will not remain constants of motion. One can evolve 
the binding energy and the angular momentum
via balance equations connecting the far-zone flux with the near-zone,
deduced from the time derivatives of the
source multipole moments. This will be done in the following section.

\section{Energy and Angular Momentum Loss}\label{sec:dissipative}
\label{Sec::RadiationReaction}

% For the radiative dynamics, 2PN point-mass contributions and the
% leading-oder terms for the spin-orbit, spin(1)- spin(2), spin-squared 
% interactions and tail contributions \cite{Rieth:Schafer:1997} are used.
For the radiative dynamics, taking the instantaneous parts only, 
	2PN point-mass contributions and the leading-oder terms for the spin-orbit,
	spin(1)- spin(2), and spin-squared interactions are used.
We later transform into ADM coordinates and canonical spin variables 
(see e.g. \cite{Steinhoff:2011} and references therein).

\subsection{Source multipole moments in harmonic coordinates, using covariant SSC}
We begin this section by collecting the relevant source
terms of the far-zone gravitational field.
% Kidder \cite{Kidder:1995} states that each spin term appears
% together with one power of velocity,
% such that, for leading-order effects only, 
% we do not spin-squared amplitudes.
It is to be mentioned that we can rewrite spin contributions in terms of 
spin tensors rather than in terms of spin vectors, 
\begin{equation}
 \label{Eq::SpinTensorToSpinVector}
\tilde{S}_{a}^{i} = 
% \frac{1}{2} \rm{S1}(\rm{k},\rm{m})
%    \rm{epsilong}(i,-\rm{k},-\rm{m})
\frac{1}{2}
\spinHC{a}{k}{m}
   \epsilon^{i k m}
\,, \hspace{2cm} a=\{1,2\}
\end{equation}
The inversion is realised via
\begin{equation}
\spinHC{a}{i}{j}=\tilde{S}_a^{k} \epsilon_{ijk}.
\end{equation}
% {Groessenordnung herausstellen!}
% 
The far-zone field of the gravitational wave reads \cite{Blanchet:Damour:Iyer:1995},
defining ${\cal R}$ to be the distance observer--center of mass of the binary,
and $N_{A_l}:=N_{a_1}N_{a_2}...N_{a_l}$ to be an $l$-fold tensor product of the
line-of-sight vector $N_{a_i}$,
\begin{eqnarray}
\fl
h_{ij}^{\rm TT}{}_{\rm inst} 
&=&
\frac{2\,G}{c^4 {\cal R }}
\Biggl\{
% \sum_{}^{} ... 
\stackrel{(2)}{I}_{\!\!ij}
+
\frac{1}{c}
\left[
\frac{1}{3}
N_{a} \!\!\! \stackrel{(3)}{I}_{\!\!ija}
+
\frac{4}{3}
\epsilon_{ab(i} \!\!\! \stackrel{(2)}{J}_{\!\!j)a}N_{b}
\right]
+
\frac{1}{c^2}
\left[
\frac{1}{12}
N_{ab} \!\!\! \stackrel{(4)}{I}_{\!\!ijab}
+
\frac{1}{2}
\epsilon_{ab(i} \!\!\! \stackrel{(3)}{J}_{\!\!  j)ac}N_{bc}
\right]
\neanl \fl && %========================
+
\frac{1}{c^3}
\left[
\frac{1}{60}
N_{abc} \!\!\! \stackrel{(5)}{I}_{\!\! ijabc}
+
\frac{2}{15}
\epsilon_{ab(i} \!\!\! \stackrel{(4)}{J}_{\!\! j)acd}N_{bcd}
\right]
+
\frac{1}{c^4}
\left[
\frac{1}{360}
N_{abcd} \!\!\! \stackrel{(6)}{I}_{\!\! ijabcd}
+
\frac{1}{36}
\epsilon_{ab(i} \!\!\! \stackrel{(5)}{J}_{\!\! j)acde}N_{bcde}
\right]
\Biggr\}
\,,
\end{eqnarray}
The point-mass multipoles appearing above at 2PN are taken from \cite{Gopakumar:Iyer:1997},
the linear-in-spin terms from \cite{Kidder:1995},
and the leading order spin-squared contribution from \cite{Porto:Ross:Rothstein:2010}.
Note that \cite{Gopakumar:Iyer:1997} applied the harmonic gauge and \cite{Kidder:1995}
the covariant spin supplementary condition.
{\em In the following, $\cInv{1}$ is regarded as a bookkeeping parameter. 
To check out the relative orders of magnitude,
we employ the scaling according to \ref{SubSec::Scaling}
% {\cite{Hergt:Steinhoff:Schafer:2011}} which reduces all terms to
% distances as $x \sim c^0$, $p\sim c^{-3}$, $S_A \sim c^{-3}$ and $m_A \sim c^{-2}$.
and look for the powers of inverse $c$ in such a way that the units of
spinless and spin dependent contributions are the same.
We remind of the fact that a time derivative
goes along with a factor $\frac{Gm}{c^3}$.
}
Written in terms of spin tensors satisfying the covariant spin supplementary condition, they read
\begin{eqnarray}
\label{Eq::Mass2}
\fl
{I}_{ij} &=& 
% \colgray{\mu} \, 
{\frac{1}{c^4}}
{\eta \, G^2 \, m^3} \, {\rm STF}_{ij}\biggl \{ x^{ij}
        \biggl [1
       + \frac{1}{42\,c^2}\;\left( (29-87\eta)v^2 -
           (30-48\eta)\frac{1}{r}\right )
\neanl \fl
       &&+ \frac{1}{c^4}\left (\frac{1}{504} (253-1835\eta +
           3545\eta^2) v^4\right.
        + \left.\frac{1}{756} (2021-5947\eta -4883\eta^2)
           \frac{1}{r}\,v^2\right.\neanl \fl
       &&- \left.\frac{1}{756} (131-907\eta +1273\eta^2)
           \frac{1}{r}\,\dotrel^{2}\right.
        - \left.\frac{1}{252} (355+1906\eta-337\eta^2)
           \frac{G^2m^2}{r^2}\right )\biggr ]\neanl \fl
       &&- x^{i}v^{j}\left[\frac{r\dotrel}{42\,c^2}
         (24-72\eta)\right.
        + \left.\frac{r\dotrel}{c^4} \left( \frac{1}{63}
           (26-202\eta +418\eta^2) v^2\right.\right.\neanl \fl
       &&\left.\left.+\frac{1}{378} (1085-4057\eta -1463\eta^2)
           \frac{1}{r}\right )\right] \neanl \fl
       &&+ v^{ij}\left[ \frac{r^2}{21\,c^2} 
       (11-33\eta)\right.
        + \left.\frac{r^2}{c^4} \left (\frac{1}{126} (41-337\eta
           +733\eta^2)v^2\right.\right.\neanl \fl
       &&\left.\left.+\frac{5}{63} (1-5\eta
           +5\eta^2)\dotrel^2\right.\right.
        + \left.\left. \frac{1}{189}\, (742-335\eta -985\eta^2)
           \frac{1}{r}\right )\right]
\neanl \fl
&& + 
{
% \left(
{\frac{\delta_S}{c^2}} \frac{4}{3} \eta ^2 G^2 {m}^3
% \right)
 \,
{\rm STF}_{ij} \,
(2
   x_{i} v^{k}-{v}_{i}
   x^{k})
	\left(
	\spinHC{1}{j}{k}+\spinHC{2}{j}{k}
	\right)
}
\neanl \fl && %=======================
+ {\frac{\delta_S^2}{c^2}}{\frac{1}{2} {\rm STF}_{ij} G^2 m^3 
\biggl({\CqOne} \left(\left(\sqrt{1-4 \eta
   }+3\right) \eta -\sqrt{1-4 \eta }-1\right) 
	\tilde{S}_1^{i}
	\tilde{S}_1^{j}
}
\neanl \fl && \qquad \qquad %=======================
{
+{\CqTwo} \left(-\left(\sqrt{1-4 \eta }-3\right)
   \eta +\sqrt{1-4 \eta }-1\right) 
	\tilde{S}_2^{i}
	\tilde{S}_2^{j}\biggr)}\biggr \}
\,,
\eanl \fl %============================================
% {\rm with ~~}
\tilde{S}_1^{i} \tilde{S}_1^{j} &=&
\frac{1}{2} \delta_{ij}
\left[
\spinHC{1}{m}{k}
\spinHC{1}{m}{k}
\right]
-
\spinHC{1}{i}{k}
\spinHC{1}{j}{k}\,,
\neanl \fl %============================================
 {I}_{ijk} &=& 
% \colgray{   -\left(\mu\, \frac{\delta_m}{m} \right)}
{-{\frac{1}{c^6}}\eta \, \sqrt{1-4 \eta } \,  G^3 {m}^4}
          {\rm  STF}_{ijk}\Biggl\{
        x^{ijk}\, \Bigl[ 1 + \frac{1}{6\,c^2}
           \Bigl( (5-19\eta )v^2
  -(5-13\eta ) \frac{1}{r} \Bigr) \Bigr]
\neanl \fl
       &&- x^{ij}v^k\left [\frac{r\dotrel}{c^2}
          (1-2\eta )\right ]
       + x^iv^{jk}\left [\frac{r^2}{c^2} 
       (1-2\eta )\right ] \Biggr\}
\,,
\eanl \fl %============================================
{
{I}_{ijkl}
}
	&=& 
% \colgray{{\mu}}
\,
{\frac{1}{c^8}}{(1-3 \eta ) \eta  G^4 {m}^5 }
 \,{\rm STF}_{ijkl}
             \biggl\{ 
	 x^{ijkl}\,\left.\biggl[(1-3\eta ) 
            \right.  
\neanl \fl && %====================================
            +\frac{1}{110\,c^2} 
	     \biggl( (103-735\eta
            +1395\eta^2) v^2  
        - (100-610\eta +1050\eta^2)
            \frac{1}{r}  \biggr)\biggr] 
\neanl \fl
        &&- v^ix^{jkl}\left \{\frac{72\,r\,{\dot r}}{55\,c^2} 
	     (1-5\eta +5\eta^2) \right \}
        + v^{ij}x^{kl}\left \{\frac{78\,r^2}{55\,c^2}
	(1-5\eta +5\eta^2) \right \}\biggr\}
\,,
\eanl \fl %===============================================
\label{Eq::Mass5}
{I}_{ijklm} &=& 
% \colgray{-\left( \mu \,\frac{\delta_m}{m} \right)\,( 1-2\eta)}
{   \frac{1}{c^{10}}    }{\sqrt{1-4 \eta } \eta  (2 \eta -1) G^5 {m}^6}
             {\rm STF}_{ijklm}\left \{ x^{ijklm}\right \}
\,,
\eanl \fl
\label{Eq::Mass6}
{I}_{ijklmn} &=& {   \frac{1}{c^{12}  }  } \mu (1-5\eta +5\eta^2 )
            {\rm STF}_{ijklmn}\left \{  x^{ijklmn}\right \}
\,,
\eanl \fl %===============================================
%===============================================
\label{Eq::Current2}
 {J}_{ij} &=& 
% \colgray{-\left(\mu\,\frac{\delta_m}{m} \right)}
{   \frac{1}{c^5}  }  {\sqrt{1-4 \eta } \eta  G^2 m^3}
             {\rm  STF}_{ij} \epsilon_{jab} \Biggl\{
 x^{ia}v^b\, \Bigl[1 + \frac{1}{28\,c^2}
           \Bigl( (13-68\eta )v^2 
        + (54+60\eta )\frac{1}{r}\Bigr) \Bigr] \neanl \fl
       &&+ v^{ib}x^a \left [\frac{r\,{\dot r}}{28\,c^2} (5-10\eta )
            \right ] \Biggr\}
\neanl \fl
&& + {\frac{{\delta_S} }{{c^5}} \frac{3}{8} \, \eta \, G^2 \, m^3 \,
{\rm  STF}_{ij}
\Biggl\{
   x^{i}
   \epsilon^{j k m}
   \left[
\bigl(\sqrt{1-4 \eta }+1 \bigr)
\spinHC{1}{k}{m}
+
\bigl(\sqrt{1-4 \eta }-1 \bigr)  
\spinHC{2}{k}{m}
\right]
\Biggr\}
}
\,, 
\eanl \fl %===============================================
 {J}_{ijk} &=&
% \colgray{  -\left(\mu\,\frac{\delta_m}{m} (1-2\eta )\right )}
{\frac{1}{c^7} }{\eta \, (3 \eta -1) \, G^3 \, m^4}
            {\rm STF}_{ijk}\left \{
            \epsilon_{kab}\, x^{aij}v^b \right \}
+{\frac{1}{c^9}} \eta\,
G^3 { m } ^4 {\rm STF}_{ijk} \biggl\{ 
   -\frac{2}{9} \left(5 \eta ^2-5 \eta +1\right)
   \epsilon_{k m n}
   x^i v^{m} x^{n} v^{j}
   \dot{r}{r}
\neanl \fl && %===============================================
-\frac{7}{45} \left(5 \eta ^2-5 \eta +1\right)
   { r ^2 }
   \epsilon_{kmn}
 v^{i} v^{j} {v}^{m} x^{n}
\neanl \fl && %===============================================
+\frac{2}{9 r} \left(43 \eta ^2+8 \eta -7\right)
   {\epsilon}_{k m n}
   v^{m} x^{n}  x^{i} x^{j}
\neanl \fl && %===============================================
-\frac{1}{90} \left(925 \eta
   ^2-385 \eta +41\right) 
   \epsilon_{kmn}
   x^{i} v^{m} x^{n}
   x^{j}
   \VelSq
\biggr\}
\,, \eanl \fl %===============================================
 {J}_{ijkl} &=&  
% \colgray{\left ( \mu\, (1-5\eta +5\eta^2)\right )}
{  \frac{1}{c^{9}  }   }{\sqrt{1-4 \eta } (1-2 \eta ) \, \eta \, G^4 m^5}
 \,
             {\rm STF}_{ijkl}\left \{
            \epsilon_{lab}\, x^{aijk}v^b\right \}
\,, \eanl \fl 
\label{Eq::Current5}
 {J}_{ijklm} &=&
% \colgray{ \mu\, (1-5\eta +5\eta^2)} \,
{  \frac{1}{c^{11}  }   }{\eta \, (1-5\eta +5\eta^2) \, G^5 \, m^6} \,
             {\rm STF}_{ijklm}\left \{
            \epsilon_{mab}\, x^{aijkl}v^b\right \}
\,,
\end{eqnarray}
where the symbol ${\rm STF}_{i_1...i_k}$ labels the symmetric and trace-free part
of an expression with respect to its indices $i_1,\dots,i_k$.
We do have access to higher-order accelerations in the ADMTT gauge, as we
use Poisson brackets to compute time derivatives of separations and spins.
We need a contact transformation that arbitrates between
the covariant spin supplementary condition and canonical spin variables on the one hand and
between harmonic and ADM gauge on the other. We give the relevant 
transformation terms in the subsequent section.

\subsection{Transformation from harmonic to ADM and from covariant to canonical spin variables}
\newcommand{\ADMCanGauge}{{\rm A}} % TODO: export into macros.tex
\newcommand{\HarmCovGauge}{{\rm HC}}
The coordinates transform from ADM to harmonic gauge (subscript ``\ADMCanGauge'') according to \cite{Gopakumar:Iyer:1997},
and from the canonical spin variables to covariant spin supplementary condition (subscript ``\HarmCovGauge'') as in, e.g. \cite{Kidder:1995, Damour:Jaranowski:Schafer:2008:1},
\begin{eqnarray}
\label{Eq::Gauge_X}
\fl
x_{\HarmCovGauge}^i &=&
x_{\ADMCanGauge}^i
-
{ 
\frac{1}{2 \, c^2} \delta_S \, \eta \, v_{\ADMCanGauge}^{k}
   (\spin{1}{i}{k}+\spin{2}{i}{k}
)
}
\neanl \fl
&&
+
\frac{1
}{8 \, c^4 \, \rel^3_{\ADMCanGauge}}
\Bigl\{
x_{\ADMCanGauge}^{i}
   \left(
-\eta \, 
% \scpm{ \vct{v} }{ \vct{x} }
% \scpm{ \vct{v} }{ \vct{x} }
\rel^2_{\ADMCanGauge} \, \dotrel^2_{\ADMCanGauge}
+5 \eta  \rel^2_{\ADMCanGauge} 
% \scpm{\vct{v}  }{  \vct{v}   }
v^2_{\ADMCanGauge}
+24 \, \eta \, \rel_{\ADMCanGauge}+2 \rel_{\ADMCanGauge}
\right)
-18 \, \eta 
   \rel^2_{\ADMCanGauge} \, v^i_{\ADMCanGauge}
% \scpm{  \vct{v}  }{ \vct{x} }
\rel_{\ADMCanGauge} \, \dotrel_{\ADMCanGauge}
\Bigr\}
\,,
\eanl \fl %=====================================
\label{Eq::Gauge_T}
t_{\ADMCanGauge} &=& t_{\HarmCovGauge} + \frac{1}{c^4} \, \eta \, \dotrel_{\ADMCanGauge}
\,,
\eanl \fl %=====================================
{v}_{\HarmCovGauge}^{i}
&=&
\diffq{x_\HarmCovGauge^{i}}{t_\HarmCovGauge}
=
\left( \diffq{x_{\HarmCovGauge}^i }{t_{\ADMCanGauge}} \right)
\cdot
\left( \diffq{t_{\ADMCanGauge}}{t_{\HarmCovGauge}} \right)
\neanl \fl
&=&
v_{\ADMCanGauge}^{i}
+
% \frac{\cInv{2}  \, {\delta_S} \,
%    \eta \, \nunit{n} \,
%    (\sOne{j} +\sTwo{j}) 
%    \epsilon^{i}_{~ n,j}
% }{2 \rel^2}
% 
{
{ \frac{\delta_S}{c^2} } \frac{ \eta }{2 \,\rel^2_{\ADMCanGauge}}
 \nunit{k}{}_{\ADMCanGauge}
  \left[ 
	\spin{1}{i}{k}+\spin{2}{i}{k}
	\right]
}
\neanl \fl
&&+
\frac{1 }{8 \, c^4}
\Biggl\{
\nunit{i}{}_{\ADMCanGauge}
 \left(\eta 
   \left[
\frac{
 3 
\dotrel^2_{\ADMCanGauge}-7 
\vel{2}_{\ADMCanGauge}
}{\rel_{\ADMCanGauge}}
-\frac{38
}{\rel^2_{\ADMCanGauge}}\right]
-\frac{4}{\rel^2_{\ADMCanGauge}}\right)
\dotrel_{\ADMCanGauge}
% \neanl \fl
% && 
% \qquad \qquad
+\vel{i}_{\ADMCanGauge}
   \left(\eta  
\left[
\frac{9 \dotrel^2_{\ADMCanGauge}-5 \vel{2}_{\ADMCanGauge} }{\rel_{\ADMCanGauge}}
+\frac{34}{\rel^2_{\ADMCanGauge}}
\right]
+\frac{2}{\rel^2_{\ADMCanGauge}}\right)
\Biggr\}
\,.
% \Biggr)
\end{eqnarray}
% An additional transformation will be necessary for the multipoles to match with the
% canonical spin variables:
% \begin{equation}
% bla
% \end{equation}
% Combining both will generate a set of all necessary transformations.
% \begin{eqnarray}
% Satz von Trafos.
% \end{eqnarray}
In the appendix of \cite{Tessmer:Hartung:Schafer:2010}, the transformation
of $\vct{v}$ was incorrectly implemented at 2PN point mass level. The gauge
term in {Eq. (A2) of \cite{Damour:Gopakumar:Iyer:2004}} has to be reproduced,
and finally reads
% \colred{alt:
% \begin{eqnarray}
% \fl
% {^{\bf g}}\xi_{\times,+}^{\rm (0+2)PP + PP} &=&
% % G^2 m^3
%  \eta \left\{
% \frac{\OutXPnn}{4 \, \rel^3}  \left(\eta  \left(-55 \rel
%    \RadVel^2+5 \rel \VelSq+24\right)+2\right)
% \right.
% \neanl \fl  %======================================
% && +
% \left.
% \frac{\OutXPnv}{2 \,\rel^2}  \RadVel \left(\eta  \left(3
%    \rel \RadVel^2-7 \rel \VelSq-20\right)-4\right)
% +\frac{\OutXPvv}{2 \, \rel^2} \left(\eta  \left(9 \rel
%    \RadVel^2-5 \rel \VelSq+34\right)+2\right)
% \right\}
% \,.
% \end{eqnarray}
% }

\noindent
% \colora{Anmerkung fuer uns: Das ist der korrekte Term, der die $\xi$'s und NICHT das htt transformiert --
% dieses hier hat die Form von Gopu:Iyer:2002 Gl. (2.17), aber dort scheint ein $\mu$ vergessen worden zu sein -- Vgl. hier mit
% Gopakumar:Iyer:1997, Gl (5.3)! Die zus. ``2'' im Nenner kommt vom GW-Vorfaktor $\frac{2G\mu}{c^4 R}$.}
\begin{eqnarray}
\fl
{^{\bf g}}\xi_{\times,+}^{\rm (0+2)PP + PP} &=&
\frac{1}{2\rel}
\biggl\{
%  \nunit{i} \nunit{j}
\OutXPnn
\left(\frac{5 \eta  \left(\VelSq-11
   \RadVel^2\right)}{2 \rel}+\frac{12 \eta
   +1}{\rel^2}\right)
\neanl \fl  %======================================
&&
+
2 \OutXPnv
\left(
\frac{1}{2} \eta  \RadVel \left(3
   \RadVel^2-7 \VelSq\right) 
%    (\vel{i} \nunit{j}+\nunit{i}\vel{j})
  -\frac{2 (5 \eta +1) \RadVel
%    (\vel{i} \nunit{j}+\nunit{i}\vel{j})
   }{\rel}
\right)
% \neanl \fl  %======================================
% &&
+ 
%    \vel{i}\vel{j}
\OutXPvv
   \left(\eta  \left(9 \RadVel^2-5 \VelSq\right)+\frac{34
   \eta +2}{\rel}\right)
\biggr\}
\,.
\end{eqnarray}%
The operator ${\cal P}$ projects
the radiation field $h^{\rm TT}_{ij}$
onto the polarisation tensors,
the subscripts $n$ and $v$ means contraction of free indices with
those of the unit normal vector or the velocity, respectively --
see Section 7 of \cite{Tessmer:Hartung:Schafer:2010},
and for convenience, the superscript $(0+2)$PP+PP stands for
the 2PN point particle contribution solely of the leading-order
gravitational wave polarisation $\xi_{+,\times}$ with velocities and
distances modified according to Eqs. (\ref{Eq::Gauge_X}) and (\ref{Eq::Gauge_T}).
All the distances and velocities in Eqs. (\ref{Eq::Mass2})--(\ref{Eq::Current5})
have to be understood as harmonic variables with covariant
spin supplementary condition, and we express the results for
energy and angular momentum loss in ADM coordinates
from now on for the remaining sections of the paper.

The next section applies expressions of the far-zone
flux of energy and angular momentum of the binary
and provides differential equations of the binding energy

\subsection{Differential equations for the orbital elements}

In this section, we derive (orbital averaged) differential equations
for the loss of orbital energy and angular momentum in terms
of the energy itself and the radial eccentricity $e_r$. 

The balance equations for $\ang_i$ and $\NRG$ 
through 2PN read \cite{Thorne:1980}
\begin{eqnarray}
\fl
 \left( \frac{\rmd \NRG}{\rmd t} \right)_{\rm fz}
&=&
\frac{G}{c^5}
\Biggl\{
	  \frac{1}{5}\!\stackrel{(3)}{I}_{\!\! ij} \! \stackrel{{(3)}}{I}_{\!\! ij}
+\frac{1}{c^2}
\left[
		\frac{1}{189}\! \stackrel{(4)}{I}_{\!\!ijk} \! \stackrel{(4)}{I}_{\!\!ijk}
	+	\frac{16}{45}\! \stackrel{(3)}{J}_{\!\!ij}  \! \stackrel{(3)}{J}_{\!\!ij}
\right]
+\frac{1}{c^4}
\left[
		\frac{1}{9072}\!\! \stackrel{(5)}{I}_{\!\!ijkm} \! \stackrel{(5)}{I}_{\!ijkm}
	+	\frac{1}{84}\!\! \stackrel{(4)}{J}_{\!\!ijk} \! \stackrel{(4)}{J}_{\!ijk}
\right]
\Biggr\}
% \neanl \fl && +{\left(\frac{\rmd \NRG}{\rmd t}\right)_{\rm tail}}
\,,
\eanl \fl
%===============================
%===============================
%===============================
 \left( \frac{\rmd \ang_i}{\rmd t} \right)_{\rm fz}
&=&
\frac{G}{c^5}
\epsilon_{ipq}
\Biggl\{
\frac{2}{5} \!\stackrel{(2)}{I}_{\!\!pj} \! \stackrel{(3)}{I}_{\!\!qj}
% ============================
+ \frac{1}{c^2} \left[ 
	\frac{1}{63}	\!\stackrel{(3)}{I}_{\!\!pjk} \! \stackrel{(4)}{I}_{\!\!qjk}
+	\frac{32}{45}	\!\stackrel{(2)}{J}_{\!\!pj} \! \stackrel{(3)}{J}_{\!\!qj}
\right]% ============================
+ \frac{1}{c^2} \left[ 
	\frac{1}{2268}	\!\stackrel{(4)}{I}_{\!\!pjkl} \! \stackrel{(5)}{I}_{\!\!qjkl}
+	\frac{1}{28}	\!\stackrel{(3)}{J}_{\!\!pjk} \! \stackrel{(4)}{J}_{\!\!qjk}
\right]
\Biggr\}
% \neanl \fl && +{\left(\frac{\rmd \ang_i}{\rmd t}\right)_{\rm tail}}
\,,
\end{eqnarray}
where a superscript in round brackets, $(n)$, denotes the $n^{\rm th}$ time derivative.
Note: the spin-squared contribution to $I^{ij}$ are static in the sense that they contain 
neither velocities nor distances, and as one inserts the equations of motion it becomes 
clear that they do not contribute.
Without specifying the direction of the spins, the results for the instantaneous parts
%(superscript ''inst")
read
% (in ADM coordinates and canonical spin variables)
\begin{eqnarray}
\fl
\left(\frac{\rmd \NRG}{\rmd t}\right)_{\rm fz}%^{\rm inst}
&=&
\eta\,\Biggl(
-\frac{88 \dotrel^2}{15
   \rel^4}+\frac{32 \VelSq}{5 \rel^4}
% \neanl \fl %====================================
% &&
+ \frac{1}{c^2} \Biggl\{\frac{2 \dotrel^4 (687-620 \eta )}{35
   \rel^4}
+\dotrel^2 \left[\frac{\frac{5872}{105}-\frac{16 \eta }{7}}{\rel^5}+\frac{4
   (1392 \eta -1487) \VelSq}{105 \rel^4}\right]
-\frac{32 (4 \eta -1)}{105\rel^6}
\neanl \fl %====================================
&&
+\frac{64 (\eta -17) \VelSq}{21 \rel^5}
+\frac{\left(\frac{314}{21}-\frac{568\eta }{35}\right)
 v^4}{\rel^4}
\Biggr\}
\neanl \fl %=====================
&&+
%  \frac{1}{c^2} 
\Biggl\{
\biggl[ \frac{4}{15
   \rel^5} \left(9 \dotrel^2 \left(-2 \eta +3 \sqrt{1-4 \eta
   }+3\right)+\left(36 \eta -37 \left(\sqrt{1-4 \eta }+1\right)\right) \VelSq\right)
\neanl \fl && %=====================
+\frac{16 \left(2 \eta -3 \sqrt{1-4 \eta }-3\right)}{15 \rel^6}
\biggr]
\times
% \neanl \fl && %=====================
\left(
\nunit{i} \vel{j} \spin{1}{i}{j}
\right)
\neanl \fl && %=====================
-\biggl[\frac{4}{15
   \rel^5} \left(9 \dotrel^2 \left(2 \eta +3
   \sqrt{1-4 \eta }-3\right)+\left(-36 \eta -37 \sqrt{1-4 \eta }+37\right) \VelSq\right)
\neanl \fl && %=========================
-\frac{16 \left(2
   \eta +3 \sqrt{1-4 \eta }-3\right)}{15 \rel^6}
\biggr]
\times
\left(
 \nunit{i} \vel{j} \spin{2}{i}{j}
\right) \Biggr\}\frac{\delta_S}{c^2}
\neanl \fl && %=====================
%=====================
%=====================
%=====================
%=====================
+
\frac{\delta_S^2}{c^2 \rel^6}
\Biggl\{
-\frac{1}{5} \left(-2 \eta +\sqrt{1-4 \eta}+1\right) \left((272 \CqOne+3)
\RadVel^2-168 \CqOne
\VelSq
\right)
\left(
\nunit{i}
\nunit{j}
% \spin{1}{-i}{k}
\SpinUD{1}{~}{(k)}{(i)}{~}
% \spin{1}{-j}{-k}
\SpinUD{1}{~}{~}{(j)}{(k)}
\right)
\neanl \fl && %============
+\frac{2}{5} (58
   \CqOne+1) \left(-2 \eta +\sqrt{1-4 \eta }+1\right)
\RadVel
\left(
\nunit{i} 
\vel{k}
\spin{1}{i}{m}
\spin{1}{k}{m}
\right)
\neanl \fl && %============
-\frac{1}{15} (72
   \CqOne+1) \left(-2 \eta +\sqrt{1-4 \eta }+1\right)
\left(
\vel{i}
\vel{k}
\spin{1}{i}{m}
\spin{1}{k}{m}
\right)
\neanl \fl && %============
+\frac{2}{15} \left(-2 \eta
   +\sqrt{1-4 \eta }+1\right)
\biggl((78 \CqOne+3) 
	\RadVel^2
	+(1-72 \CqOne)
	\VelSq
\biggr)
\left(
\spin{1}{i}{k}
\spin{1}{i}{k}
\right)
\neanl \fl && %============
+\frac{1}{5}
   \left(2 \eta +\sqrt{1-4 \eta }-1\right) 
\biggl((272\CqTwo+3)\RadVel^2 - 168 \CqTwo \VelSq \biggr) 
\left(
\nunit{i}\nunit{m} 
\spin{2}{i}{k}
\spin{2}{m}{k}
\right)
\neanl \fl && %============
-\frac{2}{5} (58\CqTwo+1) \left(2 \eta +\sqrt{1-4 \eta }-1\right)   \RadVel
\left(
\nunit{i} 
\vel{k}
\spin{2}{i}{m}
\spin{2}{k}{m}
\right)
\neanl \fl && %============
+\frac{1}{15} (72
   \CqTwo+1) \left(2 \eta +\sqrt{1-4 \eta }-1\right)
\left(
\vel{i} \vel{k}
\spin{2}{i}{m}
\spin{2}{k}{m}
\right)
\neanl \fl && %============
-\frac{2}{15} \left(2 \eta
   +\sqrt{1-4 \eta }-1\right)
\left(
\spin{2}{i}{k}
\spin{2}{i}{k}
\right)
   \left((78 \CqTwo+3) 
\RadVel^2
+(1-72 \CqTwo)
   \VelSq \right)
\neanl \fl && %============
+\frac{4}{5}
   \eta  \left(168 \VelSq -269 \RadVel^2\right) 
\left(
\nunit{i}
\nunit{k}
\spin{1}{i}{m}
\spin{2}{k}{m}
\right)
% \neanl \fl && %============
+\frac{228}{5} \eta 
   \RadVel \,
\left(
\nunit{i} 
\vel{k}
\spin{2}{i}{m}
\spin{1}{k}{m}
\right)
\neanl \fl && %============
+\frac{228}{5} \eta 
\RadVel \,
\left(
\nunit{i}
\vel{k}
\spin{1}{i}{m}
\spin{2}{k}{m}
\right)
% \neanl \fl && %============
-\frac{284}{15} \eta 
\left(
\vel{i}
\vel{k}
\spin{1}{i}{m}
\spin{2}{k}{m}
\right)
\neanl \fl && %============
+\frac{8}{15} \eta  \left(75
   \RadVel^2-73
   \VelSq \right)
\left(
\spin{1}{i}{k}
\spin{2}{i}{k}
\right)
\Biggr\}
\neanl \fl && %=====================
%=====================
%=====================
%=====================
%=====================
+\frac{1}{c^4} \Biggl\{
-\frac{
4 \dotrel^6 \left(8404 \eta ^2-20234 \eta +2501\right)}{315 \rel^4}
+\dotrel^4 \biggl[\frac{4 (3 \eta  (2524 \eta -5069)+2018) \VelSq}{105
   \rel^4}
\neanl \fl && %=====================
-\frac{8 (\eta  (14290 \eta -49757)+33510)}{945 \rel^5}\biggr]
+\dotrel^2
   \biggl[
\frac{8 (\eta  (2165 \eta-6903)+4987) \VelSq}{105 \rel^5}
\neanl \fl && %=====================
-\frac{2 (6 \eta  (896 \eta +415)+105185)}{945 \rel^6}
% \neanl \fl && %=====================
-\frac{4 (2 \eta  (3146 \eta -5139)+1719)
   v^4}{105 \rel^4}\biggr]
\neanl \fl && %=====================
-\frac{16 (4 \eta -1) (14 \eta -253)}{945
   \rel^7}+\frac{2 (156 \eta  (28 \eta +321)+276937) \VelSq}{2835 \rel^6}
\neanl \fl && %=====================
-\frac{8
   (\eta  (1393 \eta -4292)+4446) v^4}{315 \rel^5}+\frac{4 (\eta  (4430 \eta
   -5497)+1692) v^6}{315 \rel^4}
\Biggr\}
\Biggr)\,,
\end{eqnarray}

\begin{eqnarray}
\fl
 \left( 
\frac{\rmd \ang^i}{\rmd t} 
\right)%^{\rm inst}
_{\rm fz}
&=&
\frac{16}{5} \eta \, \nunit{k} v_m
   \epsilon^{i}_{~ k m}
\Biggl[
\Biggl\{
-\frac{3 \RadVel^2}{2
   \rel^2}+\frac{1}{{\rel}^3}+\frac{{\VelSq}}{{\rel}^2}
\Biggr\}
% \neanl \fl && %=======================
+\frac{1}{c^2}
% \frac{16}{5} \eta  \nunit{k} v^{m}  \epsilon^{i}_{~km}
\Biggl\{
\frac{\left(\frac{95}{56}-\frac{45 \eta }{7}\right)
   \RadVel^4}{\rel^2}
+\RadVel^2
   \left(\frac{\frac{197 \eta
   }{84}+\frac{31}{7}}{\rel^3}+\frac{(277 \eta -74)
   \VelSq}{28 \rel^2}\right)
\neanl \fl && %=======================
+\frac{2 \eta -745}{84 \rel^4}
+\frac{(-95 \eta -58) \VelSq}{42\rel^3}
+\frac{(307-548 \eta ) v^4}{168\rel^2}
\Biggr\}
+\frac{1}{c^4}
% \frac{16}{5} \eta  \nunit{k} v^{m}  \epsilon^{i}_{~km}
   \Biggl\{   -\frac{5 \left(97 \eta ^2-163 \eta +39\right)
   \RadVel^6}{36 \rel^2}
\neanl \fl && %=======================
+\RadVel^4
   \left(\frac{-9695 \eta ^2+18529 \eta -22312}{1008
   \rel^3}+\frac{5 \left(3136 \eta ^2-3361 \eta +715\right)
   \VelSq}{336 \rel^2}\right)+\RadVel^2
   \Biggl(\frac{4587 \eta ^2-22519 \eta +8499}{504
   \rel^4}
\neanl \fl && %=======================
+\frac{\left(2551 \eta ^2+1266 \eta +21853\right)
   \VelSq}{1008 \rel^3}+\frac{\left(-15637 \eta
   ^2+12653 \eta -2246\right) v^4}{336
   \rel^2}\Biggr)+\frac{1386 \eta ^2+48915 \eta
   +168094}{4536 \rel^5}
\neanl \fl && %=======================
+\frac{\left(-3428 \eta ^2+12321
   \eta -10525\right) \VelSq}{504
   \rel^4}+\frac{\left(4022 \eta ^2-3641 \eta +165\right)
   v^4}{1008 \rel^3}+\frac{\left(12894 \eta
   ^2-12355 \eta +2665\right) v^6}{1008
   \rel^2}
   \Biggr\}
\Biggr]
\neanl \fl && %========================
+
{
\frac{16\eta}{5} \frac{\delta_S}{c^2}
}
\Biggl\{
v^i \Biggl(\left( \hat{\Delta}_i \nunit{i} \right) \left(\frac{15
   \sqrt{1-4 \eta } \RadVel^3}{4 \rel^3}+\RadVel \left(-\frac{33
   \sqrt{1-4 \eta }}{8 \rel^4}-\frac{9 \sqrt{1-4 \eta } \VelSq}{4
   \rel^3}\right)\right)
\neanl \fl && %========================
+\left( \hat{\Delta}_i v^{i} \right)
   \Biggl(-\frac{3 \sqrt{1-4 \eta } \RadVel^2}{2 \rel^3}
% \neanl \fl && %========================
+\frac{23
   \sqrt{1-4 \eta }}{8 \rel^4}
+\frac{3 \sqrt{1-4 \eta } \VelSq}{4
   \rel^3}\Biggr)
% \neanl \fl && %========================
   +\frac{1}{\rel^3}\left(\frac{7 \eta }{4}-\frac{3}{2}\right)
   \RadVel^2 \left( \hat{\Sigma}_i v^{i} \right)
\neanl \fl && %========================
+\left(\frac{\left(\frac{15}{4}-\frac{5 \eta }{2}\right)
   \RadVel^3}{\rel^3}+\RadVel \left(\frac{\frac{35 \eta
   }{12}-\frac{33}{8}}{\rel^4}+\frac{\left(\frac{5 \eta }{4}-\frac{9}{4}\right)
   \VelSq}{\rel^3}\right)\right) \left( \hat{\Sigma}_i \nunit{i} \right)
% \neanl \fl && %========================
+\left(\frac{\frac{23}{8}-\frac{31 \eta
   }{12}}{\rel^4}+\frac{\left(\frac{3}{4}-\frac{2 \eta }{3}\right)
   \VelSq}{\rel^3}\right) \left( \hat{\Sigma}_i v^{i} \right)
\Biggr)
\neanl \fl && %==================
+\nunit{i} \Biggl(
\left( \hat{\Delta}_i \nunit{i} \right) \left(\RadVel^2 \left(\frac{3 \sqrt{1-4 \eta }}{4
   \rel^4}-\frac{15 \sqrt{1-4 \eta } \VelSq}{4
   \rel^3}\right)+\frac{\sqrt{1-4 \eta }}{4 \rel^5}+\frac{25 \sqrt{1-4
   \eta } \VelSq}{8 \rel^4}+\frac{9 \sqrt{1-4 \eta } \VelSq^2}{4
   \rel^3}\right)
\neanl \fl && %==================
+\RadVel \left(\frac{3 \sqrt{1-4 \eta } \VelSq}{4
   \rel^3}-\frac{21 \sqrt{1-4 \eta }}{8 \rel^4}\right)
   \left( \hat{\Delta}_i v^{i} \right)
+\Biggl(\RadVel^2
   \left(\frac{\left(\frac{5 \eta }{2}-\frac{15}{4}\right)
   \VelSq}{\rel^3}+\frac{3}{4
   \rel^4}\right)+\frac{\frac{1}{4}-\frac{\eta
   }{6}}{\rel^5}+\frac{\left(\frac{25}{8}-\frac{17 \eta }{6}\right)
   \VelSq}{\rel^4}
\neanl \fl && %==================
+\frac{\left(\frac{9}{4}-\frac{5 \eta }{4}\right)
   \VelSq^2}{\rel^3}\Biggr) \left( \nunit{i} \hat{\Sigma}_i \right)
   -\frac{5 \eta  \RadVel^3 
   \left( \hat{\Sigma}_i v^i \right) }{4 \rel^3}
   +\RadVel \left(\frac{\frac{29 \eta
   }{12}-\frac{21}{8}}{\rel^4}+\frac{\left(\frac{\eta }{4}+\frac{3}{4}\right)
   \VelSq}{\rel^3}\right) 
\left( \hat{\Sigma}_i v^i \right)
\Biggr)
\neanl \fl && %==================
+\Delta^i \left(\RadVel^2 \left(\frac{25
   \sqrt{1-4 \eta }}{12 \rel^4}+\frac{3 \sqrt{1-4 \eta } \VelSq}{4
   \rel^3}\right)-\frac{\sqrt{1-4 \eta }}{4 \rel^5}-\frac{25 \sqrt{1-4
   \eta } \VelSq}{12 \rel^4}-\frac{3 \sqrt{1-4 \eta } v^4}{4
   \rel^3}\right)
\neanl \fl && %==================
+\Sigma^i \left(\frac{5 \eta  \RadVel^4}{4
   \rel^3}+\RadVel^2 \left(\frac{\frac{25}{12}-\frac{9 \eta
   }{4}}{\rel^4}+\frac{\left(\frac{3}{4}-2 \eta \right)
   \VelSq}{\rel^3}\right)+\frac{\frac{\eta
   }{6}-\frac{1}{4}}{\rel^5}+\frac{\left(\frac{7 \eta }{3}-\frac{25}{12}\right)
   \VelSq}{\rel^4}+\frac{\left(\frac{2 \eta }{3}-\frac{3}{4}\right)
   v^4}{\rel^3}\right)
\Biggr\}
\neanl \fl && %==================
+
\frac{\delta_S^2}{c^2}
\Biggl\{
% \colblu{\rm spin1}^2 \Biggl[
% \neanl \fl && %==================
\frac{1}{5 \rel^5}
\left(-2 \eta +\sqrt{1-4 \eta }+1\right) 
\eta  \epsilon^{i}_{~ m k}
   \biggl[
   \nunit{m} \biggl(\hat{S}_1^{k} \left((6 \CqOne+1) \SoneVel-6
   \CqOne \RadVel \SoneNunit  \right)
\neanl \fl && %==================
+v^{k} \left((24 \CqOne+1)
   \SoneSq-90 \CqOne \SoneNunit^2\right)\biggr)
   +(12 \CqOne+1) \SoneNunit {\hat S}_1^{m} v^{k}
   \biggr]
\neanl \fl && %==================
-\frac{6}{5 \rel^4} \CqOne \left(-2\eta +\sqrt{1-4 \eta }+1\right) \eta
   v^{k}
   \epsilon^{i}_{~ m k} 
   \biggl[
   {\hat S}_1^{m} \left(\SoneNunit
   \left(5 \RadVel^2-3 \VelSq\right)-\RadVel
   \SoneVel\right)
\neanl \fl && %==================
   -\nunit{m} \left(5 \SoneNunit^2 \left(7 \RadVel^2-2
   \VelSq\right)+\SoneSq \left(2 \VelSq-5 \RadVel^2\right)-15 \RadVel
   \SoneNunit \SoneVel+\SoneVel^2\right)
   \biggr] 
% \Biggr]
\neanl \fl && %==================
+
% \neanl \fl && %==================
\frac{6}{5 \rel^4} \CqTwo \eta  \left(2 \eta +\sqrt{1-4 \eta }-1\right) v^{k}
   \epsilon^{i}_{~ m k} \biggl[S_2^m
   \left(\StwoNunit
   \left(5 \RadVel^2-3 \VelSq\right)-\RadVel
   \StwoVel\right)
\neanl \fl && %==================
   -\nunit{m} \left(5 \StwoNunit^2 \left(7 \RadVel^2-2
   \VelSq\right)+\StwoSq \left(2 \VelSq-5 \RadVel^2\right)-15 \RadVel
   \StwoNunit \StwoVel+\StwoVel^2\right) \biggr]
\neanl \fl && %==================
-\frac{\eta }{5 \rel^5}
   \left(2 \eta +\sqrt{1-4 \eta }-1\right) \epsilon^{i}_{~ m k}
   \biggl[\nunit{m} \biggl(\hat{S}_2^k ((6 \CqTwo+1) \StwoVel-6
   \CqTwo \RadVel \StwoNunit)
\neanl \fl && %==================
   +v^{k} \left((24 \CqTwo+1)
   \StwoSq-90 \CqTwo \StwoNunit^2\right)\biggr)
+(12 \CqTwo+1) \StwoNunit
   {\hat S}_2^m v^{k}\biggr]
\neanl \fl && %==================
+
\eta ^2 \Biggl\{_1
   \frac{1}{5 \rel^5}
   \Biggl[_5
   2 
   \epsilon^{i}_{~ m k} \nunit{m} \biggl(-6
   \RadVel \StwoNunit \hat{S}_1^{k}+\hat{S}_2^k (5 \SoneVel-6
   \RadVel \SoneNunit)+5 \StwoVel {\hat S}_1^{k}
\neanl \fl && %==================
   -180 \SoneNunit
   \StwoNunit v^{k}+46 \SoneStwo v^{k} \biggr)+11
   v^{k} (\StwoNunit S_1^{m}+\SoneNunit
   {\hat S}_2^m)
   \Biggr]_5
\neanl \fl && %==================
   -\frac{1}{5 \rel^4}
   12 \epsilon^{i}_{~ m k} v^{k} 
   \Biggl[_4
   \nunit{m} \biggl(\SoneNunit
   \left(\StwoNunit \left(20 \VelSq-70 \RadVel^2\right)+15 \RadVel
   \StwoVel\right)
\neanl \fl && %==================
   +2 \SoneStwo \left(5 \RadVel^2-2
   \VelSq\right)+\SoneVel (15 \RadVel \StwoNunit-2
   \StwoVel)\biggr)
\neanl \fl && %==================
   +{\hat S}_1^{m} \left(\StwoNunit \left(5 \RadVel^2-3
   \VelSq\right)-\RadVel \StwoVel\right) + {\hat S}_2^m
   \left(\SoneNunit \left(5 \RadVel^2-3 \VelSq\right)-\RadVel
   \SoneVel \right)
   \Biggr]_4 
   \Biggr\}_1
\end{eqnarray}
The tail terms at leading order can be taken from \cite{Rieth:Schafer:1997}.
We use the quasi-Keplerian parameterisation from the previous sections to express the time dependent terms 
in terms of the eccentric anomaly $\EccAno$ when we specify to aligned spin vectors and orbital angular momentum.
The time average of energy loss is done using the following relation,
\begin{equation}
\label{Eq::dNRGdt}
 \left< \frac{\rmd \NRG}{\rmd t}  \right>_{\rm T}
 =
 \int_{\EccAno=0}^{\EccAno=2\pi}
 \frac{\rmd \NRG}{\rmd t}(\EccAno) \left[ \frac{\rmd t}{\rmd \EccAno} \right] \rmd \EccAno\,,
\end{equation}
where the term in square brackets results from the Kepler equation.
We can insert the above integration limits because the Kepler equation
possesses fixed points at $\EccAno=n\pi$, ($n\in \mathds{Z}$).
The squares of the velocity and the radial velocity as functions of $\EccAno$
enter via 
\begin{eqnarray}
\dot{\rel}&=& \frac{\rmd \rel}{\rmd \EccAno} \frac{\rmd \EccAno}{\rmd t} \,, \\
v^2 &=& \dot{\rel}^2 + \rel^2 \dot{\phi}^2 \,.
\end{eqnarray}

The orbital averaged differential equations show total agreement up to all orders
(purely including the instantaneous parts)
appearing in \cite{Gopakumar:Iyer:1997}, and symbolically read
\begin{eqnarray}
 \left< \frac{\rmd |E|}{\rmd t} \right>
 &=&
\left<
   {\cal L}_{\rm N}
+{\cInv{2}} {\cal L}^{\rm PP}_{\rm 1PN}
% +{\cInv{3}} {\cal L}^{\rm tail}_{\rm LO}
+{\cInv{2}}  {\cal L}^{\rm SO}_{\rm LO}
+{\cInv{2}}  {\cal L}^{\rm SS, S^2}_{\rm LO}
+{\cInv{4}}  {\cal L}^{\rm PP}_{\rm 2PN} 
\right> \,, \\
\left< \frac{\rmd e_r}{\rmd t} \right>
 &=&
\left<   {\cal D}_{\rm N}
+{\cInv{2}} {\cal D}^{\rm PP}_{\rm 1PN}
% +{\cInv{3}} {\cal D}^{\rm tail}_{\rm LO}
+{\cInv{2}}  {\cal D}^{\rm SO}_{\rm LO}
+{\cInv{2}}  {\cal D}^{\rm SS, S^2}_{\rm LO}
+{\cInv{4}}  {\cal D}^{\rm PP}_{\rm 2PN} 
\right> \,,\\
\frac{\rmd \MeAno}{\rmd t} &=& \MeMo (|E|,e_r) \,.
\end{eqnarray}
The symbol
$\langle \dots \rangle$ means average over {\em one orbital (radial) time period}.
The spin-orbit terms match with
\cite{Zeng:Will:2007}; the S(1)S(2) terms match \cite{Wang:Steinhoff:Zeng:Schafer:2011},
where we checked the aligned spins case.
% The tail contributions ${\cal L}^{\rm tail}_{\rm LO}$ and ${\cal D}^{\rm tail}_{\rm LO}$
% are added manually to the above equations in the source code. 
% Due to the lengthy form of the ${\cal L}$ and ${\cal D}$ coefficients, we will not provide them here.
The explicit expressions read
\begin{eqnarray}
\label{Eq::dEdt_avraged}
\fl
 \left< \frac{\rmd \NRG}{\rmd t} \right>
&=&
\frac{32 \left(37 e_r^4+292 e_r^2+96\right) \eta  \NRG^5}{15 \ORS^7}
% \neanl \fl && %==========================================
+\frac{1}{c^2}
\Biggl\{
\frac{8 \eta  \NRG^6}{105 \ORS^9}
\bigl(
	e_r^6 (518 \eta -5377)-378
	e_r^4 (27 \eta +290)
\neanl \fl && %==========================================
-8 e_r^2 (6419 \eta +12828)-13440
   \eta +208
\bigr)
\Biggr\}
\neanl \fl && %==========================================
+\frac{\delta_S}{c^2}
\frac{\eta  \NRG^6}{\ORS^{11}}
\Biggl\{
\left(\chi_1-\chi_2\right)
{\sqrt{1-4 \eta }}
 \biggl(-\frac{32}{15} \Ang \left(195
   e_r^6+3810 e_r^4+5480 e_r^2+784\right)  \NRG
\neanl \fl && %==========================================
-\frac{64}{15 \Ang} \left(111 e_r^6+1497 e_r^4-352
   e_r^2-1256\right) e_r^2
+\frac{256 \sqrt{2\NRG}}{15 \ORS}
 \left(201 e_r^6+427
   e_r^4-484 e_r^2-144\right) 
\biggr)
\neanl \fl &&  %==========================================
+\left(\chi _1+\chi _2\right)
   \biggl(\frac{32}{15} \Ang \NRG \left(3 e_r^6 (66
   \eta -65)+6 e_r^4 (616 \eta -635)+8 e_r^2 (638 \eta
   -685)+704 \eta -784\right)
\neanl \fl &&  %==========================================
+\frac{32 e_r^2 (\eta -2)}{15\Ang}
	\left(111 e_r^6+1497 e_r^4-352 e_r^2-1256\right)
-\frac{128 (\eta -2)
   \sqrt{2\NRG}}{15 \ORS}
 \biggl(201 e_r^6+427e_r^4
\neanl \fl && %==========================================
-484 e_r^2-144 \biggr) 
\biggr)
\Biggr\}
\neanl \fl && %==========================================
+\frac{\delta_S^2}{c^2}
\frac{\eta  \NRG^6}{\ORS^{11}}
\Biggl\{
\chi_1^2
 \biggl(\frac{1}{\Ang^2}
\biggl[
\frac{16}{15} \CqOne e_r^2 \left(111
   e_r^6+1497 e_r^4-352 e_r^2-1256\right) \sqrt{1-4
   \eta }
\neanl \fl && %==========================================
-\frac{16}{15} \CqOne e_r^2 \bigl(111
   e_r^6+1497 e_r^4
% \neanl \fl && %==========================================
-352 e_r^2-1256 \bigr) (2 \eta
   -1)
\biggr]
\neanl \fl && %==========================================
+\NRG \biggl(\sqrt{1-4 \eta }
   \Bigl(\frac{32}{15} \CqOne \left(177 e_r^6+4368
   e_r^4+7768 e_r^2+1344\right)+\frac{8}{5} \bigl(9
   e_r^6
\neanl \fl && %==========================================
+138 e_r^4+152
   e_r^2+16\bigr)\Bigr)-\frac{32}{15} \CqOne \left(177
   e_r^6+4368 e_r^4+7768 e_r^2+1344\right) (2 \eta
   -1)
\neanl \fl && %==========================================
-\frac{8}{5} \left(9 e_r^6+138 e_r^4+152
   e_r^2+16\right) (2 \eta -1)\biggr)\biggr)
\neanl \fl && %==========================================
+\chi _2^2
   \Biggl(
\frac{1}{\Ang^2}
\biggl[
-\frac{16}{15} \CqTwo \bigl(111 e_r^6+1497
   e_r^4-352 e_r^2-1256\bigr) e_r^2 (2 \eta
   -1)
-\frac{16}{15} \CqTwo \biggl(111 e_r^6+1497
   e_r^4
\neanl \fl && %==========================================
-352 e_r^2-1256 \biggr) e_r^2 \sqrt{1-4 \eta
   }
\biggr]
+\NRG \biggl(\sqrt{1-4 \eta }
   \biggl(-\frac{32}{15} \CqTwo \left(177 e_r^6+4368
   e_r^4+7768 e_r^2+1344\right)
\neanl \fl && %==========================================
-\frac{8}{5} \left(9
   e_r^6+138 e_r^4+152
   e_r^2+16\right)\biggr)-\frac{32}{15} \CqTwo \left(177
   e_r^6+4368 e_r^4+7768 e_r^2+1344\right) (2 \eta
   -1)
\neanl \fl && %==========================================
-\frac{8}{5} \left(9 e_r^6+138 e_r^4+152
   e_r^2+16\right) (2 \eta -1)\biggr)
\Biggr)
% \neanl \fl && %==========================================
+\chi _2 \chi _1
   \biggl(\frac{64  e_r^2 \eta }{15
   \Ang^2}
\left(111 e_r^6+1497 e_r^4-352
   e_r^2-1256\right)
\neanl \fl && %==========================================
+\frac{32}{15} \left(681 e_r^6+17058
   e_r^4+30616 e_r^2+5328\right) \eta 
   \NRG \biggr)
\Biggr\}
\neanl \fl && %==========================================
%==========================================
+\frac{1}{c^4}
\frac{\eta  \NRG^6 }{\ORS^9}
\Biggl\{
% \left(
\frac{16}{15 \Ang^2} \left(111
   e_r^4+1608 e_r^2+1256\right) e_r^2 (11 \eta
   -17)
\neanl \fl && %==========================================
-\frac{16 \Ang^2 \eta  (\eta+4) \NRG^2}{15 \left(e_r^2-1\right)^2}
\bigl(425e_r^6+2540 e_r^4+2024 e_r^2+192 \bigr)
\neanl \fl && %==========================================
+\frac{96 \sqrt{2 \, \NRG}}{5 \Ang}
   \left(37 e_r^6+597 e_r^4+580 e_r^2+32\right) (2
   \eta -5) 
\neanl \fl && %==========================================
-\frac{4 \NRG}{2835
   \left(e_r^2-1\right)}
 \biggl(9
   e_r^8 \left(3108 \eta ^2-78882 \eta +283685\right)
-18e_r^6 \bigl(23247 \eta ^2+777 \eta  (192 \ORS-305)
\neanl \fl && %==========================================
-372960
   \ORS-2653303 \bigr)+12 e_r^4 \biggl(8253 \eta ^2-162 \eta 
   (19096 \ORS-47315)+7733880 \ORS+
\neanl \fl && %==========================================
2417662 \biggr)+32
   e_r^2 \left(201285 \eta ^2-63 \eta  (13896
   \ORS-21293)+2188620 \ORS+129133\right)
\neanl \fl && %==========================================
+32 \left(47628
   \eta ^2-153513 \eta +253937\right)\biggr)
% \right)
\Biggr\}
\,,
\end{eqnarray}

\begin{eqnarray}
\label{Eq::derdt_avraged}
\fl
\left< \frac{\rmd e_r}{\rmd t}\right>
&=&
-\frac{16 \NRG^4 e_r \left(121
   e_r^2+304\right) \eta }{15
   \left(1-e_r^2\right)^{5/2}}
\neanl \fl && %==========================================
+\frac{1}{c^2}
\Biggl\{
\frac{4 e_r \eta  \NRG^5}{105 \ORS^7}
   \left(e_r^4 (22221-3388 \eta )+36 e_r^2 (651 \eta
   +5020)+216 (252 \eta +67)\right)
\Biggr\}
\neanl \fl && %==========================================
+
\frac{\delta_S}{c^2}
\Biggl\{
\frac{8 \sqrt{2} e_r \eta  \NRG^{11/2} }{15 \ORS^8}
   \left(-623 e_r^4+2664 e_r^2+11896\right)
   \sqrt{1-4 \eta } 
   (\chi_1-\chi_2)
\neanl \fl && %==========================================
+\frac{8 \sqrt{2}
   e_r \eta  \NRG^{11/2}
}{15 \ORS^8}
 (\chi_1+\chi_2)
   \left(e_r^4 (92 \eta -623)+e_r^2 (2664-3948 \eta
   )-7680 \eta +11896\right)
\Biggr\}
\neanl \fl && %==========================================
+
\frac{\delta_S^2}{c^2}
\Biggl\{
\chi_1^2 \NRG^6
   \Biggl(
\frac{32 \CqOne e_r}{15 \ORS^9} \left(143
   e_r^4+2298 e_r^2+2900\right) \eta  (2 \eta
   -1)
% \neanl \fl && %==========================================
+\frac{12 e_r}{\ORS^9}
   \left(e_r^4+12 e_r^2+8\right) \eta  (2 \eta
   -1)
\neanl \fl && %==========================================
+\sqrt{1-4 \eta } \biggl[
-\frac{32\CqOne e_r \eta }{15 \ORS^9}
\left(143 e_r^4+2298
   e_r^2+2900\right)
-\frac{12 e_r}{\ORS^9}
\left(e_r^4+12 e_r^2+8\right) \eta
\biggr]
\Biggr)
\neanl \fl && %==========================================
-\frac{16 \chi_1 \chi_2}{15
   \ORS^9}
   e_r \left(1099 e_r^4+17844
   e_r^2+22840\right) \eta ^2 \NRG^6
\neanl \fl && %==========================================
+\chi_2^2 \NRG^6 \Biggl(
\frac{32
   \CqTwo e_r }{15
   \ORS^9}
\left(143 e_r^4+2298
   e_r^2+2900\right) \eta  (2 \eta -1)
+\frac{12 e_r}{\ORS^9} 
   \left(e_r^4+12 e_r^2+8\right) \eta  (2 \eta
   -1)
\neanl \fl && %==========================================
+\sqrt{1-4 \eta } \Biggl[
\frac{32 \CqTwo}{15 \ORS^9}
   e_r \left(143 e_r^4+2298 e_r^2+2900\right)
   \eta 
+\frac{12 e_r\eta   }{\ORS^9}
   \left(e_r^4+12 e_r^2+8\right)  \Biggr]
\Biggr)
\Biggr\}
\neanl \fl && %==========================================
-\frac{1}{c^4}
\frac{2 e_r \eta  \NRG^6}{945 \ORS^{10}}
\Bigl\{
9
   e_r^6 \left(3388 \eta ^2 \ORS-129131 \eta 
   \ORS+81312 \eta +367614
   \ORS-203280\right)
\neanl \fl && %==========================================
+e_r^4 \left(-755496 \eta ^2
   \ORS+9 \eta  (780827 \ORS+41664)+26914259
   \ORS-937440\right)
\neanl \fl && %==========================================
+12 e_r^2 \left(18753 \eta ^2
   \ORS+72 \eta  (63520 \ORS-3409)-2461036
   \ORS+613620\right)
\neanl \fl && %==========================================
+8 \left(300636 \eta ^2
   \ORS-293445 \eta  \ORS+229824 \eta +543431
   \ORS-574560\right)
\Bigr\}
\,.
\end{eqnarray}

\section{Conclusions}\label{sec:conclusions}
In this article, we have completed a previous work on (anti-)aligned spins in
a compact binary system
\cite{Tessmer:Hartung:Schafer:2010} by 
% next-to-next-to-leading order
\nnlo linear-in-spin effects and by spin dependent radiation reaction effects.
We provided expressions for the decay of a set of orbital elements,
namely the binding energy and the radial eccentricity $(|E|, e_r)$;
the reader may rewrite the above expressions to other sets of integrals,
e.g. to use $(\MeMo, e_t)$ instead of $(|E|, e_r)$.
The results for spin-squared far-zone flux at leading order show up to be conform with \cite{Poisson:1998}, where
the 
term of interest 
%leading-order spin-squared radiation reaction terms 
for the
binding energy come purely from the equations of motion and the
higher-order multipole moments rather that the leading-order
quadrupole term in the mass quadrupole.
% 
% All is needed to do so is to re-express the listed orbital elements and losses 
% perturbatively in terms of other ones to the desired order of spin or 
% post-Newtonian orders.
The 3PN point-mass contributions to the energy loss are not included, but can
be taken directly from the literature for a further publication, as well as the spin
contributions, as soon as they are available.

% It is up to the reader to complete the discussion about
% conservation of alignment by proving conservation of
% constraints in \cite{Tessmer:Hartung:Schafer:2010} to
% higher orders in spin-orbit and S(1)S(2) interactions,
% where in the case each spin vector points
% in the same direction as $\vct{\ang}$ nothing has to be shown.
For the discussion of the conservation of orbital angular momentum see 
\ref{subsec:erhaltungL}, and regarding the conservation of the spin
orientations, see \ref{subsec:parallelspins}. Several integrals necessary
for the QKP at formal 3PN order are provided in \ref{sec:quasikeplerdetails}.

A subsequent publication will discuss the approximate solutions
to the conservative dynamics in general orbits and arbitrarily orientated
spin axes under the influence of
spin(1)-spin(2) and spin-orbit interactions.
For a naive insight, one could -- for simplicity -- assume that the
time of observation of the binary is of the order of several
orbital revolutions (rather than the much larger spin precession
time scales) and assume the spin dependent terms to be approximately
constant. Then one is able to employ the methods for a calculation of
eccentric orbits from the literature which have been used here.
\ack
The authors wish to thank Jan Steinhoff for helpful discussions.
This work is partly funded by the DFG (Deutsche Forschungsgemeinschaft) through
SFB/TR7 ``Gravitationswellenastronomie'' and 
the Research Training Group GRK 1523 ``Quanten- und Gravitationsfelder''
and by the DLR (Deutsches Zentrum f\"ur Luft- und Raumfahrt)  through ``LISA Germany''. 

\appendix
\section{Dimensionless Quantities}\label{Sec::DimlessQuantities}
\label{SubSec::Scaling}

{\bf Everything} appearing in our prescription and the code is evaluated in scaled (dimensionless) quantities. The scaling is as follows:
\begin{eqnarray}
t		&=& \frac{G \, m}{c^3} \bar{t} \,, \\
E		&=& \bar{E} \mu c^2 \,, \\
% \vAng	&=& \crpm{\bar{\vct{x}}}{ \bar{\vct{v}} } \frac{G \, m}{c^2 } (\mu c)
% 				\hspace{1cm} \rm{(timescale cancels with velocity-timescale)}	\,,\\
r		&=& \bar{r} \frac{G \, m}{c^2 } \,, \\
\omega 	&=& \bar{x}^{3/2}\left(\frac{G\, m}{c^3}\right)^{-1} \,,\\
\vct{p}	&=& \bar{\vct{p}} \, c \mu \,, \\
\vct{S}_1	&=& \bar{\vct{S}}_1 \frac{G\, m_1^2}{c} \,, \\
\vct{S}_2	&=& \bar{\vct{S}}_2 \frac{G\, m_2^2}{c} \,,
\end{eqnarray}
where $\omega = {\rm d}\phi/{\rm d}t$ is the (unscaled) angular velocity.
The bars are, from now on, omitted: ''bared`` quantities on the right-hand sides
are understood as to be used in each case\footnote{If the spins are not aligned
or anti-aligned to $\vAng$, they precess due to the spin-orbit Hamiltonians.
In this case, the conservation of $\vct{J} = \vAng + \vSone + \vStwo$ can only
be applied in this form if the spins are scaled
the same way $\vAng$ is scaled! This will not affect the discussion regarding
the conservation of spins and $\vAng$ in \ref{subsec:erhaltungL}.}.
$|E|$, which will be mostly used later as the binding energy will increase
due to radiation reaction.
For a discussion about {\em formal} and {\em physical}
counting of the spin orders, see Ref. \cite[Sect. III]{Hartung:Steinhoff:2010}.

\section{
(Non-)conservation of 
Orbital Angular Momentum
Under Spin Interactions}\label{subsec:erhaltungL}
% Die L\"osung der Bewegungsgleichungen auf Ebene der Punktmassen (ohne Spin) gr\"undet sich darauf,
% dass der Gesamtdrehimpuls des Systems mit dem Orbitaldrehimpuls zusammenf\"allt und dieser deshalb
% in Betrag und Richtung erhalten bleibt. Dies liegt daran, dass

The solution to the equations of motion on point-mass level (without spin) foot on the fact that 
orbital angular momentum coincides with the total angular momentum, which obviously remains constant
in magnitude and direction. Also, it holds
\begin{eqnarray}
 \pb{\vmom{}^2}{\vAng} &=& 0\,,\\
 \pb{\scpm{\vnunit}{\vmom{}}}{\vAng} &=& 0\,,\\
 \pb{\rel}{\vAng} &=& 0\,.
\end{eqnarray}
As the Hamiltonian on point-mass level only depends on $\vmom{}^2$, $\scpm{\vnunit}{\vmom{}}$, and $\rel$
in the center-of-mass system, $H_{\text{PT}}(\vmom{}^2,\scpm{\vnunit}{\vmom{}},\rel)$,
the following conservation laws 
$\pb{\vAng}{H_{\text{PT}}(\vmom{}^2,\scpm{\vnunit}{\vmom{}},\rel)} = \dot{\vAng} = 0$
hold and, thus, conservation of $\vAng$ in amplitude and direction.
% 
% 
% 
% Unter Ber\"ucksichtigung der Spin-Bahn-Kopplungen ist die Situation eine andere. 
% Zum einen ist der Orbitaldrehimpuls nicht mehr gleich dem Gesamtdrehimpuls, sondern es
% gilt $\vct{J} = \vAng + \vspin{1} + \vspin{2}$ und dementsprechend ist $\vAng$ im Allgemeinen nicht mehr erhalten.
% Weiterhin hat diese Hamiltonfunktion eine grunds\"atzlich andere Struktur als die Punktteilchen-Hamiltonfunktion. Sie ist durch
% 
If the spin-orbit coupling is included, this situation changes.
On the one hand, the orbital angular momentum does not equal the total angular momentum, but $\vct{J} = \vAng + \vspin{1} + \vspin{2}$.
% and $\vAng$ is not conserved anymore. 
Furthermore, this Hamiltonian has a structure completely different to the one for point-masses,
\begin{eqnarray}
\fl {\rm H}_{{\rm SO}} &=& \scpm{\vAng}{\vspin{1}}\;f_1(\vmom{}^2,\scpm{\vnunit}{\vmom{}},\rel) + 
 \scpm{\vAng}{\vspin{2}}\;f_2(\vmom{}^2,\scpm{\vnunit}{\vmom{}},\rel)\,,
\end{eqnarray}
where $f_1$ and $f_2$ (corresponding to the point-mass Hamiltonian) are only functions of the listed arguments.
Therefore, the Poisson brackets of both $f_1$ and $f_2$ versus $\vAng$ vanish exactly and only the 
contributions of $\scpm{\vAng}{\vspin{a}}$ are relevant,
\begin{eqnarray}
 \dot{\vAng}_{\rm SO} &=& \pb{\vAng}{H_{\rm SO}} = - (\vAng\times\vspin{1})\;f_1- (\vAng\times\vspin{2})\;f_2\,.
\end{eqnarray}
% Hier sieht man eindeutig, dass der Orbitaldrehimpuls pr\"azediert, aber sein Betrag erhalten bleibt.
The precessional character of the orbital angular momentum becomes obvious as one realises
\begin{eqnarray}
 \diffq{}{t}(\vAng^2)_{\rm SO} &=& 2 \scpm{\vAng}{\dot{\vAng}_{\rm SO}} = -2 \scpm{\vAng}{(\vAng\times\vspin{1})}\;f_1 
 - 2 \scpm{\vAng}{(\vAng\times\vspin{2})}\;f_2 = 0\,.
\end{eqnarray}
% Die Spin(1)-Spin(2)-Kopplung verkompliziert die Analyse dahingehend, dass nun mehr algebraisch unterschiedliche Kombinationen m\"oglich sind.
% Im Schwerpunktsystem und mit Hilfe des Spintensors ergibt sich die allgemeinste Hamiltonfunktion
Spin(1)-spin(2) couplings complicate the analysis to the fact that
more algebraically different combinations of terms become relevant.
In the center-of-mass system, taking the spin tensor in favor of the spin vector
as an aid, the following general Hamilton function appears as
\begin{eqnarray}
\fl {\rm H}_{\text{S(1)S(2)}} &=&  (\spin{1}{i}{j}\spin{2}{i}{j})\;g_1 + (\nun{i}\mom{}{j}\spin{1}{i}{j})(\nun{i}\mom{}{j}\spin{2}{i}{j})\;g_2
	      + (\nun{i}\nun{j}\spin{1}{i}{k}\spin{2}{j}{k})\;g_3 \neanl &&
\fl	      + \biggl[(\nun{i}\mom{}{j}\spin{1}{i}{k}\spin{2}{j}{k}) + (\mom{}{i}\nun{j}\spin{1}{i}{k}\spin{2}{j}{k})\biggr]\;g_4 
	      + (\mom{}{i}\mom{}{j}\spin{1}{i}{k}\spin{2}{j}{k})\;g_5\,.
\end{eqnarray}
The functions $g_1$ \dots $g_5$ are, again, general functions of $\vmom{}^2$,$\scpm{\vnunit}{\vmom{}}$ and $\rel$,
which commute with $\vAng$.
\default{The following consideration will show that the amplitude of $\vAng$ is not conserved under those interactions.}
% Die Drehimpulsbewegungsgleichung,
The equation of motion for $\vAng$ following from that is very \default{long} and will not be provided.
% die hieraus folgt ist sehr lang und wird hier nicht angegeben. 
If one asks if $\dot{\vAng}$ is perpendicular to $\vAng$ (and, thus, $\ang$ might be a conserved quantity)
one sees that two relations between the $g$-functions must hold, namely
\begin{equation}
 g_3 \stackrel{!}{=} g_5 \vmom{}^2\,,\quad g_4 \stackrel{!}{=} -g_5 \scpm{\vnunit}{\vmom{}}\label{eq:gforderung}\,,
\end{equation}
where $g_1$ and $g_2$ stay arbitrary. 
Especially, this means that $g_1$ and $g_2$ do not contribute to $\dot{\ang}$.
For the S(1)S(2) interaction at leading order
the relations
\begin{equation}
 g_1 = \cInv{2}\frac{\eta}{\rel^3}\,,\quad g_3 = -\cInv{2}\frac{3\eta}{\rel^3}\,,\quad g_5 = 0\,,
\end{equation}
hold, which contravene Eq.~\eqref{eq:gforderung} and the conservation of $\vAng$.
The above arguments are similar for the spin($a$)$^2$ coupling, and one is lead to
conclude non-conservation of $\vAng$ for general configurations as well.

The situation for a compact object moving in the field generated by another
changes substatially if the spins are (anti)parallel to $\vAng$. This will
be discussed in the subsequent lines.

\section{Conservation of parallelism of Spins and Orbital Angular Momentum}\label{subsec:parallelspins}
The scenario of aligned spins is described in the literature as a consequence of
binaries moving in a dust-rich environment, see e.g. \cite{Bogdanovic:Reynolds:Miller:2007}.
In contrast to astrophysical considerations, we are especially interested in formal aspects
of the time evolution of this condition.
As in \cite{Tessmer:Hartung:Schafer:2010} shown through NLO in der spin-orbit interaction and
LO in S(1)S(2) or spin($a$)$^2$ interaction, respectively,  the configuration of spins
aligned to the orbital angular momentum is stable if they point in the direction of $\vAng$ from
the beginning on.
The general discussion of that issue can be performed following \cite[pp. 36]{Dirac:1964}:
% Wenn man in einen System von Differentialgleichungen $n$ Zwangsbedingungen
If one imposes a number of $n$ constraints, say
\begin{eqnarray}
 C_a(q, p) &=& 0\,,\quad a=1,\dots,n\,,\label{eq:parallelconstraints}
\end{eqnarray}
on a system of differential equations,
these constraints are conserved under the system's time evolution if
one can express their first-order time derivatives in the form
\begin{eqnarray}
 \dot{C}_a(q, p) &=& \sum_{b=1}^{n} D_{ab}(q,p) C_b(q,p)\,,\label{eq:parallelconstraintsdot}
\end{eqnarray}
% (also als Linearkombination der urspr\"unglichen Zwangsbedingungen) schreiben kann.
clearly speaking: as a linear combination of the original constraints.
If this form is achieved, successive time derivatives generate only $C_a$
in combination with derivatives of $D_{ab}$
and $\dot{C}_a$ which might be rewritten with the help of Eq. \eqref{eq:parallelconstraintsdot} again.
% Dies gilt entsprechend f\"ur alle h\"oheren
% zeitlichen Ableitungen der $C_a$.
% Setzt man nun die Zwangsbedingungen \eqref{eq:parallelconstraints}
% in beliebige Zeitableitungen ein, so verschwinden sie alle.
Each time derivative of the constraints (they all contribute to
a Taylor expansion around the instant of time $t=0$) will be, using
\eqref{eq:parallelconstraintsdot}, identically zero and warrant
conservation of the constraints if they especially hold at $t=0$.

What is left to show is that in our special case of aligned spins
the time derivatives of the constraints can indeed be written
in terms of Eq.~\eqref{eq:parallelconstraintsdot}.
They are given by
\begin{eqnarray}
\label{Eq::SpinConstraints}
 \vspin{a} - \chi_a \frac{\vAng}{\ang} &=& 0\,,\quad a=1,2\,.
\end{eqnarray}
Of course, it must hold $|\chi_a| = |\vspin{a}|$, 
for the constraints to be consistent.
% damit die Zwangsbedingungen konsistent sind.
% 
The total sign in case of antiparallelism might be absorbed into $\chi_a$.
% Das Vorzeichen, was entsteht, wenn $\vspin{a}$ antiparallel zu $\vAng$ orientiert ist,
% wird in die $\chi_a$ absorbiert.
Time derivation of Eq. \eqref{Eq::SpinConstraints}, together with constant spin  lengths
through the considered post-Newtonian order (also $|\chi_a| = |\vspin{a}| = \text{const}$)
% ,(see also \Subsec{subsec:dreibeinexpansion}) 
leads to, 
one obtains
\begin{eqnarray}
 \diffq{\vspin{a}}{t} + \frac{\chi_a}{\ang}\left(\mathds{1} - \frac{\vAng\otimes\vAng}{\ang^2}\right)\sum_{b} \diffq{\vspin{b}}{t} &=& 0\label{eq:dtspinconstr}\,.
\end{eqnarray}
It has been used $\vct{J} = \vAng + \vspin{1} + \vspin{2} = \rm{const}.$
so that $\dot{\vAng}$ can be directly expressed via $\dot{\vspin{}}_{a}$.

Now the $\dot{\vspin{}}_{a}$ must be expressed in terms of the constraints.
Because of the spin's constant amplitudes, their equations of motion can be expressed
as
% 
% Da die Spins auf diesen Ordnungen ihre L\"ange nicht \"andern, kann man die Bewegungsgleichungen in diesem Fall als Pr\"azessionsgleichungen
\begin{eqnarray}
 \diffq{\vspin{a}}{t} &=& \vct{\Omega}_a \times \vspin{a}\,.
\end{eqnarray}
Now one has to classify the possible appearance of $\vct{\Omega}_a$
and if one can reconstruct the constraints themselves.
In case of the spin-orbit coupling, through NNLO (and maybe also on higher orders)
in the center-of-mass system, spins only appear in combination with $\vAng$
in the scalars form $\scpm{\vAng}{\vspin{a}}$.
It follows
\begin{eqnarray}
 \vct{\Omega}_{{\rm SO}\,a} &\sim& \vAng\,,
\end{eqnarray}
where the proportionality factor depends on $\hat{\vct{p}}$. 
Now one is allowed to add an ''active zero`` to the equations of motion,
such that one is always enabled to write them in the form
% Nun ist es zul\"assig zu den Bewegungsgleichungen, einen verschwindenden Term hinzuzuaddieren,
% so dass man auf Spin-Bahn-Ebene diese immer in die Form
\begin{eqnarray}
 \diffq{\vspin{a}}{t} &=& \vct{\Omega}_{{\rm SO}\,a} \times \left(\vspin{a} - \chi_a \frac{\vAng}{\ang}\right)\,.
\end{eqnarray}
For the spin($a$)-spin($b$) interaction 
the argumentation is not that straightforward.
Here, more possible directions which $\vct{\Omega}_a$
may point to are allowed.
 For $\vct{\Omega}_a \sim \vspin{b}$, 
one can add the ''active zero`` according to
% kann man hier eine aktive Null gem\"a\ss
\begin{eqnarray}
 \diffq{\vspin{a}}{t} &=& \vct{\Omega}_{\text{SS}\,a} \times \vspin{a} = K(\vmom{}) \vspin{b}\times\vspin{a} \neanl &&
		      = K(\vmom{}) \left(\vspin{b} - \chi_b \frac{\vAng}{\ang}\right)\times\vspin{a} 
			  + \chi_b K(\vmom{}) \frac{\vAng}{\ang} \times \vspin{a} \neanl &&
		      = K(\vmom{}) \left(\vspin{b} - \chi_b \frac{\vAng}{\ang}\right)\times\vspin{a} 
			  + \chi_b K(\vmom{}) \frac{\vAng}{\ang} \times \left(\vspin{a} - \chi_a \frac{\vAng}{\ang}\right)\,,
\end{eqnarray}
where one treats the last term above as one does with the spin-orbit equation of motion.
% wobei man den letzten Term wie die Spin-Bahn-Bewegungsgleichung behandelt. 
The only vectors left, in whose direction
$\vct{\Omega}_a$ in case of spin($a$)-spin($b$) interaction can point,
are $\vnunit$ and $\vmom{}$. 
Here, simultaneously $\vspin{b}$ has to appear in a scalar.
Both are perpendicular to $\vAng$, such that vanishing terms
can be added in the form $\scpm{\vnunit}{\vAng}$,
% Diese stehen jedoch beide senkrecht auf $\vAng$, wodurch man
% ebenso hier einen verschwindenden Beitrag einf\"ugen kann, n\"amlich bei z.B.
\begin{eqnarray}
 \scpm{\vnunit}{\vspin{b}} &=& \scpm{\vnunit}{\vspin{b}} - \frac{\chi_b}{\ang} \scpm{\vnunit}{\vAng} 
 = \scpm{\vnunit}{\left(\vspin{b}-\chi_b\frac{\vAng}{\ang}\right)}\,,
\end{eqnarray}
analogously for $\scpm{\vmom{}}{\vspin{b}}$.
Those arguments for the spin-orbit and the spin($a$)-spin($b$) coupling
are still valid for spin($a$)$^2$,
hence the conservation of all the angular momenta in case of alignment,
% Damit ist gezeigt worden, dass sowohl bis zur NNFO in der Spin-Bahn-Kopplung
% und der S(1)S(2)-Kopplung als auch bis zur NFO in der Spin($a$)$^2$-Kopplung die Zwangsbedingung von Spins, die parallel
% zum Orbitaldrehimpuls stehen, erhalten bleibt, sofern man in dieser Konfiguration startet. 
% Hier wurde nur verwendet, dass die
% Spinl\"angen und der Gesamtdrehimpuls des Systems erhalten bleiben; auch die Struktur der Wechselwirkungen in der Hamiltonfunktion
% spielte eine Rolle. 
which generalises the proof given in \cite{Tessmer:Hartung:Schafer:2010}. 

From the conservation of $\vct{J}$ one can conclude that $\dot{\vAng}=0$ via
\begin{eqnarray}
 0= \dot{\vct{J}} &=& \dot{\vAng} + \dot{\vspin{}}_1 + \dot{\vspin{}}_2 = 
 \left[\left(1-\frac{\chi_1 + \chi_2}{\ang}\right)\mathds{1} + \frac{\chi_1 + \chi_2}{\ang}\frac{\vAng\otimes\vAng}{\ang^2}\right]\dot{\vAng}\label{eq:vangcons}\,,
\end{eqnarray}
(where the left hand side of \eqref{eq:dtspinconstr} was used) and the fact,
that the matrix acting on $\dot{\vAng}$ is invertible if $\chi_1 + \chi_2 \ne \ang$).
In the almost trivial spinless case, the matrix is the unit matrix and the well-known
result for point masses emanates.
% Im spinlosen Grenzfall ist die Matrix gleich der Einheitsmatrix und es ergibt sich das bekannte Resultat f\"ur Punktmassen.

\section{Selected Details of the Quasi-Keplerian parameterisation}\label{sec:quasikeplerdetails}

In \cite{Tessmer:Hartung:Schafer:2010} the calculation of the
orbital elements through formal 2PN has been carried out, having defined
the inverse radial distance $s \defdby \frac{1}{\rel}$ and the
corresponding values $\spl$ and $\smi$ at periastron and apastron.
At formal 3PN order, some new terms appear which we like to provide to the reader.
The definite integrals $I_n^\prime(\smi,\spl)$  are necessary ingredients 
for the calculation of radial period and Periastron advance \cite[Eqs. (52), (53) and (62)]{Tessmer:Hartung:Schafer:2010}.
The integrals with variable boundaries $I_n(a_r, e_r, u, \tilde{v})$ are
used for the preliminary Kepler equation \cite[Eqs. (54) and (60)]{Tessmer:Hartung:Schafer:2010} and
the temporary orbital phase in terms of $\tilde{v}$ (Eqs. (61) and (63)).

Through 3PN, only those integrals for $n=0 \dots 7$ are relevant and will be given next.

\subsection{Integrals for Radial Period and Periastron Advance}
\label{SubSec::IntegralsA}

% Die bestimmten Integrale sind durch
The definition of the $I^{\prime}_n$ is given by
\begin{eqnarray}
 I^{\prime}_n &=& 2\int_{\smi}^{\spl} \frac{\tau^n \rm{d}\tau}{\tau^2 \sqrt{(\tau - \smi)(\spl - \tau)}}\,,
\end{eqnarray}
% gegeben und ihre L\"osung ist
and the solutions read
\begin{eqnarray}
 I^\prime_0 &=&\frac{\pi  (\smi+\spl)}{(\smi \spl)^{3/2}}\,,\eanl
 I^\prime_1 &=&\frac{2 \pi }{\sqrt{\smi \spl}}\,, \eanl
 I^\prime_2 &=& 2 \pi\,,  \eanl
 I^\prime_3 &=&\pi  (\smi+\spl)\,, \eanl
 I^\prime_4 &=&\frac{1}{4} \pi  \left(3 \smi^2+2 \smi \spl+3 \spl^2\right)\,, \eanl
 I^\prime_5 &=&\frac{1}{8} \pi  (\smi+\spl) \left(5 \smi^2-2 \smi \spl+5 \spl^2\right)\,, \eanl
 I^\prime_6 &=&\frac{1}{64} \pi  \left(35 \smi^4+20 \smi^3 \spl+18 \smi^2 \spl^2+20 \smi \spl^3+35 \spl^4\right)\,, \eanl
 I^\prime_7 &=&\frac{1}{128} \pi  (\smi+\spl) \left(63 \smi^4-28 \smi^3 \spl+58 \smi^2 \spl^2-28 \smi \spl^3+63 \spl^4\right)\,.
\end{eqnarray}
 
\subsection{Integrals for Orbital Phase and Quasi-Kepler Equation}
\label{SubSec::IntegralsB}

% Die komplizierteren Integrale mit variablen Grenzen
The more complicated $I_n$ with variable boundaries
\begin{eqnarray}
 I_n &=& \int_{s}^{\spl} \frac{\tau^n \rm{d}\tau}{\tau^2 \sqrt{(\tau - \smi)(\spl - \tau)}}\,,
\end{eqnarray}
% durch % the eccentric anomaly
expressed by
$u$, % the temporary true anomaly 
$\tilde{v}$, %, the radial eccentricity 
$e_r$, and %the radial semimajor axis
$a_r$
%  ausgedr\"uckt, sind
read
\begin{eqnarray}
 I_0 &=& a_r^2 \sqrt{1-e_r^2}\,(u - \sin u)\,, \eanl
 I_1 &=& a_r \sqrt{1-e_r^2}\,u\,, \eanl
 I_2 &=& \tilde{v}\,, \eanl
 I_3 &=& \frac{\tilde{v}+e_r \sin \left(\tilde{v}\right)}{a_r \left(1-e_r^2\right)}\,, \eanl
 I_4 &=& \frac{2 (2 + e_r^2) \tilde{v}+8 e_r \sin \tilde{v}+e_r^2 \sin \left(2 \tilde{v}\right)}{4 a_r^2 \left(1-e_r^2\right){}^2}\,, \eanl
 I_5 &=& \frac{6 \left(2 + 3 e_r^2\right) \tilde{v} +9 e_r \left(4 + e_r^2\right)\sin \tilde{v}+9 e_r^2 \sin \left(2 \tilde{v}\right)+e_r^3 \sin \left(3 \tilde{v}\right)}{12 a_r^3 \left(1-e_r^2\right){}^3}\,, \eanl
 I_6 &=& \frac{1}{96 a_r^4 \left(1-e_r^2\right){}^4}\biggl(36 \tilde{v} e_r^4+288 \tilde{v} e_r^2+24 e_r^4 \sin \left(2 \tilde{v}\right)+3 e_r^4 \sin \left(4 \tilde{v}\right)\\\nonumber &&
  \quad+288 e_r^3 \sin \left(\tilde{v}\right)+32 e_r^3 \sin \left(3 \tilde{v}\right)+144 e_r^2 \sin \left(2 \tilde{v}\right)+384 e_r \sin \left(\tilde{v}\right)+96 \tilde{v}\biggr)\,, \eanl
 I_7 &=& \frac{1}{480 a_r^5 \left(1-e_r^2\right){}^5}\biggl(900 \tilde{v} e_r^4+2400 \tilde{v} e_r^2+300 e_r^5 \sin \left(\tilde{v}\right)+50 e_r^5 \sin \left(3 \tilde{v}\right)\\\nonumber &&
  \quad+6 e_r^5 \sin \left(5 \tilde{v}\right)+600 e_r^4 \sin \left(2 \tilde{v}\right)+75 e_r^4 \sin \left(4 \tilde{v}\right)+3600 e_r^3 \sin \left(\tilde{v}\right)\\\nonumber &&
  \quad+400 e_r^3 \sin \left(3 \tilde{v}\right)+1200 e_r^2 \sin \left(2 \tilde{v}\right)+2400 e_r \sin \left(\tilde{v}\right)+480 \tilde{v}\biggr)\,.
\end{eqnarray}
Note that the definite integrals above are computed on the real axis.
It is, in contrast, also possible to compute them by integrating
in the complex plane as done in \cite{Sommerfeld:1951}.
% \remark{An Johannes: sind die variablen Integrale dort auch so berechnet worden?}
% Alle hier angegebenen Integrale sind ausreichend, 
% um die Orbitalelemente auf formaler 3PN-Ebene zu berechnen.
% Beachte, dass die hier angegebenen Integrale im Reellen
% berechnet wurden. Es ist aber ebenso m\"oglich, diese durch
% eine komplexe Integration, wie sie in \cite[Anhang 4]{Sommerfeld:1951}
% angegeben wurde, zu berechnen. 
% Die dort angegebene Prozedur ist wesentlich einfacher und weniger
% aufwendig als eine reelle Integration; dennoch spielt dieses
% Detail keine gro\ss e Rolle, da durch {\sc Mathematica} beide
% Prozeduren fast gleich schnell ausgef\"uhrt werden k\"onnen.

\section*{Bibliography}

\bibliographystyle{utphys} % arXiv 
%\bibliographystyle{hunsrt} % arXiv CQG Style
%\bibliographystyle{unsrt} % (zulaessiges) CQG Style

% \bibliographystyle{h-physrev5} % Phys Rev
%\bibliography{../references}
\providecommand{\href}[2]{#2}\begingroup\raggedright\endgroup

\end{document}